\documentclass[useAMS,usenatbib,times,letter,amssymb]{mnras}
\usepackage{epsfig,times,amssymb,amsmath,verbatim,xspace}
\usepackage{widetext}
\usepackage[usenames,dvipsnames,svgnames,table]{xcolor}

\usepackage{todonotes}
\usepackage{tabularx}

\usepackage{hyperref}
\usepackage{float}

\usepackage{hhline}
\usepackage{lineno}
\usepackage{natbib}
\usepackage{hyperref,ifthen,soul} 

\newcommand{\omegam}{\ensuremath{\Omega_\mathrm{m}}}
\newcommand{\omegab}{\ensuremath{\Omega_\mathrm{b}}}

\newcommand{\omegade}{\ensuremath{\Omega_\mathrm{de}}}
\newcommand{\as}{\ensuremath{A_\mathrm{s}}}
\newcommand{\ns}{\ensuremath{n_\mathrm{s}}}
\newcommand{\lcdm}{$\Lambda$CDM}

\newcommand{\blockfont}[1]{{\textsc{#1}}\xspace}

\newcommand{\mb}{\blockfont{MassiveBlack-II}}
\newcommand{\illustris}{\blockfont{Illustris-1}}

\DeclareRobustCommand{\PRF}[3]{#2} % set up for citation

\title[Modelling IAs: The Impact of Aspherical Satellite Distributions]{Testing the Impact of Satellite Anisotropy on Large and Small Scale Intrinsic Alignments using Hydrodynamical Simulations}

\author[Samuroff et al]{
S.~Samuroff$^1$\thanks{ssamurof@andrew.cmu.edu},
R.~Mandelbaum$^1$
and T.~Di Matteo$^1$
\\
$^1$McWilliams Center for Cosmology, Department of Physics, Carnegie Mellon University, Pittsburgh, PA 15213, USA\\
}

\begin{document}

\maketitle

\begin{abstract}
Galaxy intrinsic alignments (IAs) have long been recognised as a significant contaminant
to weak lensing-based cosmological inference. In this paper we seek to quantify the impact of a common modelling assumption 
in analytic descriptions of intrinsic alignments: that of spherically symmetric dark matter halos.
Understanding such effects is important as the current generation of intrinsic alignment models are
known to be limited, particularly on small scales,
and building an accurate theoretical description will be essential for fully exploiting the
information in future lensing data.
Our analysis is based on a catalogue of 113,560 galaxies between $z=0.06-1.00$ from \mb, a hydrodynamical simulation of box length $100 h^{-1}$ Mpc.
We find satellite anisotropy contributes at the level of 
$\geq 30-40\%$ to the small scale alignment correlation
functions. At separations larger than $1 h^{-1}$ Mpc the impact is roughly scale-independent,
inducing a shift in the amplitude of the IA power spectra of $\sim20\%$.
These conclusions are %seen to be
consistent across the redshift range and between the \mb~and 
\blockfont{illustris}
simulations. 
The cosmological implications of these results are tested using a simulated likelihood analysis.
Synthetic cosmic shear data is constructed with the expected characteristics 
(depth, area and number density) of a future LSST-like survey.
Our results suggest that modelling alignments using a halo model based upon spherical symmetry could potentially
 induce  cosmological parameter biases at the $\sim 1.5\sigma$ level for $S_8$ and $w$.
% Note from RM: I reworded this a bit; I believe that saying that these forecasts presume an
% assumption of spherical symmetry may be sufficient to indicate that they are a worst-case
% scenario.  Also, I think that quoting the contamination level to 3 sig figs is kind of overkill,
% given the uncertainties in the setup we've adopted, so I simplified those statements.

\end{abstract}

\begin{keywords}
cosmology: theory — gravitational lensing: weak – large-scale structure of Universe — methods: numerical
\end{keywords}

% ------------- 1. Introduction ------------------------------------
\section{Introduction}\label{sec:intro}

In many ways the field of cosmology has changed irrevocably in the past decade. 
With large-volume measurements from a new generation of instruments, it has finally
become possible to test the predictions of theorists to meaningful precision.
So much is this the case that the distinction between ``theorist" and ``observer"
has increasingly less meaning; 
cutting edge cosmology is now a process of using ever more powerful datasets to
constrain, test, break and ultimately rebuild our models of the Universe.
This is particularly true of weak lensing cosmology, which began to flourish
somewhat later than the study of the Cosmic Microwave Background (CMB)
as a cosmological probe.
The transition into a data-led high-precision discipline has, by necessity,
seen an increased amount of time devoted to understanding the numerous,
often subtle, sources of systematic error.
Without such efforts, as a community, our ability to constrain the cosmological
parameters encoded in large scale structure will very quickly become limited
by systematics.
For an overview of recent developments in lensing cosmology we refer the reader to
contemporary reviews on the subject \citep{kilbinger15,2018ARA&A..56..393M}.

Often new methods have been developed to circumvent limitations in our ability to
deal with certain systematics.
One good example is the case of shear measurement bias, which is inherent to the process
of inferring galaxy ellipticities. A coherent community effort 
\citep{heymans06, massey07, bridle10, kitching10, mandelbaum15}
has seen the development of several
novel techniques, which can calibrate shear biases to sub-percentage level without
the need for massive high-fidelity image simulations \citep{bernstein16, huff17, sheldon17}. 
There are hopes that a combination of innovations such as multi-object fitting \citep{y1gold},
internal self-calibration 
(see \citealt{y1keypaper, joudaki18} for practical implementations)
and improved spectroscopic overlap will bring similar advances in the case of photometric redshift 
(photo-$z$) error.

Amongst various other lensing systematics, a phenomenon known as intrinsic alignments (IAs)
poses significant theoretical challenges for future analyses. 
Generally, the term ``intrinsic alignments" covers two slightly different physical
effects;
physically close pairs of galaxies will tend to align with each other
through interaction with local large scale structure, 
producing weak positive alignment (known as II correlations; 
\citealt{catelan01, crittenden01}).
Often the dominant form of IA contamination, however, comes in
the form of GI correlations \citep{hirata04}. 
These arise due to the fact that
mass on the line of sight
simultaneously lenses background objects and tidally interacts with
nearby galaxies.
There are a good many reviews on the subject of intrinsic alignment in the
literature, to which we refer the reader for more details
\citep{troxel15, joachimi15, kirk15, kiessling15}.

Unlike many potential sources of bias, 
IAs are inherently astrophysical in nature,
rather than the result of flaws in the measurement process.
No amount of ingenuity will alter the fact that such correlations
are present in the data and enter on similar angular scales to cosmic shear.
That neglecting intrinsic alignments
when modelling 
cosmic shear or galaxy-galaxy lensing will induce significant $(\gg 1\sigma)$ biases
in one's recovered cosmological parameters is now well established
\citep{krause16, blazek17}.
In practice the most feasible way of mitigating intrinsic alignments
is to simply model them out, 
including additional parameters in any likelihood analysis, which
are marginalised over with wide priors.
Unfortunately the efficacy of such a technique is somewhat limited
by the models in question; to marginalise out the impact of IAs without residual biases,
one requires a sufficiently accurate model to describe their impact.
Due to limitations in both theory and available data, however,
no model has been demonstrated to be accurate for all galaxy types at the
level needed for future surveys. 
Understanding intrinsic alignments at the level of basic physical phenomenology,
and from a theoretical perspective is, then, crucial if the community wishes to fully exploit the 
cosmological information in the large-volume lensing data that will shortly become
available.

The most commonly used IA model, known as the 
Linear Alignment Model \citep{hirata04}
and its empirically motivated variant, 
the Nonlinear Alignment (NLA) Model \citep{bridle07}
treat intrinsic shape correlations as linear in the
background tidal field.
This model is well tested in the regime of low redshift luminous
red galaxies \citep{joachimi11,singh15},
but observational validation is somewhat lacking
in the mixed-colour samples, extending to higher redshifts, common in 
lensing cosmology.
One approach to this problem has been to develop self-consistent 
perturbative models, which include both linear (tidal alignment)
and quadratic (tidal torque) contributions.
A small handful of such models have been published
\citep{blazek15,tugendhat17,blazek17}
and implemented in practice \citep{svcosmology, y1cosmicshear}.
An alternative approach is to build an
analytic prescription for the intrinsic alignment signal using
a halo model or similar. 
Given the known limitations of perturbation theory
on the smallest scales, in the fully one-halo regime, such models
are especially attractive from a theoretical perspective. 
Similar techniques have been employed with some success to model nonlinear growth
and baryonic effects (\citealt{fedeli14}; \citealt{schneider15}; \citealt{mead15}; \citealt{mead16})
and galaxy bias \citep{peacock00,schulz06,dvornik18}.
The literature around application of such methods to IAs is, however, less extensive.
An early example is presented by \citet{smith05}, who propose a halo model
based approach to modelling the alignment of triaxial
dark matter halos.
This was followed several years later by \citet{schneider10}, 
wherein a similar method was developed to describe the power spectra of
galaxy intrinsic alignment.
Under their model galaxies are split into ``centrals'' and ``satellites'',
with the latter aligning radially towards the former within spherical halos.
Such a picture is reasonably well motivated, both by theory and 
observational evidence \citep{rood72,faltenbacher07,pereira08,sifon15,huang18}. 

Use of such models to make analytic predictions, however, requires a number of
assumptions. Generally one must assume the distribution of
satellite galaxies within dark matter halos to be spherically symmetric.
This is despite extensive observational and numerical evidence
suggesting otherwise
(see, for example, 
\citealt{west00}, \citealt{knebe04},
\citealt{bailin08}, \citealt{agustsson10}, \citealt{piscionere15}, 
\citealt{libeskind16}, \citealt{butsky16}, \citealt{welker17}.
More discussion of the physical mechanisms that generate anisotropy 
and further references can be found in \citealt{zentner05}).
This paper seeks to test this approximation using hydrodynamical simulations,
with the aim of building a physical understanding that can be propagated into
future modelling efforts;
the question has attracted some level of speculation in the past,
but has not hitherto been tested robustly.

This paper is structured as follows. 
In Section \ref{sec:theory} we describe a set of hydrodynamical simulations 
used in this analysis. These datasets are public and well documented
and so we will focus here on the relevant details and changes introduced for this
work. A brief discussion of the sample selection used to construct our galaxy
catalogues is also presented.
The process for building two- and three dimensional galaxy shape information
and a symmetrisation procedure, which is central to this work,
are set out in Section \ref{sec:measurements}.  
The latter part of this section then sets out a series of measurements designed to
capture the impact of halo anisotropy on two-point alignment statistics, and ultimately
on cosmic shear.
In Section \ref{sec:results} we present our results, alongside a series of robustness
tests intended to test the applicability of our findings.
In Section \ref{sec:cosmology} we present a simulated likelihood analysis
with the aim of assessing the cosmological
implications of modelling errors of the sort discussed in the previous
sections of the paper.   
Section \ref{sec:conclusion} concludes and offers a brief discussion of our findings.

% ------------- 2. Theory ------------------------------------
\section{A Simulacrum Universe}\label{sec:theory}

In this section we describe the mock universe realisations
used in this work
and the measurement pipeline applied to it.
For a fuller description of the simulations
see \citet{khandai15} and \citet{vogelsberger14}.
The basic statistics of the two simulations are summarised
in Table \ref{tab:simulation_comparison}.
The pipeline from \blockfont{SubFind} data to symmetrised
galaxy catalogues and correlation functions is hosted on a
public repository\footnote{\url{https://github.com/ssamuroff/mbii}}.
The processed catalogues themselves are also available for 
download from 
\url{https://git.io/iamod\_mbiicats}.

\subsection{\mb}

\mb is a hydrodynamical simulation 
in a cosmological volume of comoving dimension $L = 100 h^{-1}$ Mpc.
The mock universe's initial conditions were generated with a
transfer function generated by 
\blockfont{CMBFAST}\footnote{\url{https://lambda.gsfc.nasa.gov/toolbox/tb\_cmbfast\_ov.cfm}}
at redshift $159$. The simulation was then allowed to evolve through to $z = 0$.
The basic gravitational evolution of mass is governed by 
the Newtonian equations of motion
and the simulation run assumes a flat \lcdm~cosmology
with, 
$\sigma_8 = 0.816$, 
$\omegam = 0.275$, 
$\ns = 0.968$, 
$\omegab = 0.046$, 
$\omegade = 0.725$, 
$h = 0.701$
and
$w = -1.0$.

The simulation volume contains a total of $2 \times 1792^3$
particles (dark matter and gas); particles of dark matter and gas have
mass $1.1\times 10^7$ and $2.2\times 10^6$ $h^{-1} M_{\odot}$ respectively. 
The simulation is based on a version of \blockfont{GADGET}
(P-GADGET; see \citealt{khandai15,dimatteo12}),
which evaluates the gravitational forces between particles 
using a hierarchical tree algorithm, 
and represents fluids by means of smoothed particle hydrodynamics (SPH).

Star formation is governed by a Schmidt-Kennicutt law 
of the form 
$\dot{\rho} _{*} = \epsilon_{*} \rho/t_\mathrm{ff}$, 
where the left-hand term is the star formation rate density,
$\rho$ is the mass density of gas and
$\epsilon_{*}$ is an efficiency constant with fixed value 0.02 
\citep{kennicutt98}. The denominator $t_\mathrm{ff}$ has units of time and 
corresponds to the free fall time of the gas.
Star particles are generated from gas particles randomly with a
probability determined by the star formation rate.
The simulation run includes basic models for AGN and supernova
feedback.
The code models black holes as
collisionless particles, which acquire mass by accreting ambient gas.
As gas accretes it releases energy as photons at a rate proportional
to the rate of mass growth of the black hole;
the efficiency constant $\epsilon_r$ is assumed to have a
fixed value of 0.1 \citep{dimatteo12}. 
A fixed fraction ($5\%$) of this 
energy is then assumed to couple with nearby gas within the softening
length ($1.85$ kpc). This somewhat crude model, at least at some level, 
captures the influence of black hole feedback on the motion and thermodynamic 
state of surrounding matter.
Additionally, the simulation contains a basic model for supernova feedback.
Thermal instabilities in the interstellar medium 
are assumed to operate only above some critical density $\rho_\mathrm{th}$,
which produces two phases of baryonic matter:
clouds of cold gas surrounded by tenuous baryons
at pressure equilibrium.
Star formation occurs in the dense part,
upon which short lifetime massive
stars begin to expire, form supernovae and release fixed
bursts of energy to their immediate surroundings.
Conceptually similar to the black hole feedback model,
this has the effect of providing localised groups of
particles periodically with instantaneous boosts in energy and momentum.

\subsection{\blockfont{illustris}}

To test the robustness of our results we will also incorporate
public data from a second hydrodynamic simulation.
Though the format of the output is ultimately the same,
\blockfont{illustris}
was generated by an independent group
with a number of notable methodological differences from \mb.
We use the \blockfont{Illustris-1} dataset here,
which was generated at a 
similar but non-identical \lcdm~cosmology,
defined by
$\sigma_8 = 0.809$,
$\omegam = 0.2726$, 
$\ns = 0.963$,
$\omegab = 0.046$,
$\omegade = 0.7274$,
$h = 0.704$,
$w = -1.0$.

\begin{table}
\begin{center}
\begin{tabular}{c|cc}
\hline
                                                    & \mb         & \illustris       \\
\hhline{===}
Comoving Volume / $h^{-3}$ Mpc$^3$                  & $100^{3}$                  & $75^{3}$   \\ 
Particle Mass / $\times10^6 h^{-1} M_\odot$         & $11.0,2.2$   &  $6.3,1.3$   \\ 
Resolution /  $h^{-1}$ Mpc                          & 0.04 & 0.03   \\
\hline
\end{tabular}
\caption{Summary statistics for the two hydrodynamic simulations considered in the paper.
The row labelled `particle mass' shows dark matter, then stellar mass in units of 
$\times10^6 h^{-1} M_\odot$. 
%\rachel{Please check units carefully.  The MB-II paper (Khandai+) lists those
%  numberss for particle mass,  but with units of $h^{-1} M_\odot$, not $M_\odot$ as you've listed
%  it here.  And $h^{-1} M_\odot$ is more typical for a simulation in practice.}
}\label{tab:simulation_comparison}
\end{center}
\end{table}

The data are well documented on the data release website\footnote{\url{illustris-project.org}}
and attendant papers \citep{vogelsberger14}. 
The simulation code makes use of an approximate model for
galaxy formation, which includes
gas cooling (primordial and metal line),
stochastic star formation, stellar evolution,
kinetic stellar feedback driven by supernovae,
and supermassive black hole seeding, accretion
and merging,
in addition to a multi-modal model for
AGN feedback.
A number of tunable parameters are included in the
model and the process used to decide their values
is outlined by \citet{vogelsberger13}.

\subsection{Galaxy Catalogues}\label{sec:catalogues}

The following paragraphs describe the steps used to
construct processed object catalogues, upon which 
two-point measurements are made.
The discussion is largely generic to \mb~and \illustris.
The few points of difference between the pipeline run on
the two simulations are stated explicitly.

\subsubsection{Halos \& Subhalos}

For the purposes of this work we treat the term ``galaxy''
as a synonym for ``subhalo with a stellar component of nonzero mass''.
Particles are grouped together into halos using a friends-of-friends (FoF) 
algorithm, which uses an adaptive linking length of 
0.2 times the mean interparticle separation.
Subhalos are identified using \blockfont{SubFind}
\citep{springel01}.
The algorithm works as follows:
the local density is first calculated at the positions 
of all particles in a given FoF group,
with a local smoothing scale set
to encircle a fixed number (twenty in our case) of neighbours.
The density is estimated by kernel interpolation 
over these neighbours. 
If a particle is isolated then it forms a new density peak. 
If it has denser neighbours in multiple different structures,
an isodensity contour that
crosses the saddle point is identified.
In such cases the two structures are merged 
and flagged as a candidate subhalo
if they have above some threshold number of particles.

Baseline cuts are applied to the \mb~subhalo catalogues to
remove objects with fewer than 300 star particles 
or 1000 dark matter particles 
(equivalent to mass cuts at 
$6.6\times 10^{8} h^{-1} M_{\odot}$
and 
$1.1\times 10^{10} h^{-1} M_{\odot}$
respectively). 
Cuts on properties such as mass will likely induce some 
level of selection bias in the various properties with which they correlate.
They are, however, necessary to ensure subhalo convergence 
(see e.g.\ Appendix~B of \citealt{chisari18} and Section~2.3 of \citealt{tenneti15a}).
This is analogous to basic level cuts on flux and likelihood in shear studies on
real data; if it is not possible to derive reliable shapes (on the ensemble level) from a 
significant part of the galaxy sample, then one's ability to draw physically meaningful
conclusions is limited. 
\illustris~differs slightly from \mb~in both box size and particle
mass resolution, and so we adjust the cuts accordingly to
maintain mass thresholds at approximately the same level.
Unless explicitly stated, all results presented in this
paper assume these cuts.
The \illustris~dataset used here is smaller than \mb,
both in raw number of galaxies surviving cuts (35,349 compared with 113,560)
and in comoving number density
(0.08 against 0.11 $h^3$ Mpc$^{-3}$).
Some basic physical characteristics of the two simulations are set out in
Table \ref{tab:simulation_comparison}.

\begin{table}
\begin{center}
\begin{tabular}{c|ccccc}
\hline
Redshift & No. of Galaxies &  Satellite Frac.  & $\sigma_\mathrm{e}$ (stars) & $\sigma_\mathrm{e}$ (matter) \\   
\hhline{=|=====}
0.062 & 0.114 M & 0.351 & 0.411 & 0.354 \\
0.300 & 0.121 M & 0.359 & 0.401 & 0.341 \\
0.625 & 0.129 M & 0.366 & 0.390 & 0.330 \\
1.000 & 0.136 M & 0.380 & 0.382 & 0.320 \\
\hline
\end{tabular}
\caption{The essential characteristics of the simulated \mb~galaxy catalogues
used in this study, after baseline mass cuts. The rows represent different simulation snapshots
(labelled, from top, 85, 79, 73 and 68 in the nomenclature of the data release).
The satellite fraction is computed using a hybrid central definition 
(the most massive galaxy within the centre-most 10\% of the halo radius $R$, 
as defined for a particular halo),
which is designed to be robust to noise in the physical subhalo properties.
The final two columns show the mean per-component ellipticity dispersion (i.e.\ $[\sigma_{e1}+\sigma_{e2}]/2$)
for the projected (2D) shapes of the stellar and dark matter components of the galaxies at each
redshift. 
}
\label{tab:catalogues:statistics}
\end{center}
\end{table}

\subsubsection{Central Flagging}\label{sec:catalogues:cflag}

The processed catalogues contain three binary flags for ``central'' galaxies,
which we describe in the following paragraph.
Though classifying real galaxies as centrals or satellites is common in practice,
the quantities used to do so are, of course, observed ones.
We do not have exact analogues in the simulations for properties 
such as flux and apparent magnitude. 
Rather, we have access to fundamentally unobservable ones such
as dark matter mass and three-dimensional comoving position.
Given this, we propose two alternative criteria for identifying a
halo's central galaxy:
(a) the galaxy with the shortest physical distance 
(Euclidean separation in three dimensional comoving coordinates) from the 
minimum of the dark matter halo's gravitational potential well
and (b) the galaxy with the highest total mass (dark matter and stars) in the halo.
We compute central flags for our catalogues using both of these definitions.

Unfortunately, each has its own limitations. 
Definition (a) is a split based on a significantly noisy quantity. Given the relatively fine
mass resolution, it is not uncommon that what might naturally be described as
a halo's central galaxy (a large, high mass object close to the potential minimum)
shares the inner region with a large number of low mass objects.
The potential for misclassification is obvious; 
indeed we find that adopting (a) hinders our ability to
reproduce observable trends 
(e.g. a gradual decline in satellite fraction with stellar mass) 
due to mislabelling of high mass central objects as satellites.
Criterion (b) has similar drawbacks as a split metric. 
There is observational evidence (e.g. \citealt{johnston07, rykoff16, simet17})
that in many cases a halo's most massive galaxy
can be significantly offset from its centroid (as defined by the gravitational potential).
By visual inspection of a subset of \mb~halos, it is clear that such an offset
is not uncommon\footnote{This is true in the low redshift regime, within which the four snapshots considered in this work sit.
It is reasonable to expect that miscentering of the most massive galaxy may be less common at high redshift.},
particularly in those undergoing mergers or 
with otherwise distorted, non-spherical mass distributions.
We thus construct a third definition (c), under which a halo's central
is the most massive galaxy within a radius of $0.1R$ from
the potential minimum,
where $R = \left ( GM / 10^{6}h^2 \right )^\frac{1}{3}$
is dependent on the total FoF halo mass $M$ and (mildly) the background cosmology.
We consider (c) to be the most physically meaningful of these alternatives,
and so adopt it as our fiducial definition in for the following sections.
In Section \ref{sec:results:robustness}
we test the robustness of our results to this analysis
choice and find no qualitative change due to switching between definitions (c)
and (a).
Note that the flagging is performed prior to mass cuts, and so while every halo has a 
central galaxy under all three of these definitions, it is non-trivial that it
survives in the final catalogue.
The final satellite fractions in the four redshift slices used in this work
are shown in Table \ref{tab:catalogues:statistics}.
Note that (a) and (b) produce comparable values, with a similar
mild increase towards high redshift.
It is noticeable that the satellite fraction,
defined as the total fraction of all subhalos that are not flagged
as centrals, is
significantly lower than 0.5,
and observation that is true under all three central definitions.
Though counterintuitive given the simple
picture of halos with one central and a host of satellites,
the catalogues contain a large number of low-mass halos with
one (or fewer) subhalos. Such halos often contain no satellites and
so act to dilute the global satellite fraction. 

The same code pipeline is applied equivalently to \blockfont{SubFind} outputs
on both \mb~and \illustris to generate catalogues in a common format.
For the galaxy catalogues used in this work and example scripts
to illustrate their basic usage,
see \url{https://github.com/McWilliamsCenter/ia\_modelling\_public}. 

\section{Measurements}\label{sec:measurements}

In addition to the basic level object detection and flagging described in
the previous section, a series of additional measurements are required
before we can attempt to draw conclusions from the simulated data.
In this section we describe the process of shape measurement, converting
a collection of three dimensional particle positions associated with each 
subhalo into projected galaxy shapes. We then define a series of two-point
statistics, on which we rely in the following sections of this paper. 

\subsection{Inertia Tensors \& Shapes}\label{sec:measurements:shapes}

For any practical application of weak lensing, the salient properties of a population
of galaxies are its shape statistics. In real two-dimensional pixel data, one
typically works with ellipticities, which may be expressed in terms of two-dimensional
moments.
An analogous calculation may be performed with the three-dimensional simulated galaxies
in \mb. This is well-trodden ground, and so we will sketch out the calculation briefly
and refer the readers to a host of other papers for more detail 
\citep{chisari15, velliscig15, chisari17}. 
A detailed presentation of alignment angles, shapes, and two-point correlations
and how they depend on the chosen definition of the inertia tensor
can also be found in \citet{tenneti15b} and \citet{tenneti15a}.
Starting from a collection of stellar or dark matter particles associated with a subhalo,
we calculate the inertia tensor,
\begin{equation}
I_{ij} = \frac{1}{W} \sum_{\alpha=1}^{N_\mathrm{p}} w_{\alpha} x_{i,\alpha} x_{j,\alpha},
\end{equation}
where the Roman indices indicate one of the three spatial coordinate axes $i,j\in (x,y,z)$
and the sum runs over the $N_\mathrm{p}$ particles within the subhalo.
The positions $x_k$ are defined relative to the centroid of the subhalo.
The coefficients $w_\alpha$ are particle weights and the prefactor term is the sum
over weights  $W=\sum_\alpha w_\alpha$. 
In the simplest case, which we refer to as the basic inertia tensor, particles
are weighted equally and the sum of the (normalised) weights is simply the number of
particles in the subhalo.
This will be our default option; unless stated otherwise
the reader should assume the results presented later in this paper
use this definition of the inertia tensor.
An alternative approach, which defines the ``reduced'' inertia tensor, is to weight
by the inverse square of the radial distance from the subhalo centroid, $w_\alpha = r_\alpha^{-2}$. 
This estimator by construction downweights light on the edges of the galaxy profiles 
and in this sense produces projected ellipticities more akin to what might be measured from
real data. 
It is also true, however, that imposing circularly symmetric weighting
induces a bias towards low ellipticities; obtaining accurate measurements via this estimator
requires an explicit correction such as the iterative method of \citet{allgood06}.

Evaluating each element of the $3\times3$ inertia tensor $\mathbf{I}$ provides a simple numerical 
description of the three-dimensional shape of the galaxy.
We then decompose $\mathbf{I}$ into eigenvectors, 
which represent unit vectors defining the orientation of the
major, intermediate and minor axes of the ellipsoid
$\mathbf{A}_\mu = (A_{x,\mu}, A_{y,\mu}, A_{z,\mu})$,
$\mu \in (1,2,3)$; 
and three eigenvalues $\lambda_i$,
which quantify its axis lengths.

In order to fully capture higher order effects present in real data,
ray-tracing and light cones would be required to generate 
projected per-galaxy images (see e.g. \citealt{peterson15}).
Given the size of the dataset and the scope of this study, however,
a simple projection along the $z$ axis of the simulation box is sufficient. 
Points on the projected ellipse $\mathbf{x}$ must satisfy the equation
$\mathbf{x}^\intercal \mathbf{Q} \mathbf{x}=1$ with

\begin{equation}
\mathbf{Q}^{-1} = \sum_{\mu=1}^{3} \frac{\mathbf{A}_{\perp,\mu} \mathbf{A}^\intercal_{\perp,\mu} }{\lambda^2_\mu} - \frac{\boldsymbol{\kappa}\boldsymbol{\kappa}^\intercal}{\alpha^2},
\end{equation}

\noindent
and $\mathbf{A}_{\perp,\mu} = (A_{x,\mu}, A_{y,\mu})$. In the above we define

\begin{equation}
\alpha^2 = \sum_{\mu=1}^3 \left ( \frac{A_{z,\mu}}{\lambda_\mu} \right )^2
\end{equation}

\noindent
and 

\begin{equation}
\boldsymbol{\kappa} = \sum_{\mu=1}^3 \frac{A_{z,\mu}\mathbf{A}_{\perp,\mu}}{\lambda^2_\mu}. 
\end{equation}

\noindent
The moments of the two dimensional ellipse $\mathbf{Q}$
can then be converted into the two ellipticity components standard in lensing via

\begin{equation}
(e_1, e_2) = \frac{(Q_{xx}-Q_{yy}, 2Q_{xy})} {Q_{xx} + Q_{yy} + 2 \sqrt{|\mathbf{Q}|}}.
\end{equation}\label{eq:ellipticty}

\noindent
The above recipe mirrors similar calculations in \citet{piras18} and \citet{joachimi13}, to which we refer the reader for more detail.
Note that we have chosen a particular ellipticity definition $e = (a-b)/(a+b)$ here\footnote{The
  alternative definition, often also referred to as ellipticity, or sometimes polarization or distortion,
  in the literature is defined $\chi = (a^2-b^2)/(a^2+b^2)$. It can be recovered analogously using
  Equation~\eqref{eq:ellipticty}, but without the final term in the denominator.}.
Although these quantities can be measured equivalently using dark and visible matter constituents,
unless stated otherwise the results in the following sections use the latter only.
This is expected to give results more directly relevant to real lensing surveys,
but the impact of this decision is tested in Section \ref{sec:robustness:subhaloshapes}.

\subsection{Symmetrising the Distribution of Satellites in Halos}\label{sec:symmetrisation}

The most basic picture of the Universe sees all mass contained by discrete dark matter halos.
Each of these halos then hosts a number of subhalos, some of which contain luminous matter
and are thus considered detectable galaxies.
In the various permutations of the halo model used in the literature,
typically one assumes spherical dark matter halos and, by implication,
isotropic satellite distributions about the halo centres.
There is a significant amount of evidence from N-body (dark matter) simulations, 
however, that halos are more often than not triaxial
\citep{kasun05,bailin05,allgood06}
and the ensemble of subhalos within them has some preferred axis
\citep{west00,knebe04,zentner05,bailin08}.
This implies that satellite galaxies should be 
similarly anisotropically distributed about the centre of their host halo, 
a conclusion bourne out by the limited amount of
data from hydrodynamical simulations currently available.

\begin{figure}
\includegraphics[width=\columnwidth]{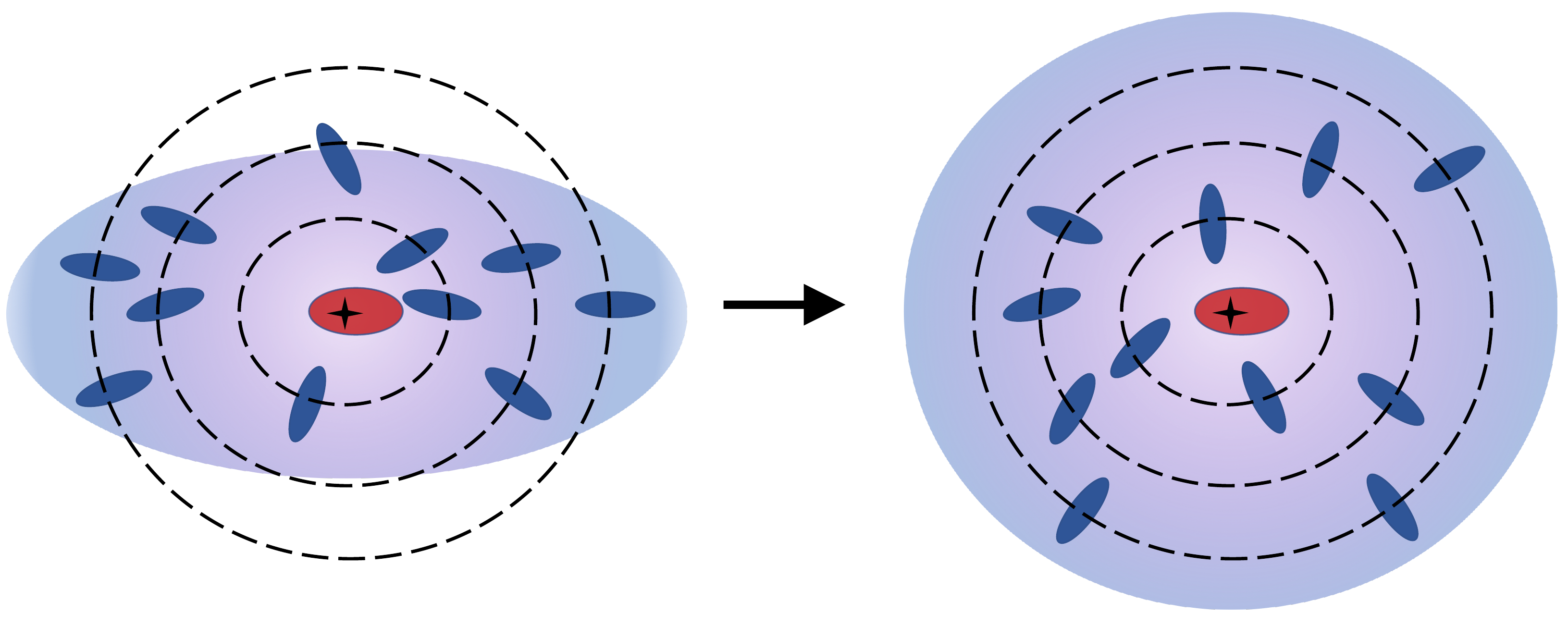}
\caption{Two-dimensional schematic diagram of the halo symmetrisation process described in Section~\ref{sec:symmetrisation}.
The background colour gradient is intended to represent the distribution of dark matter in the halo,
but is not a quantitative mapping of the halo mass profile. Galaxies (subhalos with stellar particles)
are shown as filled ellipses, and the halo centroid is shown as a black cross.
The black dashed lines show three isopleths of constant radius about the centroid. 
The central galaxy (shown in red) is unaffected, 
while satellites (blue ellipses) at each radius are redistributed randomly about the circle.
The relative orientations of the galaxies to the centroid are explicitly maintained by the rotation.
}\label{fig:symmetrisation_diagram}
\end{figure}

To test the impact of this anisotropy we create an artificially symmetrised version 
of the galaxy catalogues described in Section~\ref{sec:catalogues}.
The symmetrisation process entails identifying all satellites
associated with a particular halo and applying a random rotation about the halo centroid. 
At any particular distance from the centroid this effectively redistributes the 
satellites across a spherical shell of the same radius. 
We explicitly rotate the galaxy shapes in three dimensions, 
such that their relative orientation to the halo centre is preserved.
Without this second rotation (position, then shape) interpretation of the 
physical effects at work is difficult, as the process
washes out both the impact of halo asphericity
and the (symmetric) gravitational influence of the host halo on galaxy shape. 
A cartoon diagram of this istotropisation process is shown in Figure~\ref{fig:symmetrisation_diagram}. 
 
Though conceptually very simple, the mathematics of such a three-dimensional
rotation bears some thought. The aim is to transform the spherical coordinates 
of each satellite galaxy relative to the halo centroid as:
\begin{equation}
R \rightarrow R,\;\;\;
\theta \rightarrow \theta',\;\;\;
\phi \rightarrow \phi'.
\end{equation}

\noindent
The galaxy shape also requires an equivalent transformation in
order to maintain the relative orientation to the halo centre.
As implemented for this analysis the symmetrisation process involves the following steps:
\begin{itemize}
\item {Starting from the inertia tensor of star particles in a particular satellite galaxy,
we compute a $3\times3$ eigenvector matrix $\mathbf{A}= (\mathbf{a},\mathbf{b},\mathbf{c})$, and a set of three eigenvalues 
$\boldsymbol{\lambda} = (\lambda_1, \lambda_2, \lambda_3 )$.}
\item{Random position angles $\mathrm{cos}\theta', \phi'$ are drawn from uniform distributions
over the range $[-1/2,1/2]$ and $[-\pi,\pi]$ respectively.}
\item{A rotation matrix is constructed describing the transformation between the initial
and the new positions, such that $\mathbf{r}' = \mathbf{R}_\theta.\mathbf{r}$.
The calculation of $\mathbf{R}_\theta$ is described in Appendix \ref{app:rotation_matrix}.}
\item{The same transformation described by the rotation matrix $\mathbf{R}_\theta$ 
is applied to each of the three orientation vectors $\mathbf{a}, \mathbf{b}, \mathbf{c}$,
a process which preserves the relative orientation of the galaxy to its new radial position vector 
$\mathbf{r}'=(R,\theta',\phi')$.}
\end{itemize}

\noindent
This leaves us with, for each satellite galaxy in a particular halo, a new rotated position
$\mathbf{r}'$ and an orientation (eigenvector) matrix $\mathbf{A}'$. 
The three dimensional axis lengths
$\lambda_i$ are unchanged by the transformation.

The symmetrisation process is designed such that it respects the periodic boundary 
conditions of the simulation volume.
That is, galaxies in halos that traverse the box edge are shifted before rotation
to form a contiguous group.
In cases where rotation leaves a galaxy that was within the box
outside its edges, it is shifted back to the opposite side of the volume.

% ------------- 3. Theory ------------------------------------

\subsection{Two-Point Correlation Functions}

In the following we describe a series of two-point statistics
used as estimators for the IA contamination to cosmic shear.
All correlation functions used in this study were computed using 
the public \blockfont{halotools} 
package\footnote{v0.6; \url{https://halotools.readthedocs.io}}
\citep{hearin17}.
The most straightforward (and highest signal-to-noise) two-point measurement one could
make is that of galaxy clustering in three dimensions.
We adopt a common estimator of the form \citep{landy93}:

\begin{equation}
\xi_{gg}(r) = \frac{DD - 2DR + RR}{RR},
\end{equation}

\noindent
where $DD$, $RR$ and $DR$ are counts of galaxy-galaxy,
random-random and galaxy-random pairs within a physical separation bin
centred on $r$.
For this calculation we use 20 logarithmically spaced bins in the range
$r=0.1 - L/3 h^{-1}$ Mpc,
where $L$ is the simulation box size
(100 and 75 for \mb~and \illustris~respectively).
For the purposes of validation, 
where possible, we compared our results against analogous measurements performed using
\blockfont{TreeCorr}\footnote{v3.3.7; \url{https://github.com/rmjarvis/TreeCorr}} \citep{jarvis04}. 
For galaxy-galaxy correlations we find sub-percentage level agreement between the
two codes on all scales
(ranging from $\sim0.01\%$ on the smallest scales to $\sim0.4\%$
in the two-halo regime with the default accuracy setting (bin$\_$slop=0.1);
the discrepancy
is seen to vanish when the computational approximations of \blockfont{TreeCorr}
are deactivated.

When considering the impact of alignments on cosmological observables
($\xi_{\pm}, \gamma_t$ and the like),
it is natural to consider two-point functions of intrinsic alignment.
Beyond this broad statement, however, it is not trivial which measurement
is optimal for our interests.
We start by considering the most conceptually simple statistics,
or the three dimensional orientation correlations.
We can define two such terms,

\begin{equation}\label{eq:definition_ee}
\mathrm{EE} = \left < | \hat{\mathbf{e}}(\mathbf{x}) \cdot \hat{\mathbf{e}}(\mathbf{x}+\mathbf{r}) |^2 \right > - \frac{1}{3}
\end{equation}

\noindent
where $\hat{e}$ is a unit eigenvector 
obtained from the inertia tensor, pointing
along the major axis of the galaxy ellipsoid,
and the angle brackets indicate averaging
over galaxy pairs.
Intuitively very simple, a positive EE correlation denotes 
a tendency for the major axes of galaxies to align with those of 
other physically close-by objects in comoving three dimensional space.
Similarly:

\begin{equation}\label{eq:definition_ed}
\mathrm{ED} = \left < | \hat{\mathbf{e}}(\mathbf{x}) \cdot \hat{\mathbf{r}} |^2 \right > - \frac{1}{3}.
\end{equation}

\noindent
Here $\hat{\mathbf{r}}$ is the unit vector pointing from
one galaxy to the other.
In the case of ED positive values indicate radial shearing
(that is, a tendency for the major axis of a galaxy to align with the direction of a neighbouring galaxy).

Though three dimensional correlations are illustrative for
elucidating the mechanisms at play in the simulation they
carry a number of obvious problems. 
Not least, they are ``unobservable'' in any real sense.
Typically lensing studies rely on broad photometric filters
and thus do not have a detailed reconstruction of where galaxies
lie along the line of sight.
Even with such information, reconstructing the
shape of a galaxy in three dimensions is difficult-to-impossible.

We now consider ``projected'' correlation functions, as
a closer analogue to real observable quantities.
We do not, however, have access to lightcones and ray-tracing information
for the \mb~simulations. In the absence of the tools for a more sophisticated
approach, we obtain real-space two dimensional statistics, rather, by
projecting along the length of the simulation box.

We define the correlation function of galaxy positions and ellipticities,
$\xi_{g+}(r_\mathrm{p}, \Pi)$ as a function of 2D perpendicular separation 
$r_\mathrm{p}$ and separation along the line of sight $\Pi$.
This statistic is constructed using a modified
Landy-Szalay estimator of the form:

\begin{equation}
\xi_{g+} (r_\mathrm{p}, \Pi) = \frac{S_+D - S_+R}{RR}
\end{equation}

\noindent
(see \citealt{mandelbaum10}). By analogy one can construct a shape-shape correlation function,

\begin{equation}
\xi_{++} (r_\mathrm{p}, \Pi) = \frac{S_+S_+}{RR}.
\end{equation}

\noindent
In both cases above, $R$ represents the positions of a set of randomised positions
thrown down within the simulation volume. The terms in the numerator represent
shape correlations and are defined as

\begin{equation}
S_+D \equiv \frac{1}{2} \sum_{i\neq j} e_{+}(j|i), \;\;\;\;\; S_+S_+ \equiv \frac{1}{4} \sum_{i\neq j} e_{+}(i|j) e_{+}(j|i),
\end{equation}

\noindent
where the indices $i,j$ run over galaxies and $e_+(j|i)$ is the tangential ellipticity
of galaxy $j$,
rotated into the coordinate system defined by the separation vector with galaxy $i$.
With these three dimensional correlations in hand, obtaining the
two dimensional versions is simply a case of integrating along the line of sight. 
One has,

\begin{equation}
w_{g+} (r_\mathrm{p}) = \int_{-\Pi_\mathrm{max}}^{\Pi_\mathrm{max}} \mathrm{d}\Pi \xi_{g+} (r_\mathrm{p}, \Pi)
\end{equation}

\noindent 
and analogously 

\begin{equation}
w_{++} (r_\mathrm{p}) = \int_{-\Pi_\mathrm{max}}^{\Pi_\mathrm{max}} \mathrm{d}\Pi \xi_{++} (r_\mathrm{p}, \Pi)
\end{equation}

\noindent
Here $\Pi_\mathrm{max}$ is an integration limit, which is set by the simulation volume or the
depth of the dataset in the case of real data.
For this study we adopt a value equal to a third of the simulation box size, 
or $\Pi_\mathrm{max}=33 h^{-1}$ Mpc for \mb
and $\Pi_\mathrm{max}=25 h^{-1}$ Mpc for \illustris.
The question of how strongly and on which scales 
galaxy two point functions are affected by finite simulation
limits has been discussed extensively in the literature
(see, for example, \citealt{power06}'s Fig. 4;
also \citealt{bagla09} and the references therein).
We also test the impact of this choice directly using a theory 
calculation in Appendix \ref{app:truncation}. As seen there,
the impact of missing large scale modes on IA and galaxy
clustering projected correlation functions 
is subdominant to statistical error on scales $r_\mathrm{p}<33 h^{-1}$ Mpc.

\subsection{Quantifying the Impact of Halo Anisotropy}

To properly assess the impact of a feature in the simulations
(and thus whether it is incumbent on model-builders to account for it)
it is useful to have numerical metrics. To this end, we define a
quantity referred to as fractional anisotropy bias:

\begin{equation}\label{eq:anisotropy_bias_dfn}
f_\alpha (r, z) \equiv \frac{\left [ \alpha(r,z)-\alpha'(r,z) \right ]}{ \alpha(r,z)},
\end{equation}

\noindent
where $\alpha$ refers to a specific two-point function $\alpha \in (\xi_{gg}, \mathrm{ED}, \mathrm{EE}, w_{g+}, w_{++})$
and $\alpha'$ is its analogue, as measured on the symmetrised catalogues. 
This quantity, then, encapsulates the fractional shift in a given observable due to
symmetrisation, 
or equivalently the level of error introduced by assuming spherical halo geometry when building one's
IA model. 

The reader might note here that while the numerator 
in Equation~\eqref{eq:anisotropy_bias_dfn} might
benefit somewhat from correlated noise cancellation, division by a noisy quantity can
reintroduce it. 
We seek to minimise this effect by assessing the fractional impact on
each correlation function using
smooth fits rather than the measured correlation functions directly.
Power law relations of the form 
$\mathrm{EE} = b r^{-a}$ 
(and the equivalent for ED, $w_{g+}$ and $w_{++}$)
are fit independently to each correlation function.
Visually assessing each of these fits gives us no reason to believe additional
degrees of freedom are warranted by any of the measurements.
We obtain a fractional change in each of the alignment 
correlation function of the form
$f_\alpha = \left ( \alpha - \alpha' \right ) \times x^{a} / b$,
where $x$ is the relevant separation 
(three dimensional Euclidean $r$ or projected perpendicular $r_\mathrm{p}$)
in units of $h^{-1}$ Mpc.
In addition to this, we also fit a power law to each of the
aniostropy bias measurements of the form $f_\alpha = b_{\Delta} x^{-a_\Delta}$.
This exercise gives us the fit parameters presented in Table~\ref{tab:power_law_fits}.

\begin{table}
\begin{center}
\begin{tabular}{c|cc}
\hline
Correlation & $a_{\Delta}$         & $b_{\Delta}$       \\ %& $\alpha$ & $\beta$ \\
\hhline{===}
ED          & $0.809$     & $0.007$   \\ %& $1.135$  & $2.291$ \\ 
EE          & $1.361$     & $0.002$   \\ %& $1.465$  & $2.779$ \\
$w_{g+}$    & $0.782$     & $0.060$   \\ %& $0.999$  & $1.292$ \\
$w_{++}$    & $1.265$     & $0.004$   \\ %& $2.191$  & $3.087$ \\
\hline
\end{tabular}
\caption{Best fitting parameters for power law fits to the anisotropy bias,
obtained from the \mb~galaxy catalogue without splitting by galaxy type.
The fits are performed on the absolute residual between measurements on
the symmetrised and unsymmetrised correlation functions,
and have the form $\Delta \alpha = b_{\Delta} r^{-a_{\Delta}}$.
%The power law has the same algebraic form in all cases, but
%the first two parameters $(a,b)$ describe the (unsymmetrised) correlation
%function and the latter two $(\alpha,\beta)$ describe the difference
%introduced by halo symmetrisation. 
}\label{tab:power_law_fits}
\end{center}
\end{table}

\subsection{Errors \& Covariance}

For all of the correlations discussed above we estimate the measurement uncertainties
using a jackknife algorithm;
this entails splitting the cubic simulation box into three along each axis,
then remeasuring the statistic in question repeatedly with one of the nine
sub-volumes removed.
The variance of these measurements is calculated as

\begin{equation}
\left \langle \Delta \alpha \Delta \beta \right \rangle = \frac{N_\mathrm{jk}-1}{N_\mathrm{jk}} \sum_{i=1}^{N_\mathrm{jk}} \left [ \alpha_i - \bar{\alpha} \right ] \left [ \beta_i - \bar{\beta} \right ],
\end{equation}

\noindent
where $N_\mathrm{jk}$ is the number of jackknife volumes and $\alpha_i$ and $\beta_i$
represent generic correlation functions, as measured from the full simulation volume 
minus sub-volume $i$.
For statistics involving random points, we generate 
the appropriate number of samples in the full $100\times100 h^{-1}$ Mpc box
and apply the jackknife excision to this volume
before making the measurement. 
The jackknife technique provides an approximation for the true variance,
the performance of which is heavily dependent on the survey characteristics
and which of the terms contributing to the covariance are dominant 
(see \citealt{singh17} and \citealt{shirasaki17} for a demonstration of
this using SDSS mock catalogues).
Given that we do not attempt likelihood calculations or similar exercises 
where the outcome depends sensitively on 
the accuracy of the error estimates,
such methods are considered sufficient for the scope of this study.

Uncertainties on the anisotropy bias $f_\alpha(r,z)$
are obtained analogously by measuring $\alpha$ and $\alpha'$
with the same subvolume removed from the symmetrised and unsymmetrised
simulations.
Note that some level of bin shifting can be induced by the symmetrisation process,
such that the galaxy samples used to measure $\alpha$ and $\alpha'$ are non-identical.
Although this is expected to be a subdominant effect,
we test its impact as follows.
We repeat the jackknife calculation, this time using selection
masks defined using the unsymmetrised simulation. That is, the 
galaxies excluded in each jackknife realisation are identical
for measurements on the symmetrised and unsymmetrised simulations.
We find the jackknife errors obtained from this
exercise are consistent with the fiducial versions to the level
of $\sim 1-3$ percent.

% ------------- 4. Results ------------------------------------
\section{Results}\label{sec:results}

\subsection{The Impact of Halo Anisotropy on Two-Point Correlation Functions}

\begin{figure}
\includegraphics[width=\columnwidth]{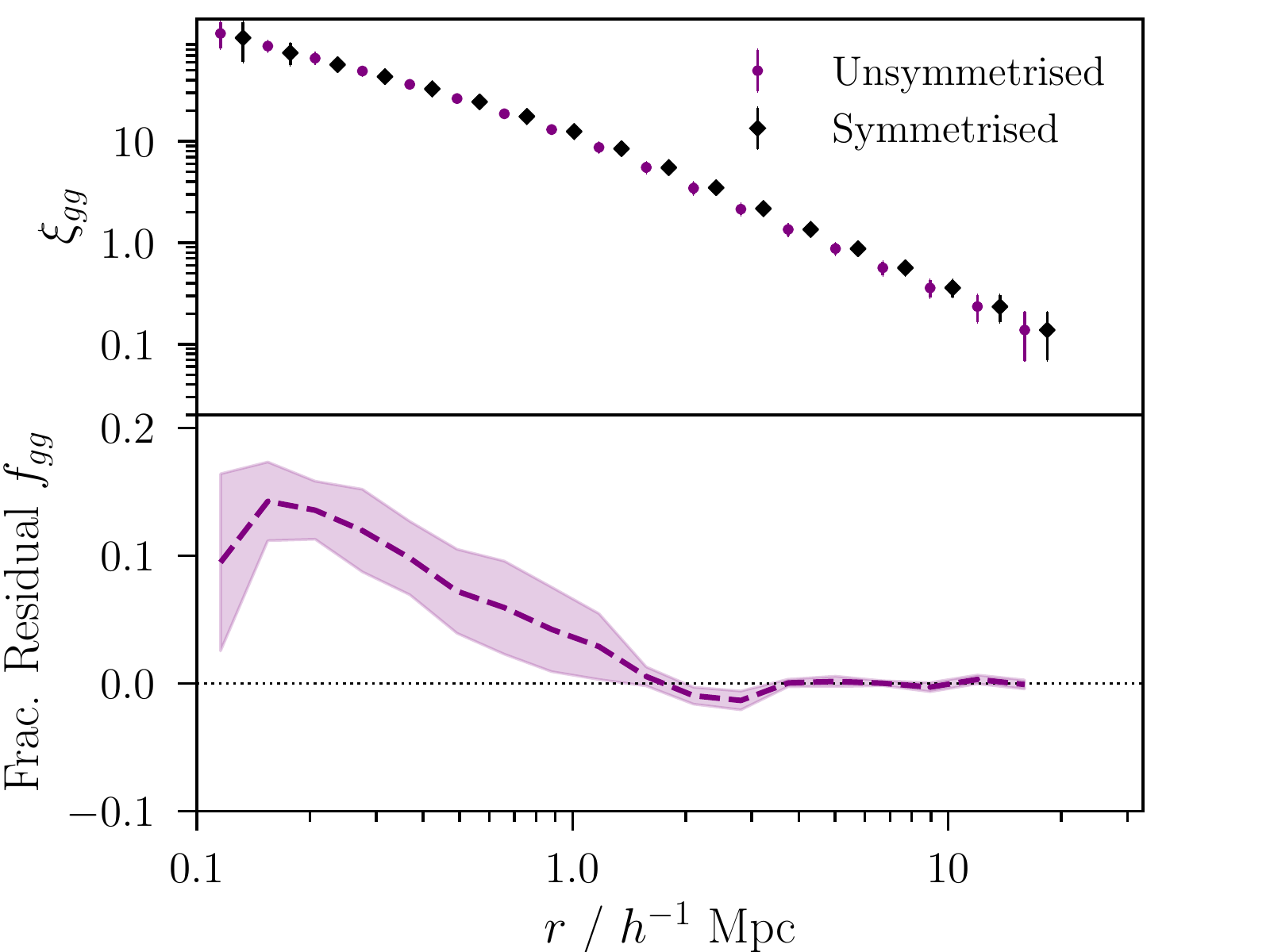}
\caption{\emph{Upper panel}: The galaxy-galaxy correlation function, as a function of three dimensional comoving separation before and after halo symmetrisation.
The purple and black points are offset horizontally left and right by the same small interval to aid clarity. 
\emph{Lower panel}: The fractional difference between the symmetrised and unsymmetrised measurements in the full sample, without satellite-central splitting. The shaded purple region shows the $1\sigma$ variance,
as estimated by jackknife resampling the simulation volume. 
}\label{fig:gg_corr}
\end{figure}

As a first exercise we present the three-dimensional galaxy-galaxy correlation $\xi_{gg}$.
There is some amount of existing literature on such measurements,
which provides a useful consistency check for our analysis pipeline.
The results, as measured before and after symmetrisation, are shown in the upper panel of
Figure~\ref{fig:gg_corr}.
The fractional difference, or fractional anisotropy bias, is shown in the lower panel. 
Consistent with the reported findings of \citet{vanDaalen12} in the Millennium Simulation, 
we find the impact to be at the level of a few
percent on one-halo scales $(r<1 h^{-1}$Mpc$)$, asymptoting to zero on large scales.
One obvious physical manifestation of halo anisotropy is as a boosted
clustering signal between satellite galaxies. 
Given that anisotropy is a property of how satellites sit within their halos, 
it is expected (and observed) that the bias in Figure~\ref{fig:gg_corr} should enter primarily
at separations corresponding to the one halo regime.

\begin{figure}
\includegraphics[width=\columnwidth]{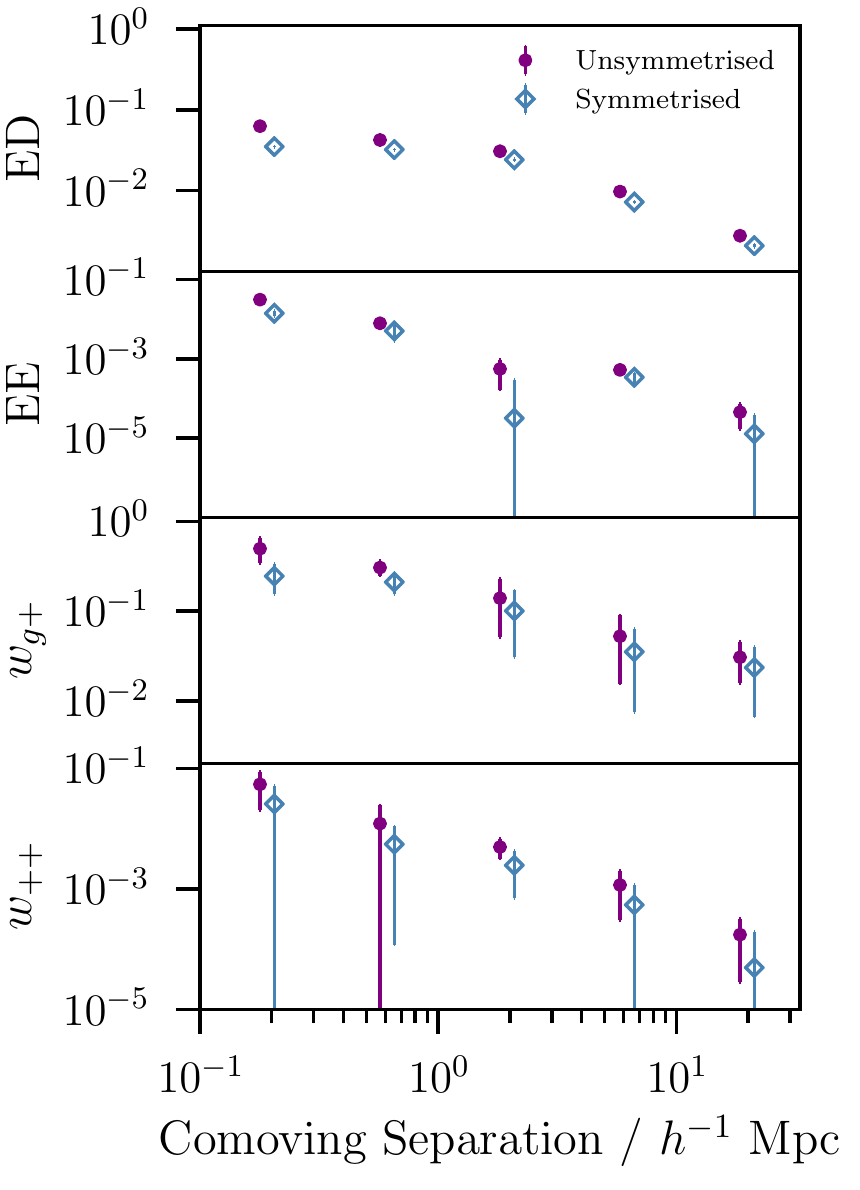}
\caption{Two-point alignment correlation functions, as measured from our fiducial \mb~galaxy catalogue.
The horizontal axis shows the separation between galaxy pairs,
either in three dimensional comoving space (in the case of EE and ED)
or as projected into a plane perpendicular to the line of sight
(for $w_{g+}$ and $w_{++}$).
Purple filled points show measurements on the unsymmetrised catalogue, which includes the effects of 
halo anisotropy. The open blue points show the same measurements on an artificially symmetrised
catalogue, in which satellite galaxies are spherically distributed around the centre of mass
of their host halo. The four statistics shown here are defined algebraically in Section
\ref{sec:measurements}.  
}\label{fig:results:allcorrs}
\end{figure}

Measuring the four IA correlations using the full catalogue
we obtain the measurements shown in Figure~\ref{fig:results:allcorrs}.
The broad trends here appear to fit with the simple physical
picture. At the most basic level, where halo symmetrisation has
an impact, it is to reduce the amplitude of the IA measurements.
The process always acts to wash out pre-existing signal, and cannot
add power.
We see all correlations, but particularly the shape-position type
statistics are most strongly affected on small scales;
modifying the internal structure of halos primarily 
affects small scale alignments, which is also intuitively correct.
It is worth remarking, however, that even on large scales ($>10 h^{-1}$
Mpc), we see some decrement in signal, which implies halo
structure is not entirely irrelevant to large scale IAs,
a principle supported by some previous studies
(e.g. \citealt{ragoneFigueroa07}).

\begin{figure*}
\includegraphics[width=1.7\columnwidth]{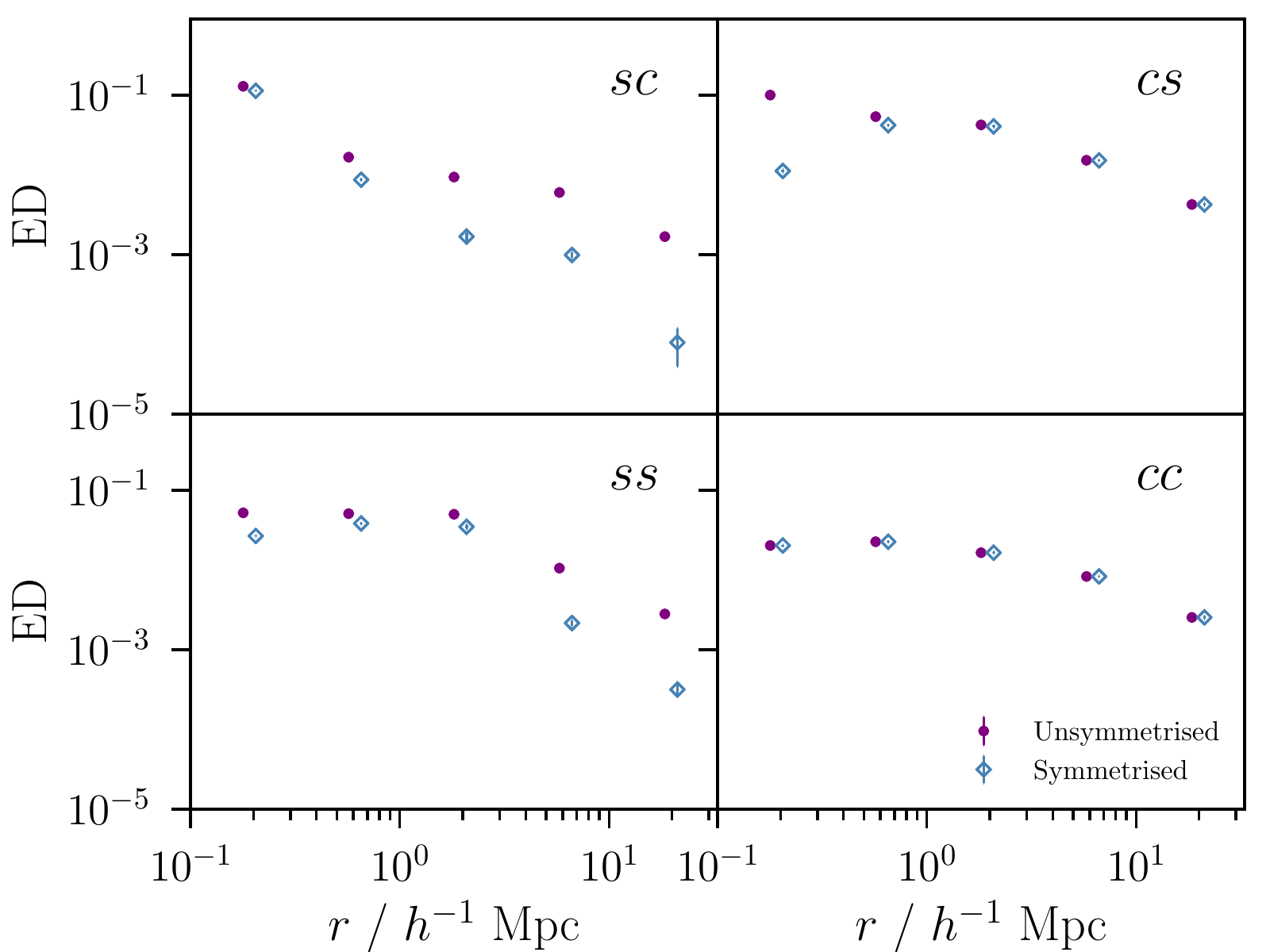}
\includegraphics[width=1.7\columnwidth]{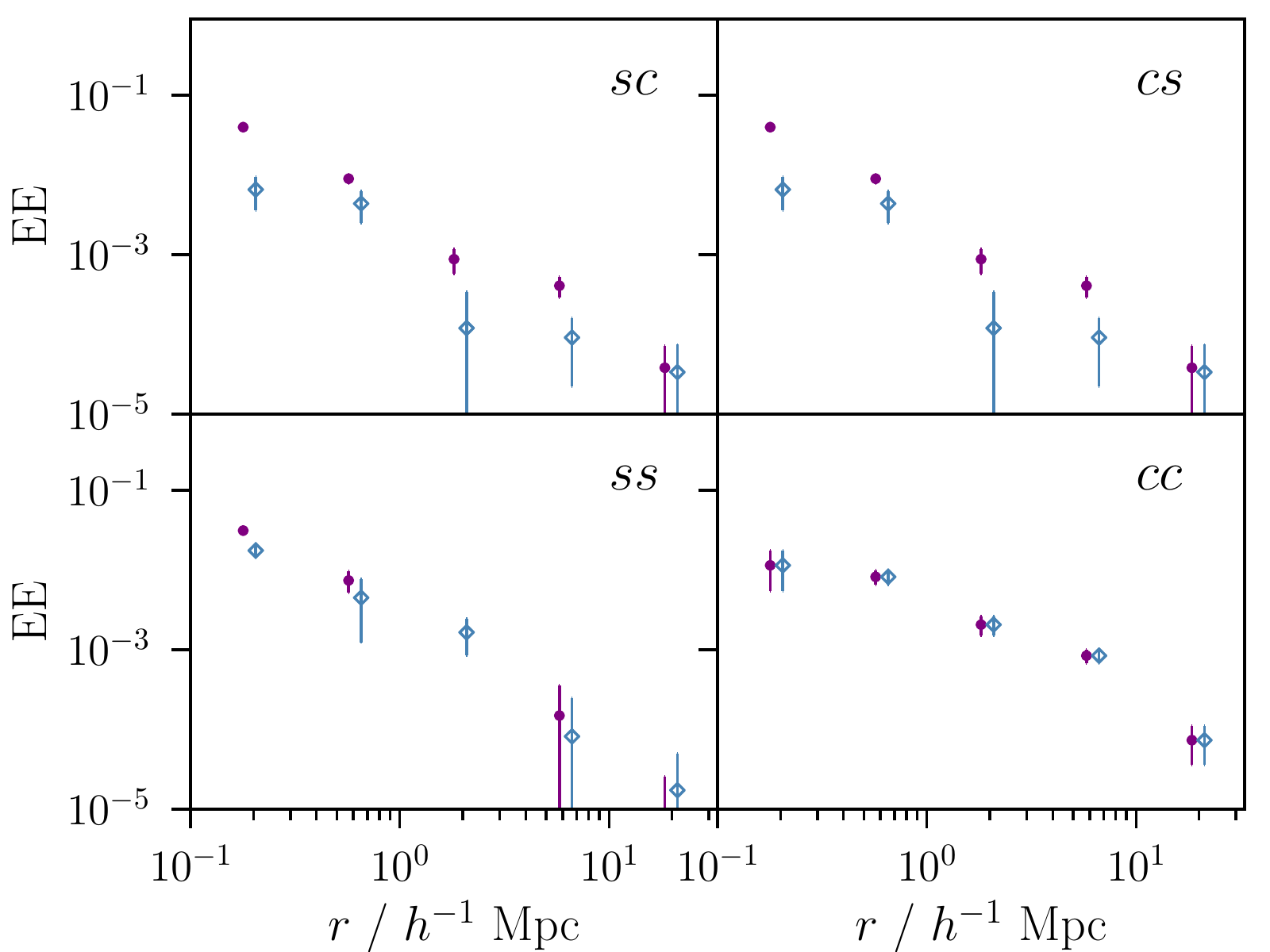}
\caption{Three-dimensional alignment correlation functions, as measured within the \mb~simulation volume
before and after halo symmetrisation. The upper panel shows the two-point correlation of galaxy shapes
and directions, while the lower shows the analogous shape-shape measurements.
The definitions of these correlations can be found in 
Equations~\eqref{eq:definition_ee} and~\eqref{eq:definition_ed}.
All are shown as a function of comoving distance at redshift zero.
Each of the four sub-panels shows a different combination of central and satellite samples
(indicated by the letters in the top right; 
ED $sc$, for example, indicates the correlation of the shapes of central galaxies 
with the relative positions of satellites).
In each case, the filled purple points show the fiducial unsymmetrised measurements and
the open blue diamonds show the result of symmetrisation by rotating satellites within each halo
about the dark matter centre of mass.
Note that the third point in the EE $ss$ correlation is negative $(\sim -0.0012\pm0.0008)$, and so not visible on a log scale.
}\label{results:main:ee_ed_cssplit}
\end{figure*}

The differences here are non-trivial and there are a number
of competing physical effects at play;
to assist in unravelling these effects we impose a catalogue-level
split into central and satellite galaxies
(using the hybrid central flag, as described in Section~\ref{sec:catalogues:cflag}).  
The various permutations of the ED and EE correlation functions 
using these two subsamples are shown 
in Figure~\ref{results:main:ee_ed_cssplit}.
The signal-to-noise on $w_{g+}$ and $w_{++}$ is sufficiently low that
the raw correlation functions are relatively unenlightening;
for completeness they are presented in Appendix~\ref{app:wgp_wpp},
but the statistical error makes it difficult to draw conclusions.
Though less useful in the sense of not having a direct analytic mapping onto the
IA contribution to cosmic shear, the relatively high signal-to-noise and three dimensional nature
of EE and ED help to build an intuitive understanding of the mechanisms at work here.
In each panel we show four sets of measurements, corresponding to all possible
combinations of satellite and central galaxies. 
Note that for the symmetric correlations (EE and $w_{++}$) reversing the order of
the two samples does not change the calculation 
(and so gives identical measurements in the upper right and upper left sub-panels).
By construction the $cc$ correlations are unchanged by symmetrisation;
we show them here for reference because they contribute to the total measured signal
shown in Figure~\ref{fig:results:allcorrs}.  

As one might expect, the ED $sc$ correlation (satellite galaxy shapes, central galaxy positions)
is unchanged on the smallest scales;
satellites point towards the centre of their host halo and the symmetrisation preserves
that relative orientation.
On larger scales the signal is weakened, eventually reaching a level consistent with zero,
as the shapes of satellites become effectively randomised relative to
the position of external halos.
On scales of a few $h^{-1}$ Mpc the unsymmetrised measurements represent a combination of
one- and two-halo contributions. 
When the latter is removed the signal is diluted but not nulled entirely.

In the case of ED $cs$ (central shapes, satellite positions) we see 
a persistent residual signal in the deep one-halo regime after symmetrisation.
Though not immediately predicted by the naive central/satellite picture,
such residual correlations could conceivably arise due to small offsets between
central galaxies and the halo centroid.
The symmetrisation, then, randomises the satellite positions relative to the \emph{centroid}
but leaves some preferred direction relative to the \emph{central galaxy} 
(which is still oriented towards the former).
We test this first by explicitly separating the one-halo and two-halo contributions to ED
to verify that the correlation in the smallest scale bin is indeed a purely one-halo effect.
Second, we artificially shift galaxies such that the central galaxy's centroid coincides
exactly with that of its host halo. Remeasuring ED $cs$ from this
counterfactual catalogue, we see the residual small scale signal vanishes.
More intuitively, on large scales the underlying tidal field will tend to align central shapes with
the positions of other halos (and hence their satellites). Tweaking the positions of
satellites within their halos does nothing to this large scale signal.

In the EE $sc$ correlation, symmetrisation eliminates correlations on the very small and large 
scales, as naively expected. 
We note there is 
a residual signal after symmetrisation
at intermediate scales, which is thought to result from
interactions between neighbouring halos (i.e. a massive halo will cause satellites within it
to orient themselves radially towards it, but also distort the shapes of galaxies in
neighbouring halos). In principle this should also be seen in EE $ss$, but the statistical
power of the simulation limits our ability to detect such effects.

\begin{figure}
\includegraphics[width=\columnwidth]{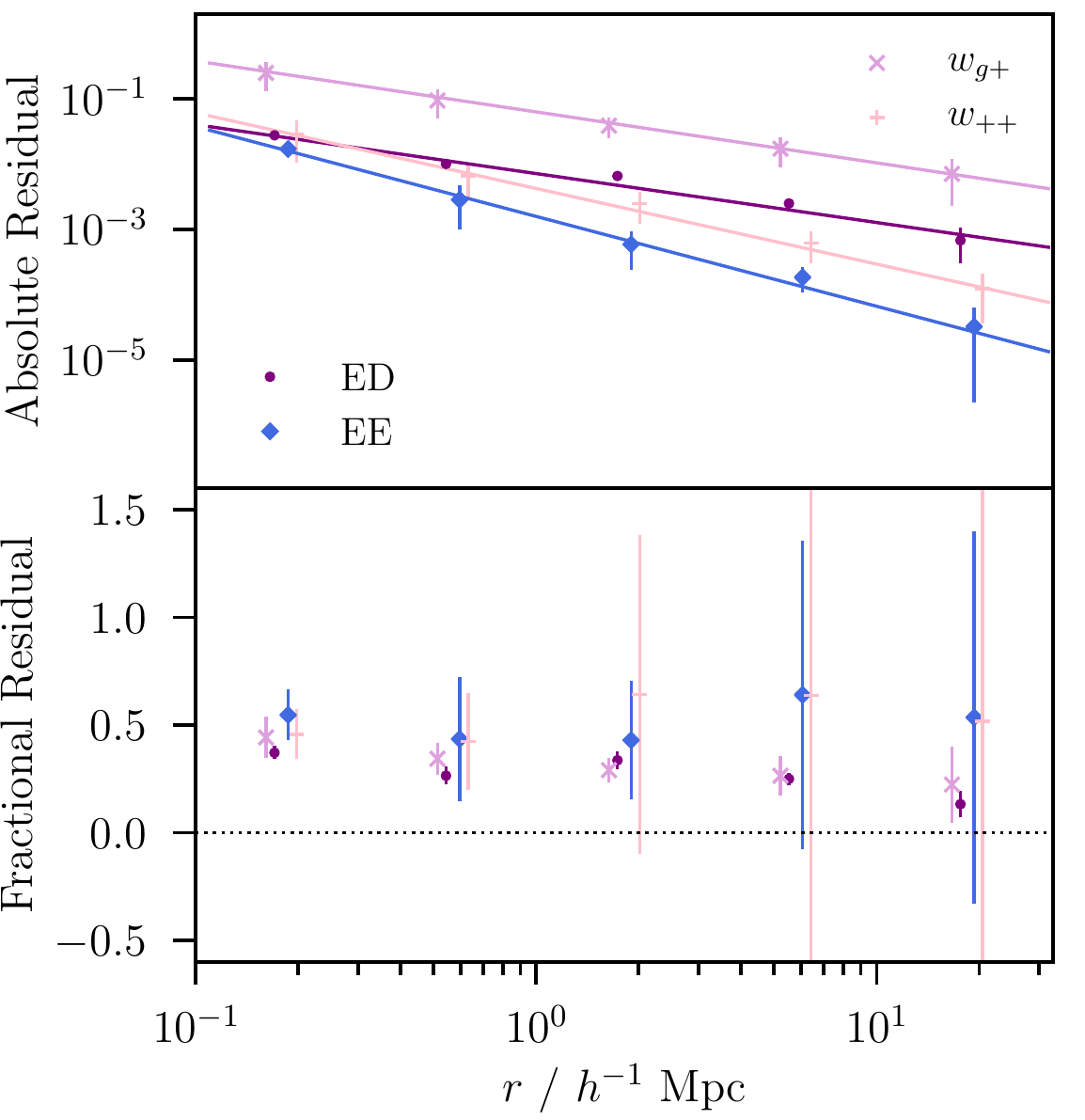}
\caption{The contribution of halo anisotropy to various alignment correlation functions. 
\emph{Upper}: the absolute difference between the two-point measurements
made on the same simulation volume before and after halo symmetrisation.
The solid lines show smooth power law fits to each residual.
\emph{Lower}: The fractional residual between symmetrised and unsymmetrised measurements,
as estimated using power law fits to the unsymmetrised correlation function.
The difference is defined such that $f =($symmetrised-unsymmetrised$)/$unsymmetrised, 
and so positive values indicate an increase in power at a particular scale relative to the
idealised treatment with perfect spherical symmetry. }\label{fig:frac_diff}
\end{figure}

As is clear from Figures~\ref{fig:results:allcorrs} and~\ref{results:main:ee_ed_cssplit}
(and even more so in the additional measurements shown in Appendix~\ref{app:wgp_wpp}),
lack of statistical power ensures the errorbars on virtually all of
the correlation functions based on \mb~are significant.
Comparisons of the sort attempted here benefit at some level from the fact that
the shape noise properties of the simulated dataset are largely unaffected by symmetrisation.
Modulo low-level bin-shifting in the one-halo regime, then, one might expect the statistical
significance of the \emph{difference} between symmetrised and unsymmetrised correlation 
functions to be greater than that on either of the measurements in isolation.   
As discussed in the previous section, we seek to minimise the impact of noise
by evaluating $f_\alpha$ using smooth fits rather than the measured correlation functions
in the denominator of Equation~\eqref{eq:anisotropy_bias_dfn}.
We show the fractional bias in the lower panel of Figure~\ref{fig:frac_diff},
along with the absolute residual in the upper panel. 

\subsection{Robustness of Results}\label{sec:results:robustness}

\subsubsection{Central Flagging \& Choice of Symmetrisation Pivot}

\begin{figure}
\includegraphics[width=\columnwidth]{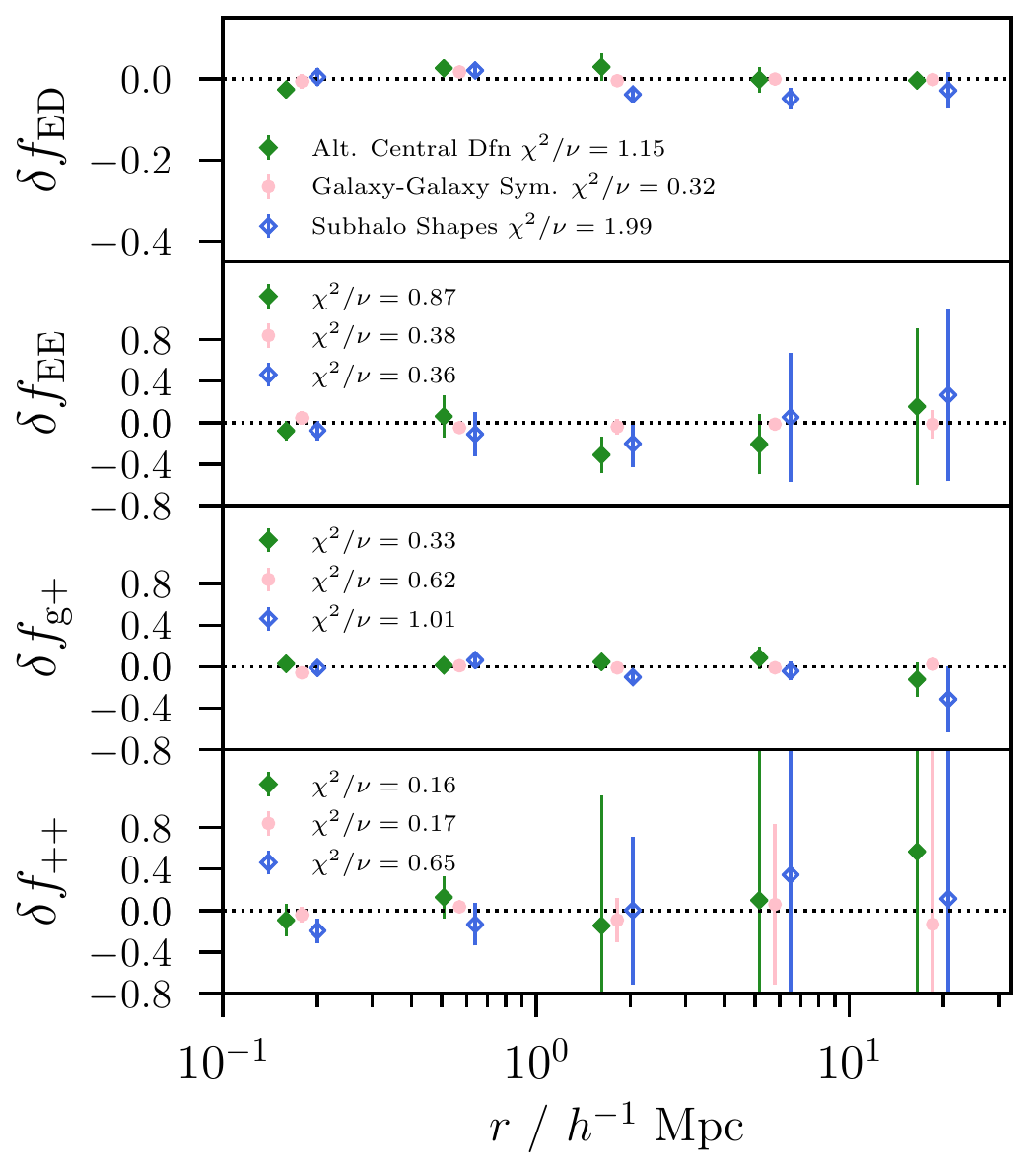}
\caption{The change in the fractional anisotropy bias in the four alignment correlations
discussed in this paper due to various analysis choices.
The pink points show the shift due to symmetrising satellite galaxies about
the central galaxy instead of the halo centre of mass. The filled green diamonds show
the impact of using a simple geometric definition for ``central" galaxies, as opposed to
the combined definition (c) in Section \ref{sec:catalogues:cflag}.
The open blue diamonds show the change induced by using ellipticities and orientations
derived from the underlying dark matter subhalo of each galaxy instead of the visible component.
The reduced $\chi^2/\nu$, where $\nu=5$ is the number of
data points in the measurement, is shown in the legend for each measurement.
A small horizontal offset is applied to the points in each bin to aid visibility.
}\label{fig:results:robustness}
\end{figure}

In this section we seek to test the explicit choices made during the course
of this analysis, with the aim of demonstrating the robustness of our
results.
The most obvious (and controllable) such choices regard the halo symmetrisation
process;
we chose to spin galaxies about the dark matter potential minimum.
Similarly, the galaxies labelled as ``centrals" are chosen by defining
a fixed boundary at $0.1R$ and identifying the most massive galaxy within
that sphere (see Section~\ref{sec:catalogues:cflag}).
This quantity has no clear mapping onto observables, and the
correspondence to the objects classified as centrals in real data is non-trivial.
We thus rerun our pipeline twice using slight modifications to these features.
That is, we (a) perform the symmetrisation of satellites using the central galaxy
of a halo as the pivot instead of the potential minimum
and then (b) flag central galaxies by minimising the Euclidean distance
from the potential minimum within each halo,
rather than the more complicated definition described.
The results are shown by the pink and green points (circles and filled diamonds) 
in Figure~\ref{fig:results:robustness}, which quantify the difference in anisotropy bias when
switching to the modified analysis configuration.

As is apparent here, neither choice has a significant impact on our results.
The reduced $\chi^2$ values, shown in each panel, give us no reason
to suspect the deviations from zero are anything more systematic than statistical noise.  
We also test the impact of switching from the basic inertia tensor to a reduced version,
wherein the stellar particles used to compute the moments of a subhalo are weighted by
the inverse square of the radial distance from the centre of mass.
This induces a more significant difference, at the level of $\delta f \sim -0.4$.
The interpretation of this result is not totally straightforward.
Dividing by the radial distance in effect imposes circular weighting, 
which will bias the resulting projected ellipticity low.
Although the difference in the weight of
the wings of a galaxy relative to its core impacts how strongly it
is affected by symmetrisation,
differences in the effective shape bias will also produce such effects.
The magnitude and direction of the shift noted here is approximately
the same as that seen across several mass bins 
by \citet{tenneti15b} (see their Fig. 5; compare black and blue solid lines).
This, while not definitive, does suggest that other factors
contributing to the difference in the $f_\alpha$ are likely subdominant to
the shape bias.

\subsubsection{Comparing \mb~\& \blockfont{Illustris}}

Though grouped under the umbrella term ``hydrodynamical simulations",
the choices that go into building a dataset such as \mb~can significantly
affect its observable properties. 
A small handful of comparable simulations exist in the literature,
most notably 
\blockfont{Horizon-AGN} \citep{dubois16},
\blockfont{Illustris} 
(\citealt{vogelsberger14}; 
and as of December 2018 also \blockfont{IllustrisTNG} \citealt{nelson18}),
\blockfont{EAGLE} \citep{schaye15} and
\blockfont{cosmo-OWLS} \citep{lebrun13}.

\begin{figure}
\includegraphics[width=\columnwidth]{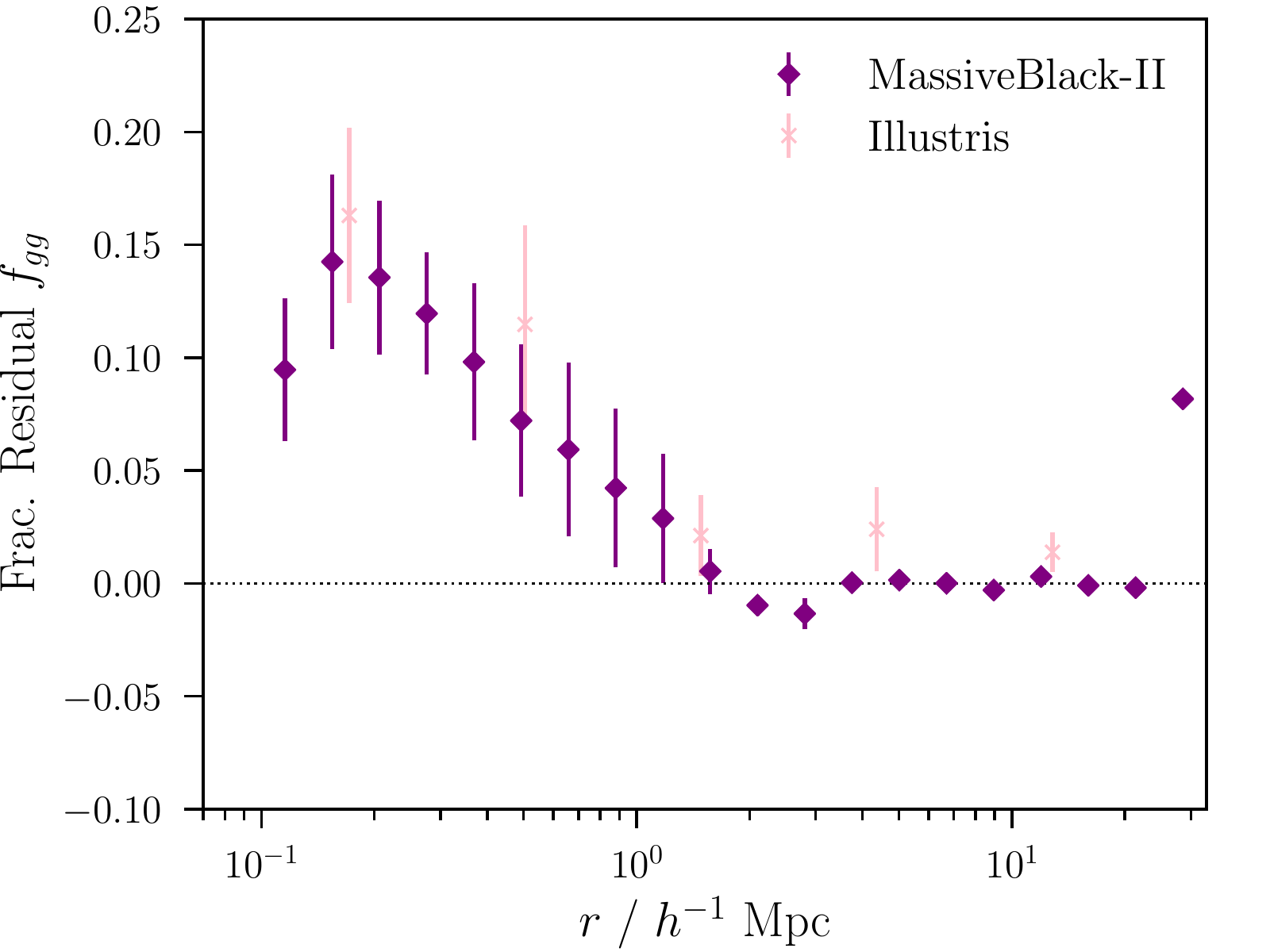}
\includegraphics[width=\columnwidth]{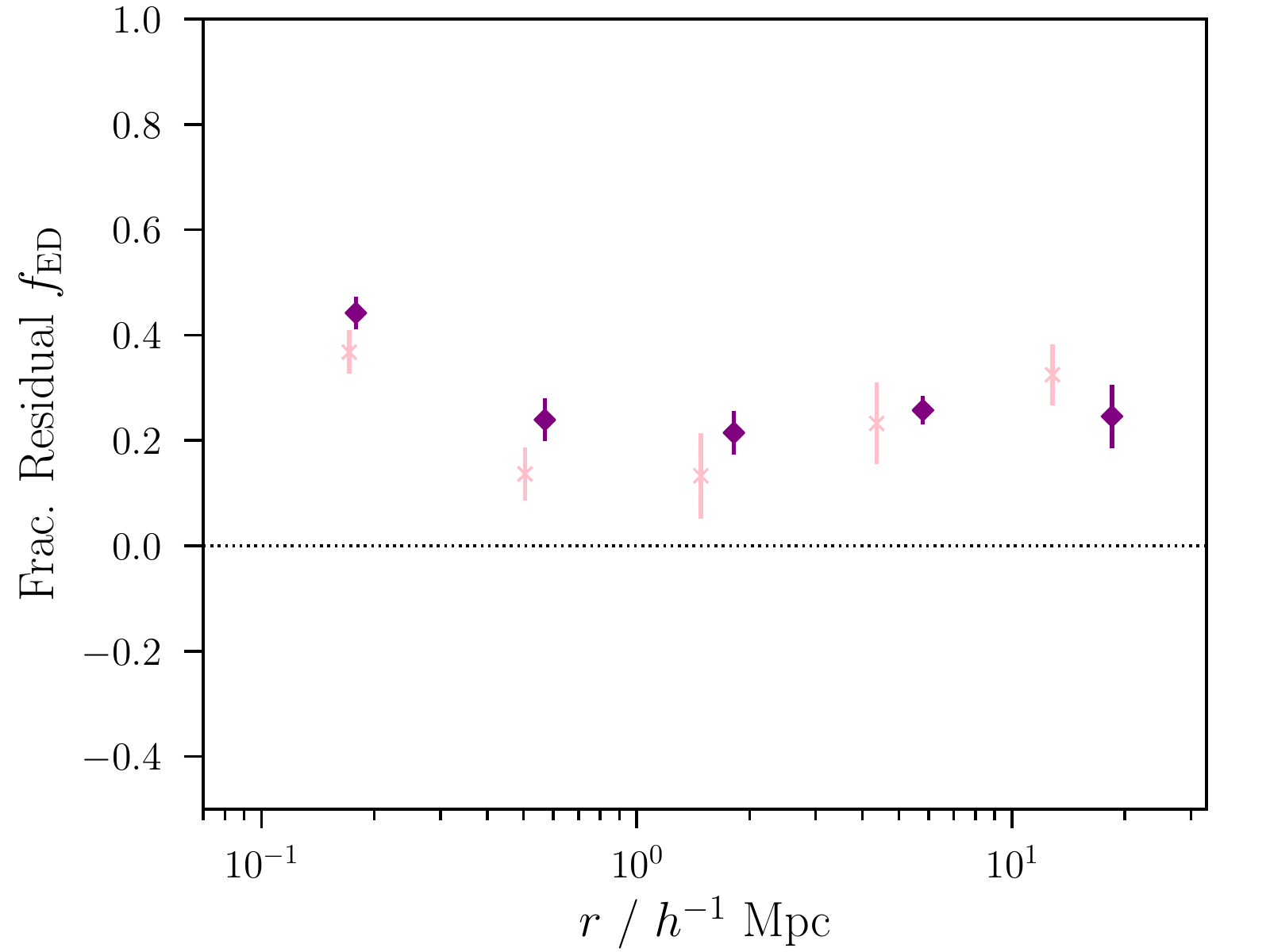}
\includegraphics[width=\columnwidth]{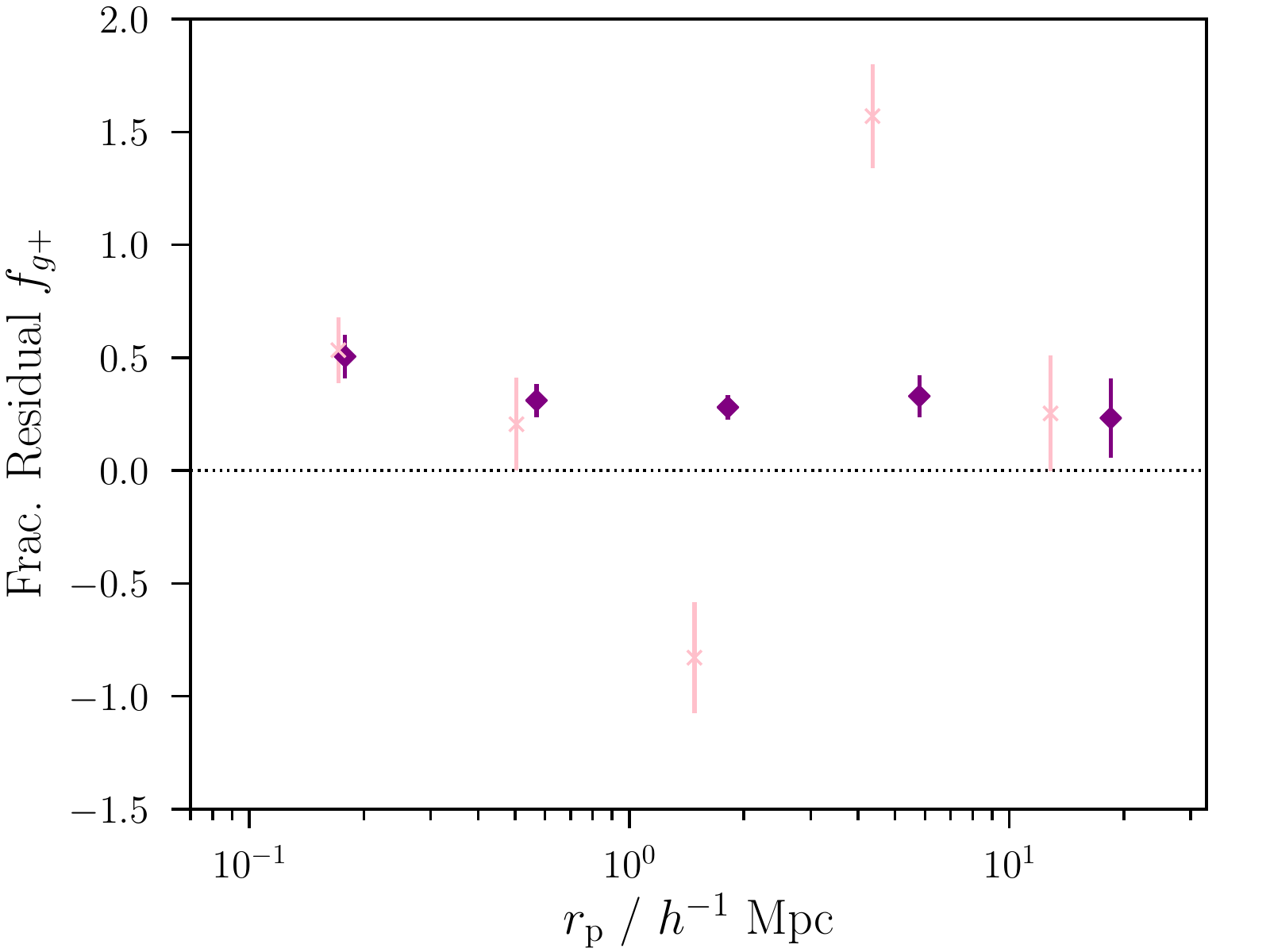}
\caption{A comparison of anisotropy bias entering various correlation functions,
as measured on the \mb~and \illustris~galaxy
catalogues. We show here (from top) 3D galaxy-galaxy, 
3D ED and
projected galaxy-tangential shape ($w_{g+}$).
In each panel the purple
diamonds indicate \mb, while the pink crosses show measurements based on \illustris.
Note that we omit the EE and $w_{++}$ correlations here;
though these measurements were made on \illustris, the signal-to-noise
is sufficiently low as to make the comparison unenlightening.  
  }\label{fig:results:illustris_comp}
\end{figure}

In the recent past a number of studies have set out to
explore the behaviour of intrinsic alignment between galaxies
in these mock universes (see 
\citealt{codis15a, chisari15, velliscig15, tenneti16, chisari16})
with not entirely consistent results.
\citet{chisari15}, for example,
report a dual IA mechanism
in early-type and late-type galaxies,
with the latter aligning tangentially about the former,
and early-type (spheroidal) galaxies tending to point 
towards each other.
Addressing the same question using the public 
\mb~and \blockfont{Illustris} data, however,
\citet{tenneti16}
find no such duality in either simulation.
No conclusive answer has been provided as to why these
datasets disagree,
although there has been speculation \citep{chisari16}
that it is a product of differing prescriptions for
small scale baryonic physics
(see also \citealt{soussana19}, whose findings appear to support this idea).
Similarly, it has been noted in \blockfont{Horizon-AGN}
that as the 
mass of the host halo declines, there is a transition from
parallel alignment between galaxy spins and the direction of the closest
filament to anti-alignment.
Such a shift is expected based on
observational data \citep{tempel13},
and tidal torque theory
\citep{codis15b}, but has not been reported in
\mb~\citep{chen15}. 
More recently, 
and somewhat in tension with earlier results, 
\citealt{krolewski19} report a transition from alignment 
to anti-alignment in dark matter subhalo shapes in both
\blockfont{Illustris} and \mb. Notably, in the former
case a very similar sign flip is seen in stellar shapes,
but this feature is not seen in the latter.
Though we highlight the discrepancies here, 
it is worth bearing in mind that despite
differing considerably in aspects of their methodology 
there is consistency between the bulk of IA measurements 
on hydrodynamical simulations, at least up to a
constant amplitude offset.

Given that we have no \emph{prima facie} reason to 
believe one simulation over another, we will consider such
variation as a source of systematic uncertainty
and seek to constrain it as best we can.
To this end we apply our pipeline to the public \blockfont{Illustris-1}
dataset. This requires some small changes in our treatment of cuts and periodic
boundary conditions to account for differences in
mass resolution and box size, but the comparison is otherwise straightforward.
We show the fractional anisotropy bias in the resulting correlation functions in Figure~\ref{fig:results:illustris_comp}.

The simplest case, that of galaxy-galaxy clustering, is shown in the upper left-hand panel. 
The two measurements are consistent to within statistical precision.
It is worth bearing in mind here that we should not expect the two to be identical
as the galaxy samples differ slightly in comoving number density and background cosmology.
Within the bounds of our jackknife errorbars the impact of halo symmetrisation is 
likewise consistent in the IA correlations and enters on the same scales in the two datasets.
Though one can see minor differences in the alignment correlation functions themselves
(not shown here)
our results do not indicate any clear discrepancy in the anisotropy bias between 
\mb~and \illustris.
There are known issues with the two simulations;
the AGN feedback prescription in \illustris, for example, is known to overestimate the 
magnitude of the effect \citep{vogelsberger14}.
Likewise the absolute alignment amplitude in \mb~has been shown to
be overpredicted relative to data
\citep{tenneti16}.
Such flaws affect the stellar mass function and other basic sample statistics
but, notably, do not translate into differences in anisotropy bias.

\subsubsection{Stellar vs Dark Matter Shapes}\label{sec:robustness:subhaloshapes}

As discussed in Section \ref{sec:measurements:shapes}, use of the stellar component for
measuring galaxy shapes is a well-motivated analysis choice,
observationally speaking;
however advanced the measurement algorithm, we are constrained
to measuring the shear field using luminous matter at the
positions of galaxies.
A natural question, however, is whether the intrinsic alignment signal
(and indeed the anisotropy bias) imprinted in galaxy shapes
accurately reflects
the properties of the underlying dark matter subhalos. 
We test this by rerunning our measurements using orientations
and ellipticities derived from the dark matter particles associated with each
galaxy.
The resulting shift $\delta f_\alpha$ is shown by the open blue points 
in Figure~\ref{fig:results:robustness}. 
For each statistic the reduced $\chi^2$ per degree of freedom
is, again, shown in the legend.
As with the other tests in this section
we report no statistically significant shift.
That is, though the raw alignment statistics measured with
visible galaxy shapes and dark matter subhalo ellipticities differ,
the aniosotropy bias does not.
This is intuitively understandable, given that the positions and
orientations of dark matter subhalos mirrors relatively closely
that of visible satellite galaxies.

\subsection{Dependence on Halo Mass}

Though motivated by simulation convergence, the precise cuts applied 
to the fiducial catalogue $(n_\mathrm{m}<1000, n_\mathrm{st}<300$; 
see Section~\ref{sec:catalogues}) are
somewhat arbitrary.
Given the constraints of the dataset 
(no ray-tracing or shape measurements, no reliable photometric fluxes, 
and galaxies sitting in a limited number of well-spaced redshift slices)
constructing a realistic sample representative of a real lensing survey
is not straightforward.
One factor we can, however, test is the dependence on galaxy (subhalo) mass.
This is useful in the sense that in real data the finite flux limit will
in effect impose a lower mass cutoff at fixed redshift;
if our results are robust to the exact cutoff, it suggests more general 
applicability in the practically useful context of a shape sample in a modern
lensing survey.   

\begin{figure}
\includegraphics[width=\columnwidth]{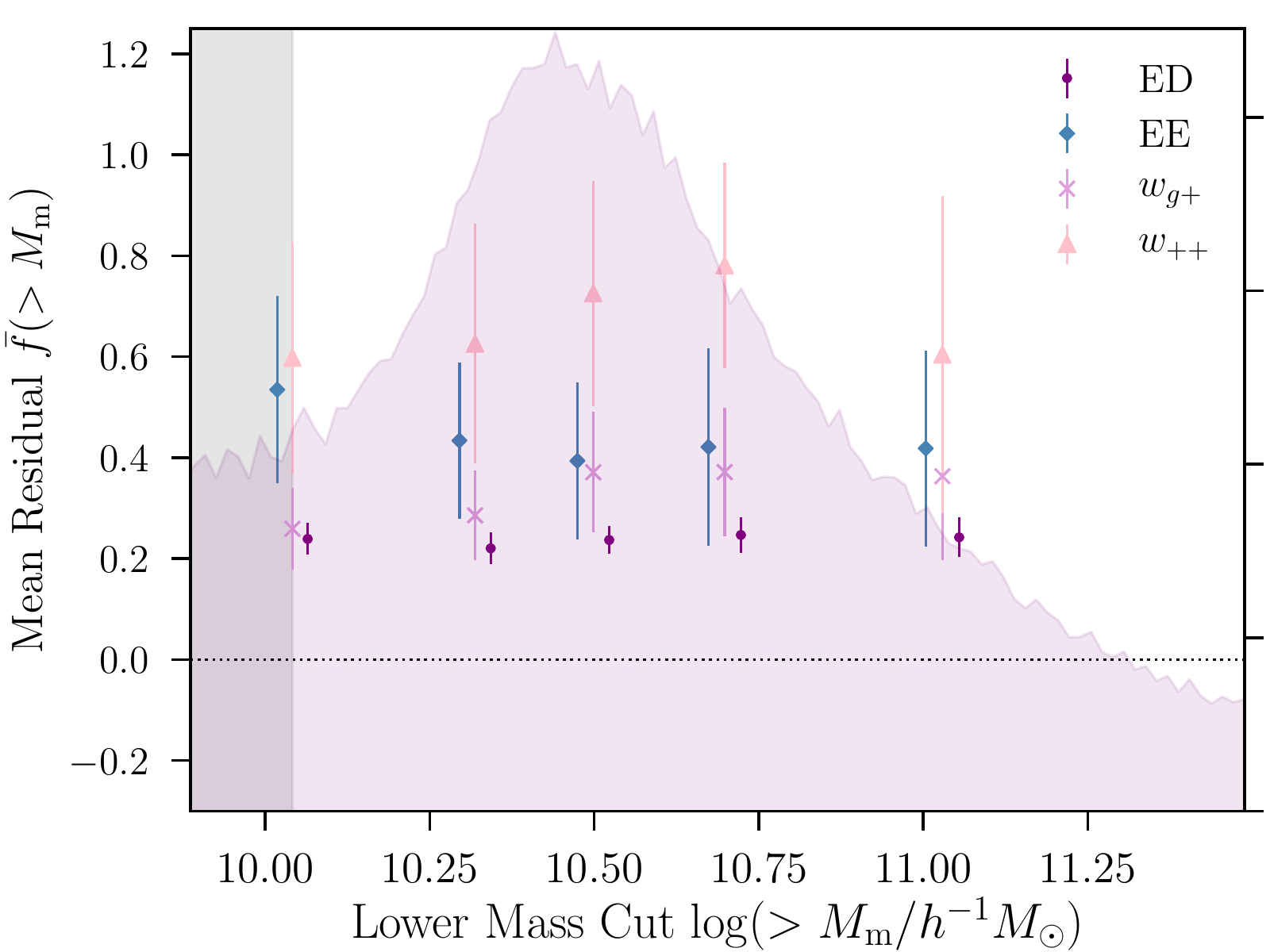}
\includegraphics[width=\columnwidth]{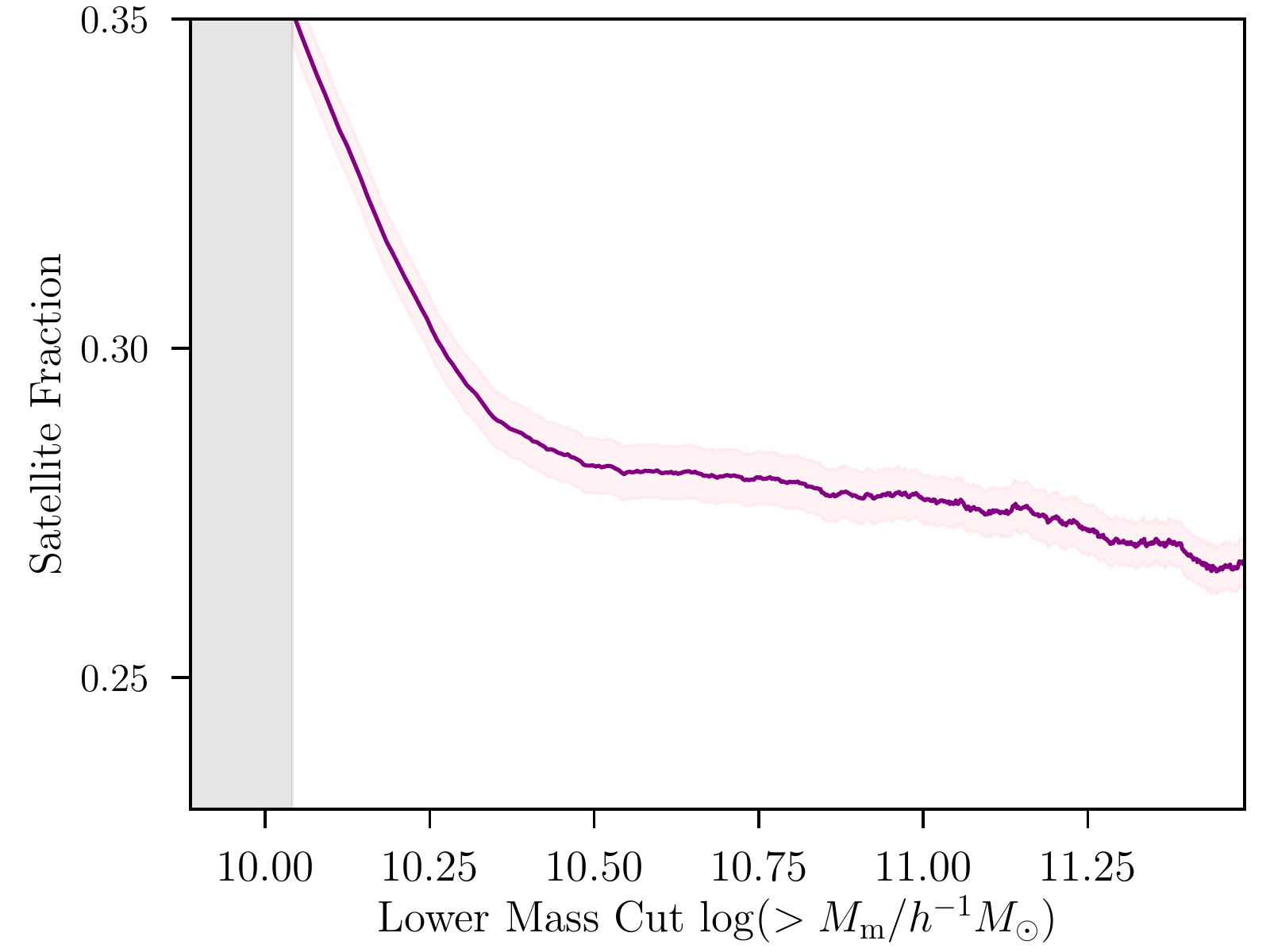}
\caption{\emph{Upper panel}: Fractional anisotropy bias as a function of the lower subhalo mass threshold
applied to the catalogue. The different point styles show the four alignment statistics
discussed in this work. In order to reduce $f_{\alpha}(r)$ into a single number,
the measured fractional difference in the correlation functions
is averaged on scales larger than $1 h^{-1}$ Mpc,
in which regime it is approximately constant.
The grey shaded region corresponds to the fiducial cut at 
$1.1 \times 10^{10} h^{-1} M_\odot$,
or 1000 dark matter particles.
Subhalos below this threshold are discarded to ensure convergence
in the three dimensional inertia tensor calculations.
A small horizontal offset has been applied to
aid visibility.
The underlain histogram shows the distribution of galaxy masses in
the uncut catalogue.
\emph{Lower panel}: The variation in satellite fraction with lower mass cut.
The purple line show the fiducial central flag definition
and the shaded band shows the $1\sigma$ jackknife uncertainty.
}\label{fig:results:fbias-vs-dmmasscut}
\end{figure}

To test this we construct a supersample of the fiducial catalogue with
a lower dark matter mass cut (300 particles).
The intrinsic alignment correlations are remeasured repeatedly with a series
of increasing mass thresholds.
The resulting fractional anisotropy biases are shown in the upper panel of 
Figure~\ref{fig:results:fbias-vs-dmmasscut}.
The measurements here are designed to illustrate the evolution
of the large scale anisotropy bias, and so are limited to scales
larger than $1 h^{-1}$ Mpc, where $f_\alpha$ is approximately
constant (see the lower panel of Figure~\ref{fig:frac_diff}).
The lack of a strong correlation here is worth remarking on; 
given the unrepresentative nature of our data,
it offers some indication that our results might be applicable beyond this
particular simulated galaxy sample.

\begin{figure*}
\includegraphics[width=\columnwidth]{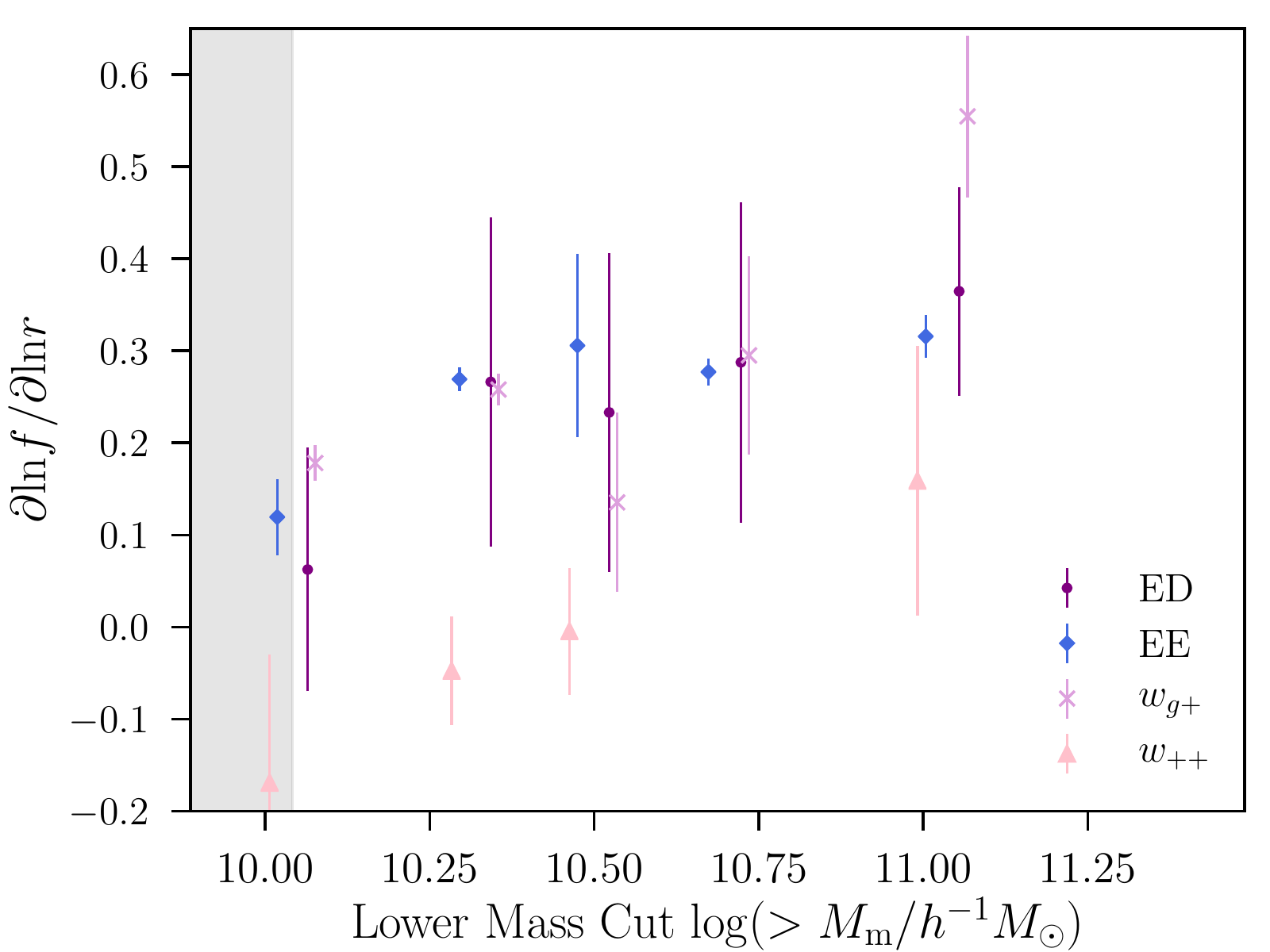}
\includegraphics[width=\columnwidth]{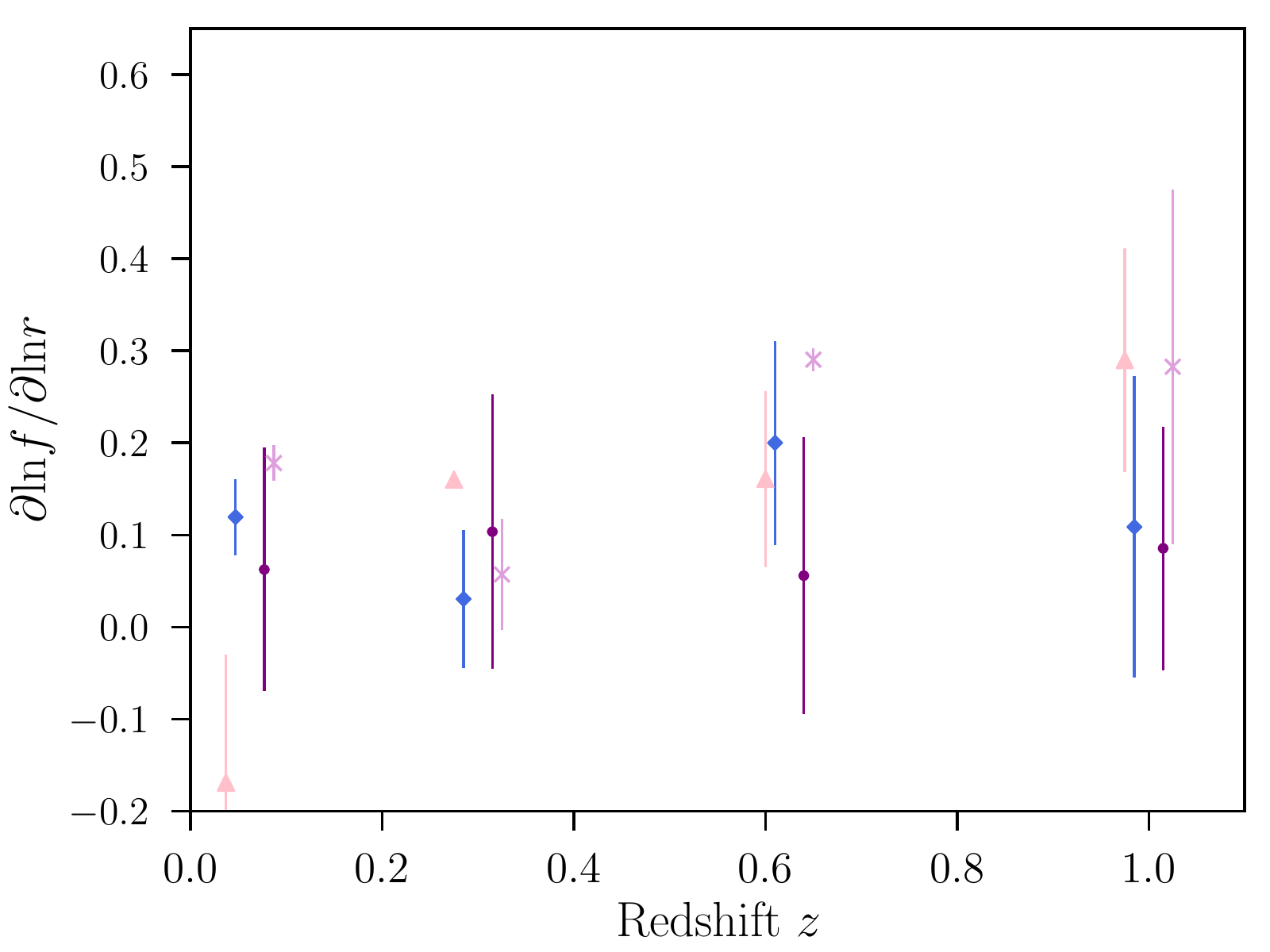}
\caption{Gradient of the small scale fractional anisotropy bias $\partial \mathrm{ln} f(r) / \partial \mathrm{ln}r$,
as a function of minimum subhalo mass (\emph{left}) and of redshift (\emph{right}).
As above, the four point styles indicate different alignment correlations. 
The shaded band in the left-hand panel shows the fiducial mass cut used in this paper.
Note that the points are slightly offset from each other along the horizontal axis to aid visibility.
}\label{fig:results:df_dr_dependencies}
\end{figure*}

To examine the evolution of the small scale anisotropy bias with halo mass we perform a
similar exercise as the above;
where before we averaged $f_\alpha(r)$ on large scales to obtain a single number,
our summary statistic is now the best fitting power law slope $\partial \mathrm{ln} f / \partial \mathrm{ln}r$,
fit on scales below $3h^{-1}$ Mpc.
We have seen that the large scale amplitude of the anisotropy bias is relatively stable
to the sample selection. The idea now is to test whether the shape of the 
scale dependent bias is similarly stable.  
This is shown in the left panel of Figure \ref{fig:results:df_dr_dependencies}.
In contrast to the previous result, we now see hints of a correlation.
All four of the correlations show a gradual but consistent increase in 
the slope of the anisotropy bias as the mass threshold is raised. 
This can be interpreted as follows. 
There is evidence in the literature that the three dimensional 
dark matter distribution of massive halos tends to flatter that that of 
smaller halos \citep{jing02, springel04, allgood06, schneider12}.
It follows, then, that the satellite distributions should be
more concentrated at small azimuthal angles
(also bourne out in the literature; see \citealt{vandenbosch16}, \citealt{huang16}), 
and so the small scale
ED $cs$ signal is stronger. This signal is strongly scale dependent,
and is also washed out entirely by symmetrisation.
This competes with ED $sc$ on small scales;
in massive halos ED $cs$ makes up a larger fraction of the 
overall alignment signal, and so the anisotropy bias is 
exhibits a stronger scale dependence. 

One can also think about the question in terms of one- and two-halo contributions.
At high masses, the small scale
anisotropy bias is dominated by intra halo central-satellite interactions, 
which fall off rapidly with separation. As the mass cut shifts downwards,
one is preferentially including galaxies in low mass halos, which induces 
a net reduction of the mean satellite fraction. Since the host halos of low mass
galaxies contain
very few satellites, the impact of the symmetrisation process enters only
at the level of two halo interactions, which scale more gradually with
separation. This is true for both shape-shape and shape-position correlations.

The above arguments provide, at least in part, an explanation for
the apparent mass dependence seen in Figure \ref{fig:results:df_dr_dependencies}.

\subsection{Redshift Dependence}

The results presented in the earlier sections of this paper are based exclusively
on the lowest simulation snapshot, at $z = 0.062$.
We now explore how our findings change as one approaches redshifts more 
typically used for cosmic shear measurements.  

\begin{figure}
\includegraphics[width=\columnwidth]{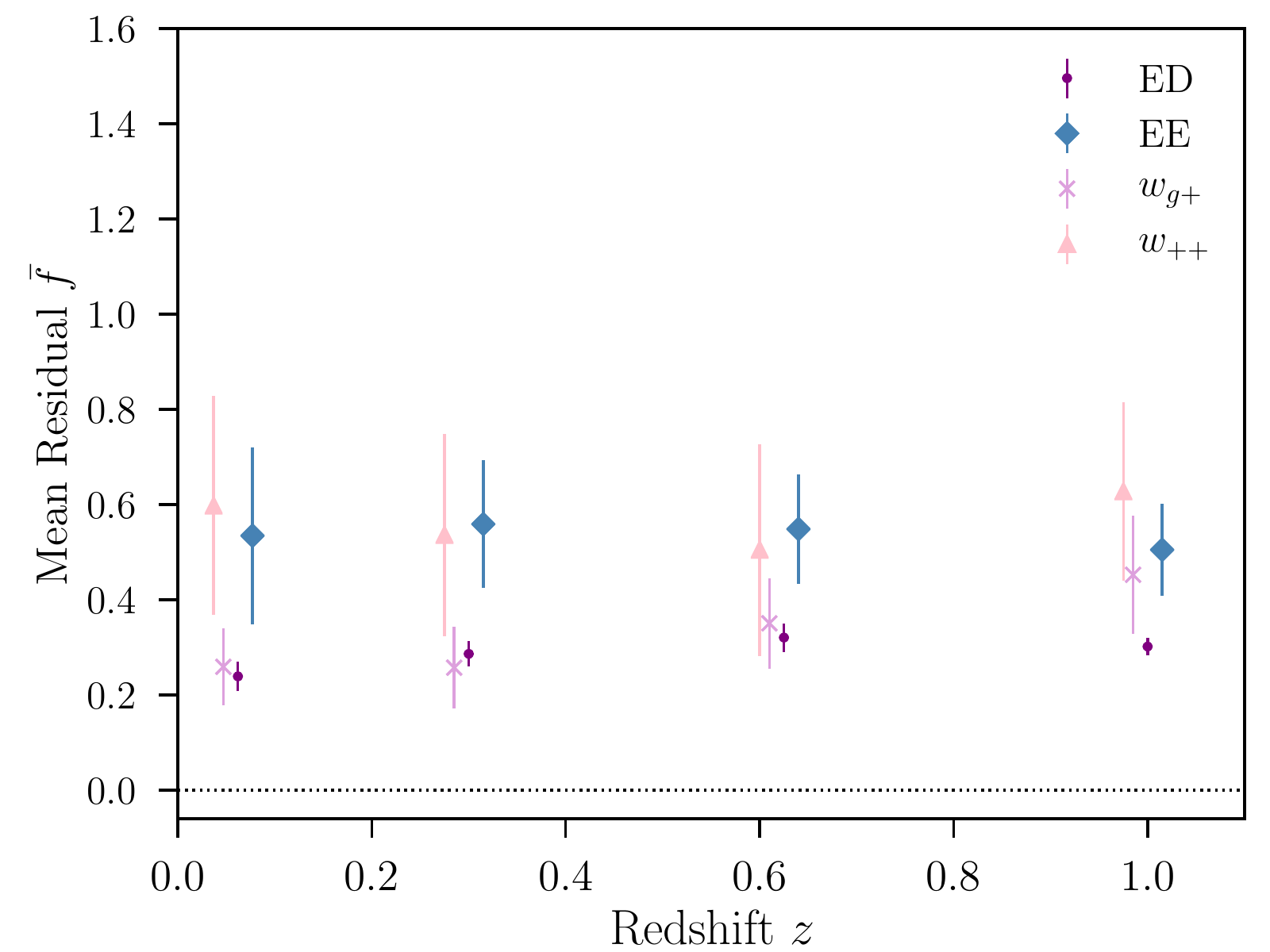}
\caption{The redshift evolution of large scale anisotropy bias.
Each bias point for a specific correlation function at given redshift
is the mean over comoving scales larger than $1 h^{-1}$ Mpc,
where $f(r)$ is approximately constant.
Note that we do not show binned values here; 
the discrete redshifts are defined by the four simulation snapshots
(see Table \ref{tab:catalogues:statistics}) 
and are exact and known precisely, meaning the horizontal errorbar is
effectively zero.
The points at each redshift are offset slightly from one another along the x-axis 
to aid visibility. }\label{fig:results:redshift}
\end{figure}

The catalogue building and symmetrisation pipeline is applied
independently to three snapshots at redshifts
$z = [0.300,0.625,1.000]$,
in addition to the fiducial dataset at $z=0.062$.
Though this will not capture correlations between measurements at
different redshifts, it allows us to make a simple comparison of the
relative importance of satellite anisotropy in different epochs.
The result is shown in Figure~\ref{fig:results:redshift}.
Note that in order to condense a fractional anisotropy bias, naturally measured
as a function of scale, into a single number we again average $f(r)$ on scales
$r>1h^{-1}$ Mpc.
This cut is intended to isolate the two-halo
regime, within which the effect of symmetrisation is roughly constant.
As before, our conclusions here are limited by the statistical power of the
dataset.
The ED correlation and, to a lesser extent, $w_{g+}$ do however offer a meaningful 
constraint. In these cases we find a potential weak positive redshift dependence over
the range in question.
Linear fits to the four points shown in Figure~\ref{fig:results:redshift} give
slopes of $m_\mathrm{ED} = 0.05 \pm 0.03$ and $m_{g+} = 0.20 \pm 0.03$ for ED
and $w_{g+}$ respectively.
No clear evolution is seen in $f_\mathrm{EE}$ and $f_\mathrm{++}$, 
though the error bars are also consistent with a moderate correlation
(of either sign).

We also fit a power law to the small scale fractional anisotropy bias $f_\alpha(r| r<3h^{-1}$ Mpc$)$
at each redshift. This exercise is entirely analogous to the earlier test for dependence on the shape
of $f(r)$ on the halo mass threshold.
The result is shown in the right-hand panel of Figure \ref{fig:results:df_dr_dependencies}.
As in Figure~\ref{fig:results:redshift}, we see no evidence of coherent evolution in 
the scale dependence of the bias with redshift.

It is worth remarking here that the catalogue selection mask is applied independently
in each redshift.
That is, the composition of the sample changes between snapshots,
and so apparent redshift dependence in the anisotropy bias can arise
either from genuine evolution in a fixed set of objects
or from differences in the ensemble properties of the galaxies
surviving mass cuts.
Disentangling these two effects would require access to \mb's merger history 
(in order to track individual subhalos between snapshots)
and is considered beyond the scope of the current analysis.

\section{Implications for Cosmology}\label{sec:cosmology}

Though assessing the impact of symmetry on the alignment correlation functions
is qualitatively useful for demonstrating a physical effect,
it says nothing about the penalties for failing to model that effect.
That is, what we ultimately wish to know is how
biased would future cosmology surveys be were they
to rely on a spherically symmetric model for intrinsic alignments.
If a spherical model inaccurately describes reality, but does not significantly
bias the cosmological information in small scale shear correlations, 
then most cosmologists would consider it sufficient.  
It is to this question we turn in the following section.
To this end we perform a simulated likelihood analysis along the lines of
\citet{y1methodology} 
using mock lensing data designed to mimic a 
future LSST-like lensing survey. The key ingredients are detailed below.
Where it is necessary to assume a fiducial cosmology, we adopt the best fitting
Planck 2018 \lcdm~constraints \citep{planck18},
$\mathbf{p}_\mathrm{cos} =
(\omegam,\as,\ns,\omegab,h) = \\
(0.311, 2.105\times10^{-9}, 0.967, 0.049, 0.677)$.

\subsection{Mock Cosmic Shear Data}

\begin{figure}
\includegraphics[width=\columnwidth]{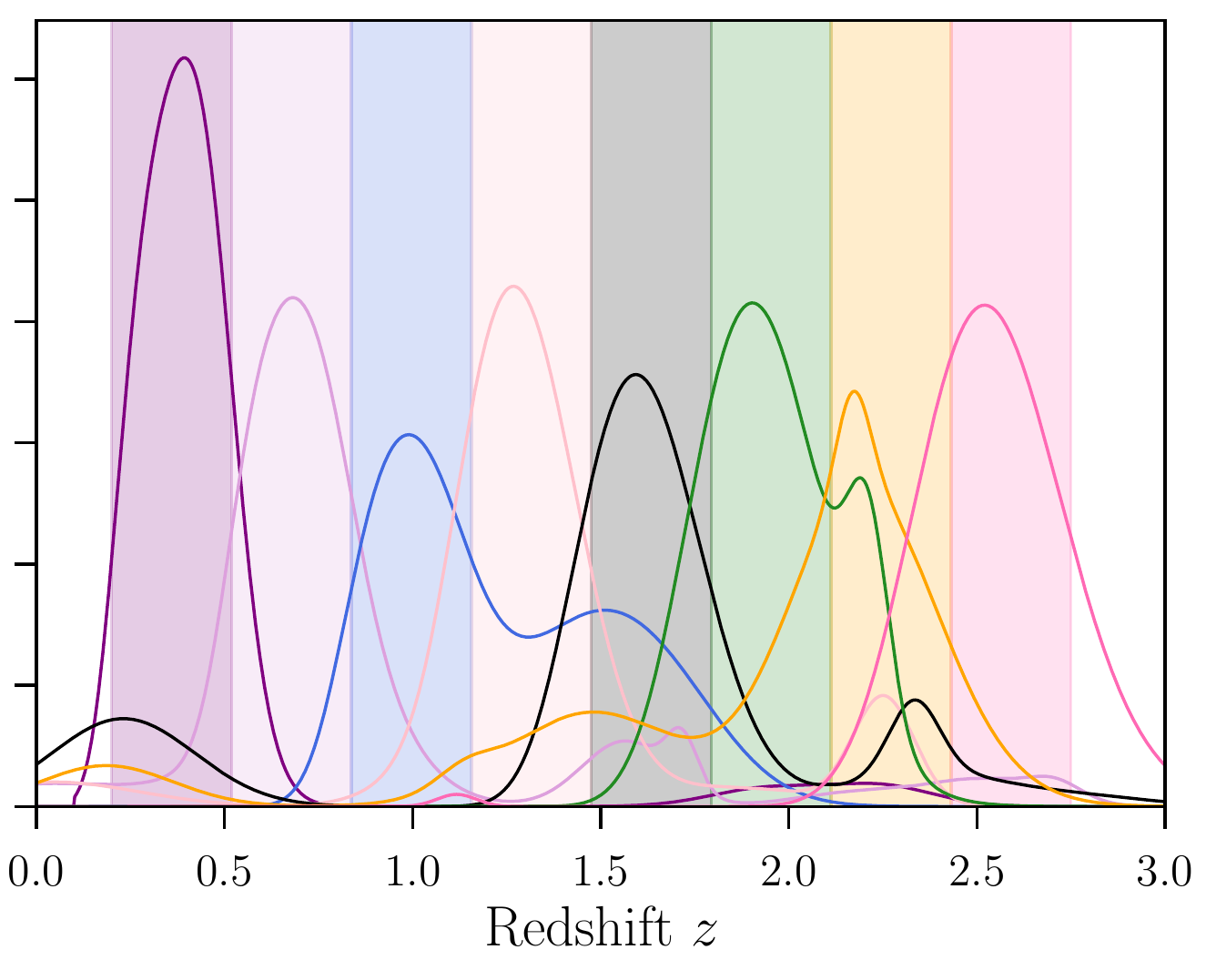}
\caption{The mock photometric redshift distributions used in our simulated likelihood analysis.
The sample is distributed among eight bins in the range
$z=[0.20-2.75]$ with number densities ${\bf n}_\mathrm{g} = (3.3,3.1,3.0,1.5,1.5,1.0,0.5,0.5)$. 
The bounds of these bins are shown by the coloured bands.
}\label{fig:forecasts:nz_wl_gg}
\end{figure}

We set up our synthetic
data vector as follows.
We first use the public Core Cosmology Library\footnote{v1.0.0; \url{https://github.com/LSSTDESC/CCL}} 
(\blockfont{CCL};
\citealt{cclpaper}) to construct source redshift distributions.
We assume eight redshift bins over the range $z=0.20-2.75$ with Gaussian photo-$z$
dispersion $\sigma_z = 0.05$. In addition, we superpose Gaussian outlier islands centred on random
points drawn from the initial smooth $p(z)$ in each bin. One should note that
this is not a rigorous quantitative error prescription 
(though it is conceptually similar to the catastrophic outlier implementation
of \citealt{hearin10}). 
The point, however, is to generate qualitatively
realistic $p(z)$ with overlapping non-analytic tails, which may be salient when considering 
the impact of intrinsic alignments. 
The resulting distributions are shown in Figure~\ref{fig:forecasts:nz_wl_gg}.
We then use the Boltzmann solver \blockfont{CAMB} \citep{lewis99} to generate a matter power spectrum
at the fiducial cosmology. 
Nonlinear corrections are calculated using 
\blockfont{halofit} \citep{takahashi12}.
With the mock $p(z)$, this is propagated through the relevant 
Limber integrals to generate $C(\ell)$s and then Hankel transformed to 
correlation functions $\xi_\pm(\theta)$.
To model the covariance of these data we use a public version of the 
\blockfont{CosmoLike}\footnote{\url{https://github.com/CosmoLike/lighthouse\_cov}}$^{,}$\footnote{\url{https://github.com/CosmoLike/cosmolike\_light}} 
code \citep{krause17},
assuming a constant shape dispersion $\sigma_e=0.25$ per bin
and a total area of 18,000 square degrees.
Given the limited aims of this exercise 
we consider a (real-space) Gaussian covariance matrix to be sufficient.
Omitting higher-order covariance contributions 
(e.g. non Gaussian and super-sample terms, and the impact of survey masks; 
see \citealt{takada13,krause16,barreira17,troxel18}) is
expected to translate into an underestimation of one's sensitivity to the smallest scales.
This in turn will reduce the impact of small-scale mismodelling. 
In this sense, the results presented in the following should be treated 
more as a lower
bound on the possible bias than a rigorous numerical prediction.

\begin{figure}
\includegraphics[width=\columnwidth]{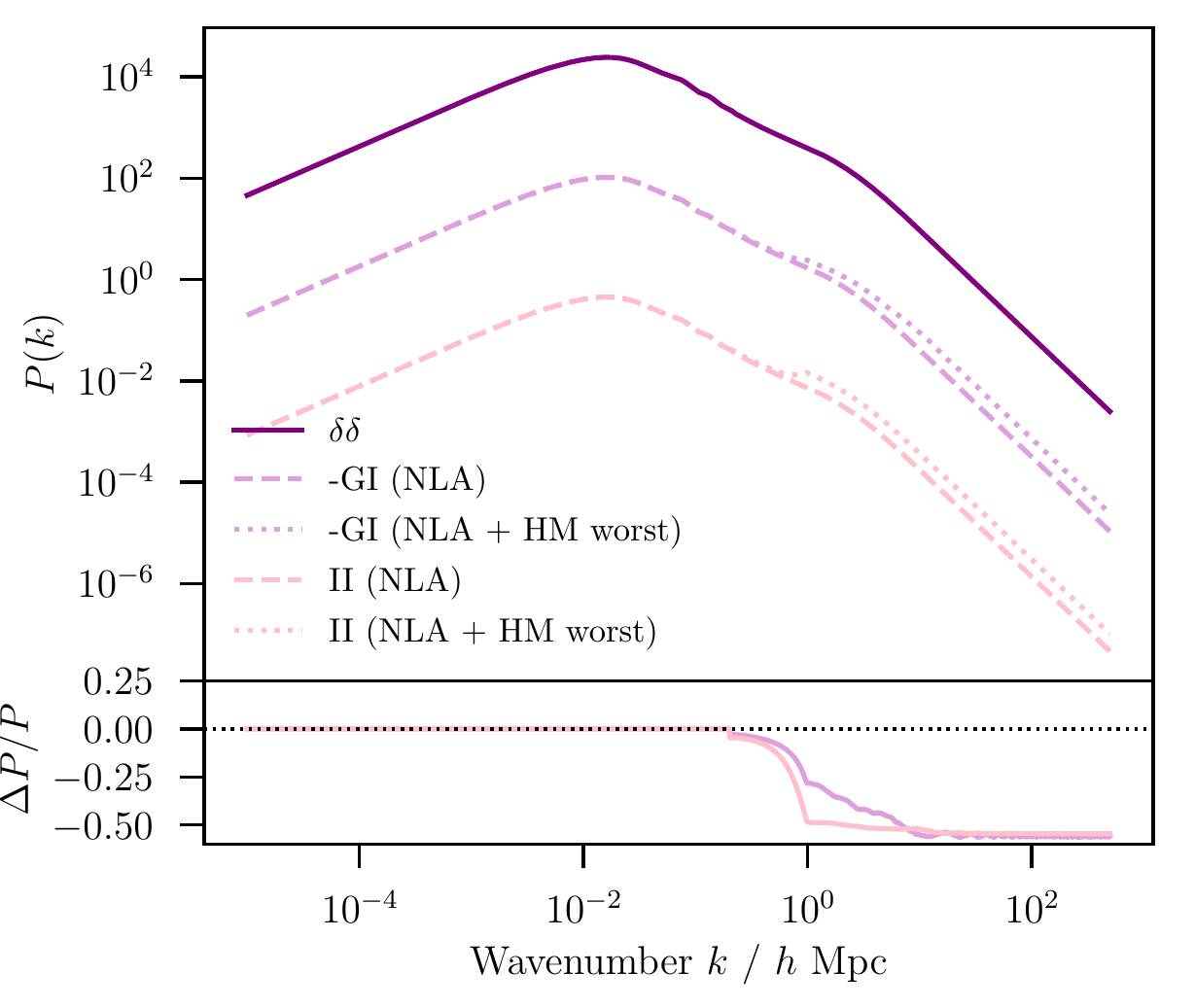}
\caption{Theory power spectra at $z=0$ used for constructing our mock lensing data.
The dashed lines show the baseline nonlinear alignment model prediction at the fiducial cosmology.
The combined power spectra including a modification at high $k$ to account for the worst-case impact
of assuming a spherical halo model are shown by the dotted lines.
For reference we show the $z=0$ matter power spectrum at the same cosmology in purple (solid).
}\label{fig:forecasts:pk_ia_nla_hm}
\end{figure}

To produce the intrinsic alignment contribution to these data we first 
compute the GI and II power spectra according to the NLA Model.
This prescription is assumed to capture the large scale IA correlations well; 
we modify it on small scales using a simple procedure designed to mimic the impact
of erroneously assuming halo symmetry.
The measurements of $w_{g+}$
and $w_{++}$ on the simulations are transformed into IA power spectrum using the relation:

\begin{equation}
P_{\delta_g \mathrm{I}}(k) = - \int \mathrm{d}r_\mathrm{p} 2 \pi r_\mathrm{p} J_2(k r_\mathrm{p}) w_{g+}(r_\mathrm{p}),
\end{equation} 

\noindent
and analogously

\begin{equation}
P_\mathrm{II}(k) = \int \mathrm{d}r_\mathrm{p} 2 \pi r_\mathrm{p} \left [ J_0(k r_\mathrm{p}) + J_4(k r_\mathrm{p}) \right ] w_{++} \left ( r_\mathrm{p} \right ),
\end{equation} 

\noindent
where $w_{++}$ and $w_{g+}$ are the (noisy) measured correlation functions from \mb
and $J_\nu$ is a Bessel function of the first kind of order $\nu$
(see Appendix A of \citealt{joachimi11} and \citealt{mandelbaum10}'s Equation 7).
These spectra are computed twice:
first using the unsymmetrised measurements of $w_{++}$ and $w_{g+}$,
and then using shifted versions corresponding to the maximum difference
allowed by the jackknife errorbars on $f_{++}$ and $f_{g+}$.
This yields difference templates $\Delta P_\mathrm{GI} (k)$ and $\Delta P_\mathrm{II} (k)$,
which are applied to the NLA GI and II power spectra generated from
theory.
Note that we use the templates derived from the lowest redshift snapshot
to modify the IA spectra at all redshifts.
Given the lack of systematic variation seen in Figures  \ref{fig:results:df_dr_dependencies} 
and \ref{fig:results:redshift} we consider this a reasonable decision. 
The resulting modified power spectra are shown in Figure \ref{fig:forecasts:pk_ia_nla_hm}.
It is worth reiterating here:
these templates are not accurate prescriptions for small scale intrinsic alignment
suitable for forward modelling, 
but simply a quantification of the maximum impact this systematic \emph{could} have,
given our results from the previous section.

\subsection{Simulated Likelihood Analysis}\label{sec:forecasts:constraints}

We perform a mock likelihood analysis on the contaminated data
(constructed using the ``NLA + HM Worst'' spectra from Figure~\ref{fig:forecasts:pk_ia_nla_hm}).
The baseline analysis includes eighteen nuisance parameters, in addition to six
cosmological parameters $\mathbf{p}_\mathrm{cos} = (\as,\ns,\omegam,\omegab,h,w)$,
following the broad
methodology set out by \citet{y1methodology}.
We adopt priors, as set out in Table~\ref{tab:forecasts:priors},
which are designed to mimic the specifications of a future LSST-like
lensing survey. 

\begin{figure}
\includegraphics[width=\columnwidth]{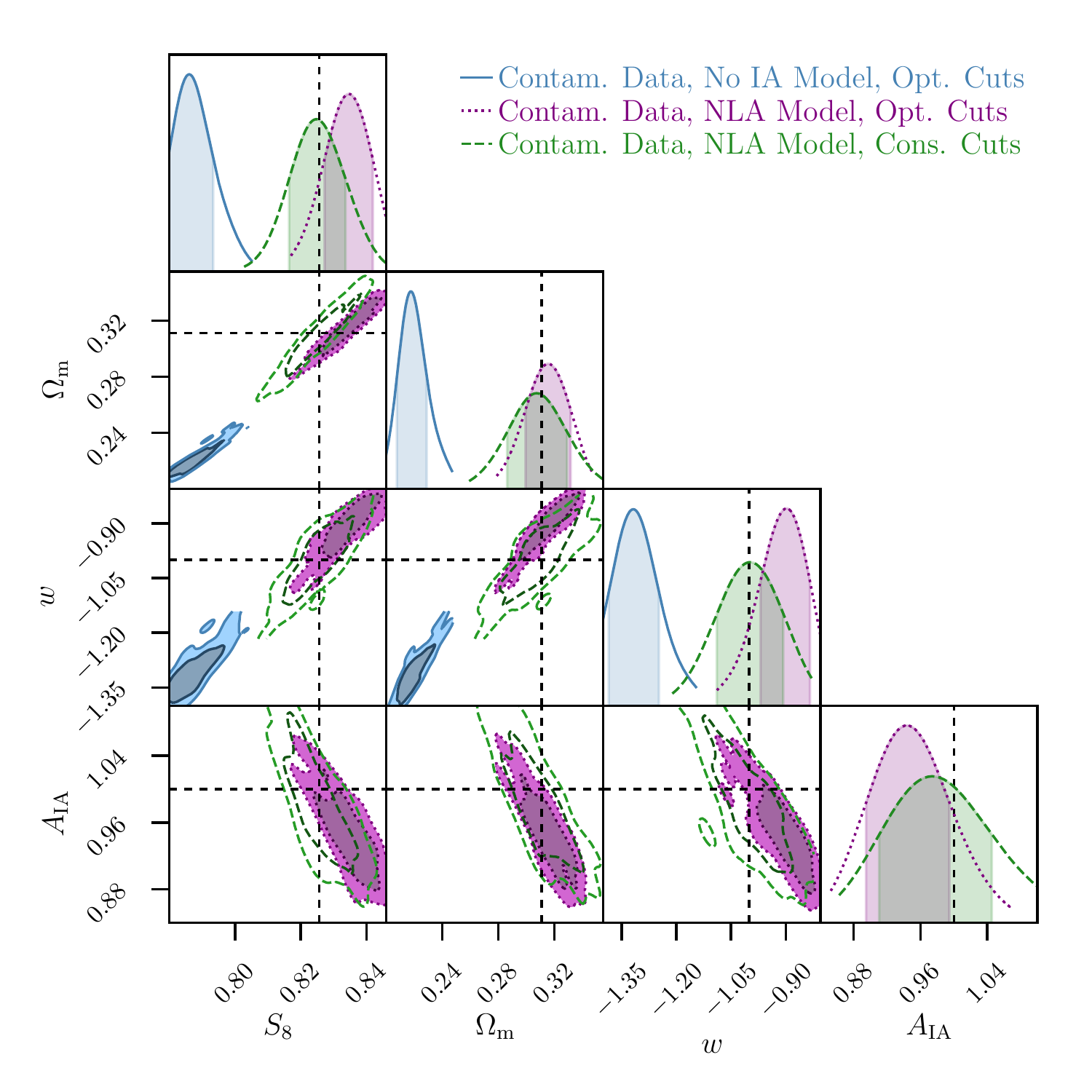}
\caption{Constraints from artificially contaminated shear data, 
under different analysis configurations.
The same mock data are used in all cases, modified in such a way 
to mimic the error caused by using a model for small scale IAs that
neglects satellite anisotropy.
The blue contours show a case in which no attempt is made to model intrinsic
alignments (i.e. $A_\mathrm{IA}$ is fixed at zero).
Purple show the same analysis, but also marginalising over a free IA amplitude
and redshift power law index.
In green (unfilled) we show the same result using stringent scale cuts, derived from
baryonic simulations (discussed in Section \ref{sec:forecasts:constraints}),
to mitigate the impact of the insufficient IA model on small scales.
}\label{fig:forecasts:constraints}
\end{figure}

\begin{table}
\begin{center}
\begin{tabular}{c|ccc}
\hline
Parameter            & Prior                                   & $\Delta p /\sigma_{p}$         & $\Delta p /\sigma_{p}$ \\
                     &                                         &  (Conservative)                      &  (Optimistic)  \\
\hhline{=|===}
$A_\mathrm{s}$       & $\mathrm{U}(1.1, 3.0 \times 10^{-9} )$  & $0.00(4)$                       & 0.84 \\
$n_\mathrm{s}$       & $\mathrm{U}(0.5,1.5)$                   & $-0.29$                          & $-3.42$ \\
$h$                  & $\mathrm{U}(0.4,0.9)$                   & 0.18                         & $-1.60$ \\
$\Omega_\mathrm{m}$  & $\mathrm{U}(0.25,0.40)$                 & $-0.16$                          & 0.22 \\
$\Omega_\mathrm{b}$  & $\mathrm{U}(0.01,0.10)$                 & $-0.11$                          & $-1.96$ \\
$w$                  & $\mathrm{U}(-4.00,-0.33)$               & $-0.01$                          & 1.42 \\
$\Omega_\nu h^{2}$   & $\delta(0.0006155)$ & 0 & 0 \\ 
\hline
$\Delta z^{i}$       & $\mathcal{N}(0.001, 0.01^2)$            & see text                       & see text \\
$m^{i}$              & $\mathcal{N}(0.001, 0.01^2)$            & see text                       & see text \\
$A_\mathrm{IA}$      & $\mathrm{U}(-6,6)$                      & $-0.33$                          & $-1.38$ \\
$\eta_\mathrm{IA}$   & $\mathrm{U}(-6,6)$                      & $-0.13$                         & $-2.10$ \\
\hline
\end{tabular}
\caption{Priors and parameter biases derived from the mock analysis described in 
this section.
The right-most two columns show the bias in the mean value of each parameter,
as a fraction of the $1\sigma$ width of the marginalised
$1\sigma$ posterior distribution, using two sets of scale cuts.
The rows in the lower part of the table are nuisance parameters (from the top, 
photometric redshift error,
shear calibration bias,
intrinsic alignment amplitude
and intrinsic alignment redshift power law index). }\label{tab:forecasts:priors}
\end{center}
\end{table}

The matter power spectrum is computed at each point in parameter space 
using \blockfont{CAMB}, with nonlinear modifications from 
\blockfont{halofit}.
To explore the posterior surface we use the \blockfont{emcee}\footnote{\url{dfm.io/emcee}}
\citep{foremanmackey12}
Metropolis-Hastings algorithm, as implemented in 
\blockfont{CosmoSIS}\footnote{\url{https://bitbucket.org/joezuntz/cosmosis}}
\citep{zuntz15}.
To ensure the chains presented in this work are fully converged 
(after burn in)
we apply the following criteria:
(a) the parameter values, plotted in order of sampling, appear visually
to be random noise about constant mean 
(i.e. with no residual direction or systematic variation in scatter)
(b) if the chain is split into equal halves, the projected 1D posterior
distributions evaulated using the two pieces do not significantly differ.
Each of the chains used has a total of $1.5-2.0$ M samples, 
of which approximately fifteen percent survive burn in.

The intrinsic alignment signal is modelled using the two-parameter NLA model 
implemented by \citet{y1cosmicshear},
which differs from the input IA model due to the small-scale modifications
described in the previous section. Note that this model includes a redshift scaling
of the form $[(z+1)/(z_0+1)]^{-\eta_\mathrm{IA}}$,
with a fixed pivot redshift $z_0=0.62$.

The scale cuts used in this analysis follow the basic methodology
of \citet{y1cosmicshear}.
That is, we compute our shear data vectors at a fiducial cosmology
first using \blockfont{halofit} alone,
and then using the same matter power spectrum, but rescaled to mimic 
the most extreme scenario of baryonic feedback in a suite of OWLS AGN simulations
\citep{vanDaalen11}.
The minimum angular scale to use in our analysis for each correlation function is
set to ensure the fractional difference in the two simulated versions of the 
data vector $|\Delta\xi^{ij}_\pm|/\xi^{\mathrm{halofit}, ij}_\pm<0.01$\footnote{This 
tolerance differs slightly from the value of $2\%$ chosen by \citet{y1cosmicshear} for DES Y1.
The greater stringency of the value adopted here 
is intended to reflect the improvement in statistical precision
of future surveys relative to the current generation.}.
This value is calculated for each bin pair $ij$ and results in 
$\theta^\mathrm{min}_{+} = 2.69$ arcminutes 
and
$\theta^\mathrm{min}_{-} = 39.91$ arcminutes
in the uppermost auto-bin correlations of
$\xi_+$ and $\xi_-$ respectively. 
For the purposes of forecasting, we will consider a second scenario
in which more (but not all) of the shear-shear data are included.
In this scenario, the minimum angular separation is
rescaled such that 
$\theta^{\mathrm{min},ij}_{\pm} \rightarrow \frac{1}{4} \theta^{\mathrm{min},ij}_{\pm}$.
In the analysis configuration described, with eight redshift bins and 50 angular bins per correlation,
this increases the number of data points included in the analysis from 1252 
to 1879.
We refer to the two sets of scale cuts respectively as ``conservative'' and ``optimistic'' cases.

For reference, the first scenario we will consider is one in which we neglect to model the impact of IAs
on any scale.
With the contaminated data described above and the optimistic scale cuts, 
this results in the blue contours in Figure \ref{fig:forecasts:constraints}.
Though it has been demonstrated elsewhere that neglecting IAs causes non-negligible
parameter biases \citep{krause16,blazek17,y1cosmicshear},
and this is clearly not a realistic analysis option,
it is illuminating as a simple baseline case,
with which to compare the size and direction of the biases in
the following.

We next imagine a slightly more realistic scenario in which the
optimistic cuts described above are employed. 
In this (purely hypothetical) setup, we assume that we can model baryonic physics
and nonlinear structure formation
perfectly; 
for small scale intrinsic alignments we imagine that we are using a halo model
that assumes spherical symmetry, in a way analogous to the way \blockfont{halofit} has
been used for nonlinear growth in the past.
This, again, is not an exact prescription but rather an illustration
of the (potentially biased) information content of the small-scale shear correlations. 
The forecast results of such an analysis are shown by the solid purple contours in 
Figure \ref{fig:forecasts:constraints}.
Unfortunately the small scale IA error in this scenario is sufficiently large to induce
cosmological biases of $> 1 \sigma$ 
(see the ``optimistic'' column in Table \ref{tab:forecasts:priors}).
Most notably, we see shifts in the best constrained parameters
$\Delta S_8=1.4 \sigma$, $\Delta w= 1.5 \sigma$. 
Although such marginal shifts might be dismissed in comparison with other
potentially larger systematics,
it is worth bearing in mind that this is an idealised model.
Indeed, for \emph{a single} systematic, a $\sim 1\sigma$ shift
is a non-trivial portion of the allowed error budget for an experiment
like LSST.
It is also worth remembering that this toy model neglects features
such as non Gaussian covariances, which will increase rather than
reduce sensitivity to small-scale mismodelling.
Interestingly, in the 24 dimensional parameter space the bias is not entirely 
(or even predominantly) absorbed into the IA model;
significant (several $\sigma$) compensatory shifts in the photo-$z$ nuisance parameters
are seen, particularly in the lowest redshift bin.
It is worth remarking here that this is precisely why 
interpretation of IA constraints is almost always non-trivial;
the interplay with photo-$z$ error is often complex and one form
of mismodelling can very easily mimic the other.
We reiterate that IAs are being modelled here,
but that the two-parameter NLA model neither matches the input IA signal
exactly, nor is flexible enough to compensate fully for the discrepancy
by absorbing it into an effective amplitude.

Finally, we consider the case in which IAs are again mismodelled,
but in combination with more stringent set of scale cuts. 
Under such an analysis we obtain the green (unshaded) contours in 
Figure \ref{fig:forecasts:constraints}.
The IA-induced parameter biases are seen to dwindle to semi-acceptable
levels 
($<1\sigma$, see the first column of Table \ref{tab:forecasts:priors}).
This result is reassuring,
but somewhat expected,
given that these cuts are conservative by construction,
and designed to compensate for what is known to be an incomplete model of the physics
entering the small scale matter power spectrum. 
One should note that the elimination (or, at least in large part, mitigation)
of bias comes with a concurrent loss of cosmological information.
The naive gain, evidenced by the visual comparison of the purple and green
contours in Figure \ref{fig:forecasts:constraints} in this case is relatively small.
To more meaningfully assess the degradation in a multidimensional parameter space
we evaluate the ratio
$R=(|\mathbf{C}_\mathrm{cos}'|^{0.5} - |\mathbf{C}_\mathrm{cos}|^{0.5})/|\mathbf{C}_\mathrm{cos}|^{0.5}$,
where $\mathbf{C}_\mathrm{cos}$ and $\mathbf{C}'_\mathrm{cos}$ are the $(6\times6)$
covariance matrices
of cosmological parameters from the two Monte Carlo chains. 
This gives a value of $R=17.7$,
which is suggestive of a greater value in the small scales than suggested from 
the projected contours in 
Figure \ref{fig:forecasts:constraints}.

% ------------- Conclusions ------------------------------------

\section{Discussion and Conclusions}\label{sec:conclusion}

We have used 113,560 galaxies from the \mb~simulation
to test the impact of halo anisotropy on the galaxy intrinsic alignment signal.
\mb~is one of a handful of high-resolution hydrodynamical simulations in existence to date
and encompasses a comoving cubic volume of length $100 h^{-1}$ Mpc.
Using artificially symmetrised copies of the galaxy catalogues, we have found a reduction
in power in the galaxy-galaxy correlation of a few percent on scales $\sim1 h^{-1}$Mpc,
increasing to $\sim 10-15 \%$ in the deep one-halo regime.
We have shown the impact on alignment correlations to be significantly greater on all physical scales.
Though our ability to quantify this effect is severely limited by the statistical precision
afforded by the finite simulation volume, our results point to a difference in measured alignment
correlations of the order of tens of percent or more, well into the two halo regime.

An IA model built on the assumption of spherical satellite distributions within dark matter halos
would, then, underestimate the strength of both GI and II intrinsic alignment correlations considerably.
This clearly has implications if we wish to use such a model to unlock the cosmological information
on small scales in future lensing surveys, below the regime for which marginalising over an
unknown amplitude parameter would be sufficient. 
We have described a series of robustness tests, 
designed to demonstrate the validity of our results beyond the immediate context of this analysis.
To the best of our ability we have shown that our findings are independent of the
various choices made in building our galaxy catalogues and the subsequent investigation based on them.
In an additional, higher level, validation exercise we have also demonstrated that applying the same
analysis pipeline to two different hydrodynamical simulations with different baryonic prescriptions
yields consistent results within the level of statistical precision. 
We have further tested for dependence on redshift of this effect, and found 
no statistically significant variation across four snapshots in the range
$z\in[0.062-1.000]$. Although this is shallower than contemporary
lensing surveys, it is a sufficient range to allow detection of even a relatively slow systematic
evolution were it to be present.
Testing the impact of the lower dark matter mass cut of the galaxy sample,
we have found no clear systematic change in anisotropy bias as the lower threshold is raised.

In a final strand of this analysis we have propagated the impact of halo anisotropy
into a set of mock cosmic shear data.
These simulated data are ``contaminated'' in such a way to mimic the impact of mismodelling
the intrinsic alignment signal using a spherically symmetric halo model.
Assuming a lensing survey with LSST-like number densities, 
and applying optimistic scale cuts, somewhat looser than those used in
the Dark Energy Survey Y1
cosmic shear cosmology analysis, 
we have found biases in $S_8$ and $w$ of 
$ 1.4\sigma$ and $ 1.5 \sigma$ respectively.
Adopting more stringent cuts following the prescription of DES Y1
(adjusted for differences in the redshift distributions)
the cosmological bias is seen to reduce to less than $1\sigma$ on all parameters,
but at the cost to the volume of the constraints in the six dimensional
cosmological parameter space of a factor  
$R=17.7$.
It is worth noting that this is a simplified toy model scenario, intended to illustrate the
effects of intrinsic alignment mismodelling alone.

Though the analysis presented here focuses on cosmic shear, 
$\xi_+$ and $\xi_-$ are not the only statistics affected by IAs.
Future cosmology studies will likely use shear as part of a joint analysis
alongside galaxy-galaxy lensing and, potentially, cluster lensing.
The extrapolation of our findings to such an analysis is non-trivial, and
a more comprehensive forecasting project would be required to
quantify the impact of halo anisotropy.
Given the scope of this paper, however, it is sufficient to
demonstrate that modelling errors of this sort induce a non-negligible
cosmological bias for at least one commonly used cosmological probe. 
Our analysis does not include non Gaussian (trispectrum) contributions to the covariance matrix,
which become significant on small scales.
This will affect the exact size of the biases presented in the previous section,
but is not expected to alter the broad conclusions of the work.

It is also worth bearing in mind that a number of other poorly understood 
effects become relevant on the smallest scales,
including nonlinear growth, baryonic feedback and beyond first-order galaxy bias.
Even if one were to build a sufficiently accurate small scale IA model, 
advances must be made in modelling or mitigating these other systematics if we are to successfully 
access the information in small scale shear and galaxy-galaxy lensing correlations.
Moreover, even given accurate models for \emph{all} small scale effects, it is quite possible that
parameter degeneracies would emerge that mimic power in the small scale intrinsic alignment
correlations.
As with all high-dimensional inference problems, this is a complicated subject that requires
careful consideration before any cosmological analysis that includes small angular scales can proceed.

\section{Acknowledgements}

The authors would like to thank 
Fran\c{c}ois Lanusse, 
Hung-Jin Huang, 
Duncan Campbell and
Aklant Bhowmick
for useful conversations and help with data access.
We would also like to thank our anonymous referee for
their thoughts on the work.
Catalogue building and processing were performed using the Coma HPC cluster, 
which is hosted by the McWilliams Center for Cosmology, Carnegie Mellon University. 
The smoothed contours in 
Figure \ref{fig:forecasts:constraints} were generated using \blockfont{chainconsumer}
\citep{hinton16}.
This research is supported by the US
National Science Foundation under Grant No. 1716131.

\DeclareRobustCommand{\PRF}[3]{#3}

\bibliographystyle{mnras}
\bibliography{refs}

\begin{thebibliography}{}
\makeatletter
\relax
\def\mn@urlcharsother{\let\do\@makeother \do\$\do\&\do\#\do\^\do\_\do\%\do\~}
\def\mn@doi{\begingroup\mn@urlcharsother \@ifnextchar [ {\mn@doi@}
  {\mn@doi@[]}}
\def\mn@doi@[#1]#2{\def\@tempa{#1}\ifx\@tempa\@empty \href
  {http://dx.doi.org/#2} {doi:#2}\else \href {http://dx.doi.org/#2} {#1}\fi
  \endgroup}
\def\mn@eprint#1#2{\mn@eprint@#1:#2::\@nil}
\def\mn@eprint@arXiv#1{\href {http://arxiv.org/abs/#1} {{\tt arXiv:#1}}}
\def\mn@eprint@dblp#1{\href {http://dblp.uni-trier.de/rec/bibtex/#1.xml}
  {dblp:#1}}
\def\mn@eprint@#1:#2:#3:#4\@nil{\def\@tempa {#1}\def\@tempb {#2}\def\@tempc
  {#3}\ifx \@tempc \@empty \let \@tempc \@tempb \let \@tempb \@tempa \fi \ifx
  \@tempb \@empty \def\@tempb {arXiv}\fi \@ifundefined
  {mn@eprint@\@tempb}{\@tempb:\@tempc}{\expandafter \expandafter \csname
  mn@eprint@\@tempb\endcsname \expandafter{\@tempc}}}

\bibitem[\protect\citeauthoryear{{Agustsson} \& {Brainerd}}{{Agustsson} \&
  {Brainerd}}{2010}]{agustsson10}
{Agustsson} I.,  {Brainerd} T.~G.,  2010, \mn@doi [\apj]
  {10.1088/0004-637X/709/2/1321}, \href
  {https://ui.adsabs.harvard.edu/abs/2010ApJ...709.1321A} {709, 1321}

\bibitem[\protect\citeauthoryear{{Allgood}, {Flores}, {Primack}, {Kravtsov},
  {Wechsler}, {Faltenbacher}  \& {Bullock}}{{Allgood} et~al.}{2006}]{allgood06}
{Allgood} B.,  {Flores} R.~A.,  {Primack} J.~R.,  {Kravtsov} A.~V.,  {Wechsler}
  R.~H.,  {Faltenbacher} A.,   {Bullock} J.~S.,  2006, \mn@doi [\mnras]
  {10.1111/j.1365-2966.2006.10094.x}, \href
  {https://ui.adsabs.harvard.edu/abs/2006MNRAS.367.1781A} {367, 1781}

\bibitem[\protect\citeauthoryear{{Bagla}, {Prasad}  \& {Khandai}}{{Bagla}
  et~al.}{2009}]{bagla09}
{Bagla} J.~S.,  {Prasad} J.,   {Khandai} N.,  2009, \mn@doi [\mnras]
  {10.1111/j.1365-2966.2009.14592.x}, \href
  {https://ui.adsabs.harvard.edu/abs/2009MNRAS.395..918B} {395, 918}

\bibitem[\protect\citeauthoryear{{Bailin} \& {Steinmetz}}{{Bailin} \&
  {Steinmetz}}{2005}]{bailin05}
{Bailin} J.,  {Steinmetz} M.,  2005, \mn@doi [\apj] {10.1086/430397}, \href
  {https://ui.adsabs.harvard.edu/\#abs/2005ApJ...627..647B} {627, 647}

\bibitem[\protect\citeauthoryear{{Bailin}, {Power}, {Norberg}, {Zaritsky}  \&
  {Gibson}}{{Bailin} et~al.}{2008}]{bailin08}
{Bailin} J.,  {Power} C.,  {Norberg} P.,  {Zaritsky} D.,   {Gibson} B.~K.,
  2008, \mn@doi [\mnras] {10.1111/j.1365-2966.2008.13828.x}, \href
  {https://ui.adsabs.harvard.edu/abs/2008MNRAS.390.1133B} {390, 1133}

\bibitem[\protect\citeauthoryear{{Barreira}, {Krause}  \& {Schmidt}}{{Barreira}
  et~al.}{2018}]{barreira17}
{Barreira} A.,  {Krause} E.,   {Schmidt} F.,  2018, \mn@doi [\jcap]
  {10.1088/1475-7516/2018/06/015}, \href
  {http://adsabs.harvard.edu/abs/2018JCAP...06..015B} {6, 015}

\bibitem[\protect\citeauthoryear{{Bernstein}, {Armstrong}, {Krawiec}  \&
  {March}}{{Bernstein} et~al.}{2016}]{bernstein16}
{Bernstein} G.~M.,  {Armstrong} R.,  {Krawiec} C.,   {March} M.~C.,  2016,
  \mn@doi [\mnras] {10.1093/mnras/stw879}, \href
  {http://adsabs.harvard.edu/abs/2016MNRAS.459.4467B} {459, 4467}

\bibitem[\protect\citeauthoryear{{Blazek}, {Vlah}  \& {Seljak}}{{Blazek}
  et~al.}{2015}]{blazek15}
{Blazek} J.,  {Vlah} Z.,   {Seljak} U.,  2015, \mn@doi [\jcap]
  {10.1088/1475-7516/2015/08/015}, \href
  {http://adsabs.harvard.edu/abs/2015JCAP...08..015B} {8, 015}

\bibitem[\protect\citeauthoryear{{Blazek}, {MacCrann}, {Troxel}  \&
  {Fang}}{{Blazek} et~al.}{2017}]{blazek17}
{Blazek} J.,  {MacCrann} N.,  {Troxel} M.~A.,   {Fang} X.,  2017, arXiv
  e-prints, \href {http://adsabs.harvard.edu/abs/2017arXiv170809247B} {p.
  arXiv:1708.09247}

\bibitem[\protect\citeauthoryear{{\PRF{Bosch}{van den}{van den} Bosch},
  {Jiang}, {Campbell}  \& {Behroozi}}{{\PRF{Bosch}{van den}{van den} Bosch}
  et~al.}{2016}]{vandenbosch16}
{\PRF{Bosch}{van den}{van den} Bosch} F.~C.,  {Jiang} F.,  {Campbell} D.,
  {Behroozi} P.,  2016, \mn@doi [\mnras] {10.1093/mnras/stv2338}, \href
  {https://ui.adsabs.harvard.edu/abs/2016MNRAS.455..158V} {455, 158}

\bibitem[\protect\citeauthoryear{{Bridle} \& {King}}{{Bridle} \&
  {King}}{2007}]{bridle07}
{Bridle} S.,  {King} L.,  2007, \mn@doi [New Journal of Physics]
  {10.1088/1367-2630/9/12/444}, \href
  {http://adsabs.harvard.edu/abs/2007NJPh....9..444B} {9, 444}

\bibitem[\protect\citeauthoryear{{Bridle} et~al.,}{{Bridle}
  et~al.}{2010}]{bridle10}
{Bridle} S.,  et~al., 2010, \mn@doi [\mnras]
  {10.1111/j.1365-2966.2010.16598.x}, \href
  {http://adsabs.harvard.edu/abs/2010MNRAS.405.2044B} {405, 2044}

\bibitem[\protect\citeauthoryear{{Butsky} et~al.,}{{Butsky}
  et~al.}{2016}]{butsky16}
{Butsky} I.,  et~al., 2016, \mn@doi [Monthly Notices of the Royal Astronomical
  Society] {10.1093/mnras/stw1688}, \href
  {https://ui.adsabs.harvard.edu/abs/2016MNRAS.462..663B} {462, 663}

\bibitem[\protect\citeauthoryear{{Catelan}, {Kamionkowski}  \&
  {Blandford}}{{Catelan} et~al.}{2001}]{catelan01}
{Catelan} P.,  {Kamionkowski} M.,   {Blandford} R.~D.,  2001, \mn@doi [\mnras]
  {10.1046/j.1365-8711.2001.04105.x}, \href
  {http://adsabs.harvard.edu/abs/2001MNRAS.320L...7C} {320, L7}

\bibitem[\protect\citeauthoryear{{Chen} et~al.,}{{Chen} et~al.}{2015}]{chen15}
{Chen} Y.-C.,  et~al., 2015, \mn@doi [\mnras] {10.1093/mnras/stv2260}, \href
  {http://adsabs.harvard.edu/abs/2015MNRAS.454.3341C} {454, 3341}

\bibitem[\protect\citeauthoryear{{Chisari} et~al.,}{{Chisari}
  et~al.}{2015}]{chisari15}
{Chisari} N.,  et~al., 2015, \mn@doi [\mnras] {10.1093/mnras/stv2154}, \href
  {http://adsabs.harvard.edu/abs/2015MNRAS.454.2736C} {454, 2736}

\bibitem[\protect\citeauthoryear{{Chisari} et~al.,}{{Chisari}
  et~al.}{2016}]{chisari16}
{Chisari} N.,  et~al., 2016, \mn@doi [\mnras] {10.1093/mnras/stw1409}, \href
  {http://adsabs.harvard.edu/abs/2016MNRAS.461.2702C} {461, 2702}

\bibitem[\protect\citeauthoryear{{Chisari} et~al.,}{{Chisari}
  et~al.}{2017}]{chisari17}
{Chisari} N.~E.,  et~al., 2017, \mn@doi [\mnras] {10.1093/mnras/stx1998}, \href
  {http://adsabs.harvard.edu/abs/2017MNRAS.472.1163C} {472, 1163}

\bibitem[\protect\citeauthoryear{{Chisari} et~al.,}{{Chisari}
  et~al.}{2018a}]{cclpaper}
{Chisari} N.~E.,  et~al., 2018a, arXiv e-prints, \href
  {https://ui.adsabs.harvard.edu/\#abs/2018arXiv181205995C} {p.
  arXiv:1812.05995}

\bibitem[\protect\citeauthoryear{{Chisari} et~al.,}{{Chisari}
  et~al.}{2018b}]{chisari18}
{Chisari} N.~E.,  et~al., 2018b, \mn@doi [\mnras] {10.1093/mnras/sty2093},
  \href {http://adsabs.harvard.edu/abs/2018MNRAS.480.3962C} {480, 3962}

\bibitem[\protect\citeauthoryear{{Codis} et~al.,}{{Codis}
  et~al.}{2015a}]{codis15a}
{Codis} S.,  et~al., 2015a, \mn@doi [\mnras] {10.1093/mnras/stv231}, \href
  {http://adsabs.harvard.edu/abs/2015MNRAS.448.3391C} {448, 3391}

\bibitem[\protect\citeauthoryear{{Codis}, {Pichon}  \& {Pogosyan}}{{Codis}
  et~al.}{2015b}]{codis15b}
{Codis} S.,  {Pichon} C.,   {Pogosyan} D.,  2015b, \mn@doi [\mnras]
  {10.1093/mnras/stv1570}, \href
  {http://adsabs.harvard.edu/abs/2015MNRAS.452.3369C} {452, 3369}

\bibitem[\protect\citeauthoryear{{Crittenden}, {Natarajan}, {Pen}  \&
  {Theuns}}{{Crittenden} et~al.}{2001}]{crittenden01}
{Crittenden} R.~G.,  {Natarajan} P.,  {Pen} U.-L.,   {Theuns} T.,  2001,
  \mn@doi [\apj] {10.1086/322370}, \href
  {http://adsabs.harvard.edu/abs/2001ApJ...559..552C} {559, 552}

\bibitem[\protect\citeauthoryear{{\PRF{Daalen}{van}{van} Daalen}, {Schaye},
  {Booth}  \& {Dalla Vecchia}}{{\PRF{Daalen}{van}{van} Daalen}
  et~al.}{2011}]{vanDaalen11}
{\PRF{Daalen}{van}{van} Daalen} M.~P.,  {Schaye} J.,  {Booth} C.~M.,   {Dalla
  Vecchia} C.,  2011, \mn@doi [\mnras] {10.1111/j.1365-2966.2011.18981.x},
  \href {http://adsabs.harvard.edu/abs/2011MNRAS.415.3649V} {415, 3649}

\bibitem[\protect\citeauthoryear{{\PRF{Daalen}{van}{van} Daalen}, {Angulo}  \&
  {White}}{{\PRF{Daalen}{van}{van} Daalen} et~al.}{2012}]{vanDaalen12}
{\PRF{Daalen}{van}{van} Daalen} M.~P.,  {Angulo} R.~E.,   {White} S.~D.~M.,
  2012, \mn@doi [\mnras] {10.1111/j.1365-2966.2012.21437.x}, \href
  {http://adsabs.harvard.edu/abs/2012MNRAS.424.2954V} {424, 2954}

\bibitem[\protect\citeauthoryear{{Dark Energy Survey Collaboration}}{{Dark
  Energy Survey Collaboration}}{2016}]{svcosmology}
{Dark Energy Survey Collaboration} 2016, \mn@doi [\prd]
  {10.1103/PhysRevD.94.022001}, \href
  {http://adsabs.harvard.edu/abs/2016PhRvD..94b2001A} {94, 022001}

\bibitem[\protect\citeauthoryear{{Dark Energy Survey Collaboration}}{{Dark
  Energy Survey Collaboration}}{2017}]{y1keypaper}
{Dark Energy Survey Collaboration} 2017, \mn@doi [\prd]
  {10.1103/PhysRevD.98.043526}, \href
  {http://adsabs.harvard.edu/abs/2017arXiv170801530D} {98, 043526}

\bibitem[\protect\citeauthoryear{{Di Matteo}, {Khandai}, {DeGraf}, {Feng},
  {Croft}, {Lopez}  \& {Springel}}{{Di Matteo} et~al.}{2012}]{dimatteo12}
{Di Matteo} T.,  {Khandai} N.,  {DeGraf} C.,  {Feng} Y.,  {Croft} R.~A.~C.,
  {Lopez} J.,   {Springel} V.,  2012, \mn@doi [\apjl]
  {10.1088/2041-8205/745/2/L29}, \href
  {http://adsabs.harvard.edu/abs/2012ApJ...745L..29D} {745, L29}

\bibitem[\protect\citeauthoryear{{Drlica-Wagner} et~al.,}{{Drlica-Wagner}
  et~al.}{2018}]{y1gold}
{Drlica-Wagner} A.,  et~al., 2018, \mn@doi [\apjs] {10.3847/1538-4365/aab4f5},
  \href {http://adsabs.harvard.edu/abs/2018ApJS..235...33D} {235, 33}

\bibitem[\protect\citeauthoryear{{Dubois}, {Peirani}, {Pichon}, {Devriendt},
  {Gavazzi}, {Welker}  \& {Volonteri}}{{Dubois} et~al.}{2016}]{dubois16}
{Dubois} Y.,  {Peirani} S.,  {Pichon} C.,  {Devriendt} J.,  {Gavazzi} R.,
  {Welker} C.,   {Volonteri} M.,  2016, \mn@doi [\mnras]
  {10.1093/mnras/stw2265}, \href
  {http://adsabs.harvard.edu/abs/2016MNRAS.463.3948D} {463, 3948}

\bibitem[\protect\citeauthoryear{{Dvornik} et~al.,}{{Dvornik}
  et~al.}{2018}]{dvornik18}
{Dvornik} A.,  et~al., 2018, \mn@doi [\mnras] {10.1093/mnras/sty1502}, \href
  {http://adsabs.harvard.edu/abs/2018MNRAS.tmp.1433D} {}

\bibitem[\protect\citeauthoryear{Faltenbacher, Li, Mao, van~den Bosch, Yang,
  Jing, Pasquali  \& Mo}{Faltenbacher et~al.}{2007}]{faltenbacher07}
Faltenbacher A.,  Li C.,  Mao S.,  van~den Bosch F.~C.,  Yang X.,  Jing Y.~P.,
  Pasquali A.,   Mo H.~J.,  2007, The Astrophysical Journal Letters, 662, L71

\bibitem[\protect\citeauthoryear{{Fedeli}}{{Fedeli}}{2014}]{fedeli14}
{Fedeli} C.,  2014, \mn@doi [\jcap] {10.1088/1475-7516/2014/04/028}, \href
  {http://adsabs.harvard.edu/abs/2014JCAP...04..028F} {4, 028}

\bibitem[\protect\citeauthoryear{{Foreman-Mackey}, {Hogg}, {Lang}  \&
  {Goodman}}{{Foreman-Mackey} et~al.}{2013}]{foremanmackey12}
{Foreman-Mackey} D.,  {Hogg} D.~W.,  {Lang} D.,   {Goodman} J.,  2013, \mn@doi
  [\pasp] {10.1086/670067}, \href
  {http://adsabs.harvard.edu/abs/2013PASP..125..306F} {125, 306}

\bibitem[\protect\citeauthoryear{{Hearin}, {Zentner}, {Ma}  \&
  {Huterer}}{{Hearin} et~al.}{2010}]{hearin10}
{Hearin} A.~P.,  {Zentner} A.~R.,  {Ma} Z.,   {Huterer} D.,  2010, \mn@doi
  [\apj] {10.1088/0004-637X/720/2/1351}, \href
  {http://adsabs.harvard.edu/abs/2010ApJ...720.1351H} {720, 1351}

\bibitem[\protect\citeauthoryear{{Hearin} et~al.,}{{Hearin}
  et~al.}{2017}]{hearin17}
{Hearin} A.~P.,  et~al., 2017, \mn@doi [doi.org/10.5281/zenodo.835898 \aj]
  {0.5281/zenodo.835898}, \href
  {http://adsabs.harvard.edu/abs/2017AJ....154..190H} {154, 190}

\bibitem[\protect\citeauthoryear{{Heymans} et~al.,}{{Heymans}
  et~al.}{2006}]{heymans06}
{Heymans} C.,  et~al., 2006, \mn@doi [\mnras]
  {10.1111/j.1365-2966.2006.10198.x}, \href
  {http://adsabs.harvard.edu/abs/2006MNRAS.368.1323H} {368, 1323}

\bibitem[\protect\citeauthoryear{{Hinton}}{{Hinton}}{2016}]{hinton16}
{Hinton} S.,  2016, \mn@doi [JOSS 10.21105/joss.00045] {10.21105/joss.00045}

\bibitem[\protect\citeauthoryear{{Hirata} \& {Seljak}}{{Hirata} \&
  {Seljak}}{2004}]{hirata04}
{Hirata} C.~M.,  {Seljak} U.,  2004, \mn@doi [\prd]
  {10.1103/PhysRevD.70.063526}, \href
  {http://adsabs.harvard.edu/abs/2004PhRvD..70f3526H} {70, 063526}

\bibitem[\protect\citeauthoryear{{Huang}, {Mandelbaum}, {Freeman}, {Chen},
  {Rozo}, {Rykoff}  \& {Baxter}}{{Huang} et~al.}{2016}]{huang16}
{Huang} H.-J.,  {Mandelbaum} R.,  {Freeman} P.~E.,  {Chen} Y.-C.,  {Rozo} E.,
  {Rykoff} E.,   {Baxter} E.~J.,  2016, \mn@doi [\mnras]
  {10.1093/mnras/stw1982}, \href
  {https://ui.adsabs.harvard.edu/abs/2016MNRAS.463..222H} {463, 222}

\bibitem[\protect\citeauthoryear{{Huang}, {Mandelbaum}, {Freeman}, {Chen},
  {Rozo}  \& {Rykoff}}{{Huang} et~al.}{2018}]{huang18}
{Huang} H.-J.,  {Mandelbaum} R.,  {Freeman} P.~E.,  {Chen} Y.-C.,  {Rozo} E.,
  {Rykoff} E.,  2018, \mn@doi [\mnras] {10.1093/mnras/stx2995}, \href
  {http://adsabs.harvard.edu/abs/2018MNRAS.474.4772H} {474, 4772}

\bibitem[\protect\citeauthoryear{{Huff} \& {Mandelbaum}}{{Huff} \&
  {Mandelbaum}}{2017}]{huff17}
{Huff} E.,  {Mandelbaum} R.,  2017, preprint, \href
  {http://adsabs.harvard.edu/abs/2017arXiv170202600H} {} (\mn@eprint {arXiv}
  {1702.02600})

\bibitem[\protect\citeauthoryear{{Jarvis}, {Bernstein}  \& {Jain}}{{Jarvis}
  et~al.}{2004}]{jarvis04}
{Jarvis} M.,  {Bernstein} G.,   {Jain} B.,  2004, \mn@doi [\mnras]
  {10.1111/j.1365-2966.2004.07926.x}, \href
  {http://adsabs.harvard.edu/abs/2004MNRAS.352..338J} {352, 338}

\bibitem[\protect\citeauthoryear{{Jing} \& {Suto}}{{Jing} \&
  {Suto}}{2002}]{jing02}
{Jing} Y.~P.,  {Suto} Y.,  2002, \mn@doi [\apj] {10.1086/341065}, \href
  {https://ui.adsabs.harvard.edu/abs/2002ApJ...574..538J} {574, 538}

\bibitem[\protect\citeauthoryear{{Joachimi}, {Mandelbaum}, {Abdalla}  \&
  {Bridle}}{{Joachimi} et~al.}{2011}]{joachimi11}
{Joachimi} B.,  {Mandelbaum} R.,  {Abdalla} F.~B.,   {Bridle} S.~L.,  2011,
  \mn@doi [\aap] {10.1051/0004-6361/201015621}, \href
  {http://adsabs.harvard.edu/abs/2011A%26A...527A..26J} {527, A26}

\bibitem[\protect\citeauthoryear{{Joachimi}, {Semboloni}, {Hilbert}, {Bett},
  {Hartlap}, {Hoekstra}  \& {Schneider}}{{Joachimi} et~al.}{2013}]{joachimi13}
{Joachimi} B.,  {Semboloni} E.,  {Hilbert} S.,  {Bett} P.~E.,  {Hartlap} J.,
  {Hoekstra} H.,   {Schneider} P.,  2013, \mn@doi [\mnras]
  {10.1093/mnras/stt1618}, \href
  {http://adsabs.harvard.edu/abs/2013MNRAS.436..819J} {436, 819}

\bibitem[\protect\citeauthoryear{{Joachimi} et~al.,}{{Joachimi}
  et~al.}{2015}]{joachimi15}
{Joachimi} B.,  et~al., 2015, \mn@doi [\ssr] {10.1007/s11214-015-0177-4}, \href
  {http://adsabs.harvard.edu/abs/2015SSRv..193....1J} {193, 1}

\bibitem[\protect\citeauthoryear{{Johnston}, {Sheldon}, {Tasitsiomi},
  {Frieman}, {Wechsler}  \& {McKay}}{{Johnston} et~al.}{2007}]{johnston07}
{Johnston} D.~E.,  {Sheldon} E.~S.,  {Tasitsiomi} A.,  {Frieman} J.~A.,
  {Wechsler} R.~H.,   {McKay} T.~A.,  2007, \mn@doi [\apj] {10.1086/510060},
  \href {http://adsabs.harvard.edu/abs/2007ApJ...656...27J} {656, 27}

\bibitem[\protect\citeauthoryear{{Joudaki} et~al.,}{{Joudaki}
  et~al.}{2018}]{joudaki18}
{Joudaki} S.,  et~al., 2018, \mn@doi [\mnras] {10.1093/mnras/stx2820}, \href
  {http://adsabs.harvard.edu/abs/2018MNRAS.474.4894J} {474, 4894}

\bibitem[\protect\citeauthoryear{{Kasun} \& {Evrard}}{{Kasun} \&
  {Evrard}}{2005}]{kasun05}
{Kasun} S.~F.,  {Evrard} A.~E.,  2005, \mn@doi [\apj] {10.1086/430811}, \href
  {https://ui.adsabs.harvard.edu/\#abs/2005ApJ...629..781K} {629, 781}

\bibitem[\protect\citeauthoryear{{Kennicutt}}{{Kennicutt}}{1998}]{kennicutt98}
{Kennicutt} R.~C.,  1998, \mn@doi [\apj] {10.1086/305588}, \href
  {http://adsabs.harvard.edu/abs/1998ApJ...498..541K} {498, 541}

\bibitem[\protect\citeauthoryear{{Khandai}, {Di Matteo}, {Croft}, {Wilkins},
  {Feng}, {Tucker}, {DeGraf}  \& {Liu}}{{Khandai} et~al.}{2015}]{khandai15}
{Khandai} N.,  {Di Matteo} T.,  {Croft} R.,  {Wilkins} S.,  {Feng} Y.,
  {Tucker} E.,  {DeGraf} C.,   {Liu} M.-S.,  2015, \mn@doi [\mnras]
  {10.1093/mnras/stv627}, \href
  {http://adsabs.harvard.edu/abs/2015MNRAS.450.1349K} {450, 1349}

\bibitem[\protect\citeauthoryear{{Kiessling} et~al.,}{{Kiessling}
  et~al.}{2015}]{kiessling15}
{Kiessling} A.,  et~al., 2015, \mn@doi [\ssr] {10.1007/s11214-015-0203-6},
  \href {http://adsabs.harvard.edu/abs/2015SSRv..193...67K} {193, 67}

\bibitem[\protect\citeauthoryear{{Kilbinger}}{{Kilbinger}}{2015}]{kilbinger15}
{Kilbinger} M.,  2015, \mn@doi [Reports on Progress in Physics]
  {10.1088/0034-4885/78/8/086901}, \href
  {http://adsabs.harvard.edu/abs/2015RPPh...78h6901K} {78, 086901}

\bibitem[\protect\citeauthoryear{{Kirk} et~al.,}{{Kirk} et~al.}{2015}]{kirk15}
{Kirk} D.,  et~al., 2015, \mn@doi [\ssr] {10.1007/s11214-015-0213-4}, \href
  {http://adsabs.harvard.edu/abs/2015SSRv..193..139K} {193, 139}

\bibitem[\protect\citeauthoryear{{Kitching} et~al.,}{{Kitching}
  et~al.}{2011}]{kitching10}
{Kitching} T.,  et~al., 2011, Annals of Applied Statistics Vol. 5, No. 3, \href
  {https://ui.adsabs.harvard.edu/\#abs/2010arXiv1009.0779K} {pp 2231--2263}

\bibitem[\protect\citeauthoryear{{Knebe}, {Gill}, {Gibson}, {Lewis}, {Ibata}
  \& {Dopita}}{{Knebe} et~al.}{2004}]{knebe04}
{Knebe} A.,  {Gill} S.,  {Gibson} B.,  {Lewis} G.,  {Ibata} R.,   {Dopita} M.,
  2004, The Astrophysical Journal, 603, 7

\bibitem[\protect\citeauthoryear{{Krause} \& {Eifler}}{{Krause} \&
  {Eifler}}{2017}]{krause17}
{Krause} E.,  {Eifler} T.,  2017, \mn@doi [\mnras] {10.1093/mnras/stx1261},
  \href {http://adsabs.harvard.edu/abs/2017MNRAS.470.2100K} {470, 2100}

\bibitem[\protect\citeauthoryear{{Krause}, {Eifler}  \& {Blazek}}{{Krause}
  et~al.}{2016}]{krause16}
{Krause} E.,  {Eifler} T.,   {Blazek} J.,  2016, \mn@doi [\mnras]
  {10.1093/mnras/stv2615}, \href
  {https://ui.adsabs.harvard.edu/\#abs/2016MNRAS.456..207K} {456, 207}

\bibitem[\protect\citeauthoryear{{Krause} et~al.,}{{Krause}
  et~al.}{2017}]{y1methodology}
{Krause} E.,  et~al., 2017, Submitted to \prd, arXiv:1706.09359, \href
  {http://adsabs.harvard.edu/abs/2017arXiv170609359K} {}

\bibitem[\protect\citeauthoryear{{Krolewski}, {Ho}, {Chen}, {Chan}, {Tenneti},
  {Bizyaev}  \& {Kraljic}}{{Krolewski} et~al.}{2019}]{krolewski19}
{Krolewski} A.,  {Ho} S.,  {Chen} Y.-C.,  {Chan} P.~F.,  {Tenneti} A.,
  {Bizyaev} D.,   {Kraljic} K.,  2019, \mn@doi [\apj]
  {10.3847/1538-4357/ab1010}, \href
  {https://ui.adsabs.harvard.edu/abs/2019ApJ...876...52K} {876, 52}

\bibitem[\protect\citeauthoryear{{Landy} \& {Szalay}}{{Landy} \&
  {Szalay}}{1993}]{landy93}
{Landy} S.~D.,  {Szalay} A.~S.,  1993, \mn@doi [\apj] {10.1086/172900}, \href
  {http://adsabs.harvard.edu/abs/1993ApJ...412...64L} {412, 64}

\bibitem[\protect\citeauthoryear{{Le Brun}, {McCarthy}, {Schaye}  \&
  {Ponman}}{{Le Brun} et~al.}{2014}]{lebrun13}
{Le Brun} A.~M.~C.,  {McCarthy} I.~G.,  {Schaye} J.,   {Ponman} T.~J.,  2014,
  \mn@doi [\mnras] {10.1093/mnras/stu608}, \href
  {http://adsabs.harvard.edu/abs/2014MNRAS.441.1270L} {441, 1270}

\bibitem[\protect\citeauthoryear{Lewis, Challinor  \& Lasenby}{Lewis
  et~al.}{2000}]{lewis99}
Lewis A.,  Challinor A.,   Lasenby A.,  2000, \mn@doi [\apj] {10.1086/309179},
  538, 473

\bibitem[\protect\citeauthoryear{{Libeskind}, {Guo}, {Tempel}  \&
  {Ibata}}{{Libeskind} et~al.}{2016}]{libeskind16}
{Libeskind} N.~I.,  {Guo} Q.,  {Tempel} E.,   {Ibata} R.,  2016, \mn@doi [\apj]
  {10.3847/0004-637X/830/2/121}, \href
  {https://ui.adsabs.harvard.edu/abs/2016ApJ...830..121L} {830, 121}

\bibitem[\protect\citeauthoryear{{Mandelbaum}}{{Mandelbaum}}{2018}]{2018ARA&A..56..393M}
{Mandelbaum} R.,  2018, \mn@doi [\araa] {10.1146/annurev-astro-081817-051928},
  \href {http://adsabs.harvard.edu/abs/2018ARA%26A..56..393M} {56, 393}

\bibitem[\protect\citeauthoryear{{Mandelbaum} et~al.,}{{Mandelbaum}
  et~al.}{2011}]{mandelbaum10}
{Mandelbaum} R.,  et~al., 2011, \mn@doi [\mnras]
  {10.1111/j.1365-2966.2010.17485.x}, \href
  {http://adsabs.harvard.edu/abs/2011MNRAS.410..844M} {410, 844}

\bibitem[\protect\citeauthoryear{{Mandelbaum} et~al.,}{{Mandelbaum}
  et~al.}{2015}]{mandelbaum15}
{Mandelbaum} R.,  et~al., 2015, \mn@doi [\mnras] {10.1093/mnras/stv781}, \href
  {http://adsabs.harvard.edu/abs/2015MNRAS.450.2963M} {450, 2963}

\bibitem[\protect\citeauthoryear{{Massey} et~al.,}{{Massey}
  et~al.}{2007}]{massey07}
{Massey} R.,  et~al., 2007, \mn@doi [\mnras]
  {10.1111/j.1365-2966.2006.11315.x}, \href
  {https://ui.adsabs.harvard.edu/\#abs/2007MNRAS.376...13M} {376, 13}

\bibitem[\protect\citeauthoryear{{Mead}, {Peacock}, {Heymans}, {Joudaki}  \&
  {Heavens}}{{Mead} et~al.}{2015}]{mead15}
{Mead} A.~J.,  {Peacock} J.~A.,  {Heymans} C.,  {Joudaki} S.,   {Heavens}
  A.~F.,  2015, \mn@doi [\mnras] {10.1093/mnras/stv2036}, \href
  {http://adsabs.harvard.edu/abs/2015MNRAS.454.1958M} {454, 1958}

\bibitem[\protect\citeauthoryear{{Mead}, {Heymans}, {Lombriser}, {Peacock},
  {Steele}  \& {Winther}}{{Mead} et~al.}{2016}]{mead16}
{Mead} A.~J.,  {Heymans} C.,  {Lombriser} L.,  {Peacock} J.~A.,  {Steele}
  O.~I.,   {Winther} H.~A.,  2016, \mn@doi [\mnras] {10.1093/mnras/stw681},
  \href {http://adsabs.harvard.edu/abs/2016MNRAS.459.1468M} {459, 1468}

\bibitem[\protect\citeauthoryear{{Nelson} et~al.,}{{Nelson}
  et~al.}{2018}]{nelson18}
{Nelson} D.,  et~al., 2018, arXiv e-prints, \href
  {http://adsabs.harvard.edu/abs/2018arXiv181205609N} {}

\bibitem[\protect\citeauthoryear{{Peacock} \& {Smith}}{{Peacock} \&
  {Smith}}{2000}]{peacock00}
{Peacock} J.~A.,  {Smith} R.~E.,  2000, \mn@doi [\mnras]
  {10.1046/j.1365-8711.2000.03779.x}, \href
  {http://adsabs.harvard.edu/abs/2000MNRAS.318.1144P} {318, 1144}

\bibitem[\protect\citeauthoryear{{Pereira}, {Bryan}  \& {Gill}}{{Pereira}
  et~al.}{2008}]{pereira08}
{Pereira} M.~J.,  {Bryan} G.~L.,   {Gill} S.~P.~D.,  2008, \mn@doi [\apj]
  {10.1086/523830}, \href {http://adsabs.harvard.edu/abs/2008ApJ...672..825P}
  {672, 825}

\bibitem[\protect\citeauthoryear{{Peterson} et~al.,}{{Peterson}
  et~al.}{2015}]{peterson15}
{Peterson} J.~R.,  et~al., 2015, \mn@doi [\apjs] {10.1088/0067-0049/218/1/14},
  \href {http://adsabs.harvard.edu/abs/2015ApJS..218...14P} {218, 14}

\bibitem[\protect\citeauthoryear{{Piras}, {Joachimi}, {Sch{\"a}fer},
  {Bonamigo}, {Hilbert}  \& {van Uitert}}{{Piras} et~al.}{2018}]{piras18}
{Piras} D.,  {Joachimi} B.,  {Sch{\"a}fer} B.~M.,  {Bonamigo} M.,  {Hilbert}
  S.,   {van Uitert} E.,  2018, \mn@doi [\mnras] {10.1093/mnras/stx2846}, \href
  {http://adsabs.harvard.edu/abs/2018MNRAS.474.1165P} {474, 1165}

\bibitem[\protect\citeauthoryear{{Piscionere}, {Berlind}, {McBride}  \&
  {Scoccimarro}}{{Piscionere} et~al.}{2015}]{piscionere15}
{Piscionere} J.~A.,  {Berlind} A.~A.,  {McBride} C.~K.,   {Scoccimarro} R.,
  2015, \mn@doi [\apj] {10.1088/0004-637X/806/1/125}, \href
  {https://ui.adsabs.harvard.edu/abs/2015ApJ...806..125P} {806, 125}

\bibitem[\protect\citeauthoryear{{Planck Collaboration}}{{Planck
  Collaboration}}{2018}]{planck18}
{Planck Collaboration} 2018, preprint, \href
  {http://adsabs.harvard.edu/abs/2018arXiv180706209P} {} (\mn@eprint {arXiv}
  {1807.06209})

\bibitem[\protect\citeauthoryear{{Power} \& {Knebe}}{{Power} \&
  {Knebe}}{2006}]{power06}
{Power} C.,  {Knebe} A.,  2006, \mn@doi [\mnras]
  {10.1111/j.1365-2966.2006.10562.x}, \href
  {https://ui.adsabs.harvard.edu/abs/2006MNRAS.370..691P} {370, 691}

\bibitem[\protect\citeauthoryear{{Ragone-Figueroa} \&
  {Plionis}}{{Ragone-Figueroa} \& {Plionis}}{2007}]{ragoneFigueroa07}
{Ragone-Figueroa} C.,  {Plionis} M.,  2007, \mn@doi [\mnras]
  {10.1111/j.1365-2966.2007.11757.x}, \href
  {https://ui.adsabs.harvard.edu/abs/2007MNRAS.377.1785R} {377, 1785}

\bibitem[\protect\citeauthoryear{{Rood} \& {Sastry}}{{Rood} \&
  {Sastry}}{1972}]{rood72}
{Rood} H.,  {Sastry} G.,  1972, \apj, p.~451

\bibitem[\protect\citeauthoryear{{Rykoff} et~al.,}{{Rykoff}
  et~al.}{2016}]{rykoff16}
{Rykoff} E.~S.,  et~al., 2016, \mn@doi [The Astrophysical Journal Supplement
  Series] {10.3847/0067-0049/224/1/1}, \href
  {https://ui.adsabs.harvard.edu/\#abs/2016ApJS..224....1R} {224, 1}

\bibitem[\protect\citeauthoryear{{Schaye} et~al.,}{{Schaye}
  et~al.}{2015}]{schaye15}
{Schaye} J.,  et~al., 2015, \mn@doi [\mnras] {10.1093/mnras/stu2058}, \href
  {http://adsabs.harvard.edu/abs/2015MNRAS.446..521S} {446, 521}

\bibitem[\protect\citeauthoryear{{Schneider} \& {Bridle}}{{Schneider} \&
  {Bridle}}{2010}]{schneider10}
{Schneider} M.~D.,  {Bridle} S.,  2010, \mn@doi [\mnras]
  {10.1111/j.1365-2966.2009.15956.x}, \href
  {http://adsabs.harvard.edu/abs/2010MNRAS.402.2127S} {402, 2127}

\bibitem[\protect\citeauthoryear{{Schneider} \& {Teyssier}}{{Schneider} \&
  {Teyssier}}{2015}]{schneider15}
{Schneider} A.,  {Teyssier} R.,  2015, \mn@doi [\jcap]
  {10.1088/1475-7516/2015/12/049}, \href
  {http://adsabs.harvard.edu/abs/2015JCAP...12..049S} {12, 049}

\bibitem[\protect\citeauthoryear{{Schneider}, {Frenk}  \& {Cole}}{{Schneider}
  et~al.}{2012}]{schneider12}
{Schneider} M.~D.,  {Frenk} C.~S.,   {Cole} S.,  2012, \mn@doi [\jcap]
  {10.1088/1475-7516/2012/05/030}, \href
  {https://ui.adsabs.harvard.edu/abs/2012JCAP...05..030S} {2012, 030}

\bibitem[\protect\citeauthoryear{{Schulz} \& {White}}{{Schulz} \&
  {White}}{2006}]{schulz06}
{Schulz} A.~E.,  {White} M.,  2006, \mn@doi [Astroparticle Physics]
  {10.1016/j.astropartphys.2005.11.007}, \href
  {http://adsabs.harvard.edu/abs/2006APh....25..172S} {25, 172}

\bibitem[\protect\citeauthoryear{{Sheldon} \& {Huff}}{{Sheldon} \&
  {Huff}}{2017}]{sheldon17}
{Sheldon} E.~S.,  {Huff} E.~M.,  2017, \mn@doi [\apj]
  {10.3847/1538-4357/aa704b}, \href
  {http://adsabs.harvard.edu/abs/2017ApJ...841...24S} {841, 24}

\bibitem[\protect\citeauthoryear{{Shirasaki}, {Takada}, {Miyatake},
  {Takahashi}, {Hamana}, {Nishimichi}  \& {Murata}}{{Shirasaki}
  et~al.}{2017}]{shirasaki17}
{Shirasaki} M.,  {Takada} M.,  {Miyatake} H.,  {Takahashi} R.,  {Hamana} T.,
  {Nishimichi} T.,   {Murata} R.,  2017, \mn@doi [\mnras]
  {10.1093/mnras/stx1477}, \href
  {http://adsabs.harvard.edu/abs/2017MNRAS.470.3476S} {470, 3476}

\bibitem[\protect\citeauthoryear{{Sif{\'o}n}, {Hoekstra}, {Cacciato}, {Viola},
  {K{\"o}hlinger}, {van der Burg}, {Sand}  \& {Graham}}{{Sif{\'o}n}
  et~al.}{2015}]{sifon15}
{Sif{\'o}n} C.,  {Hoekstra} H.,  {Cacciato} M.,  {Viola} M.,  {K{\"o}hlinger}
  F.,  {van der Burg} R.~F.~J.,  {Sand} D.~J.,   {Graham} M.~L.,  2015, \mn@doi
  [\aap] {10.1051/0004-6361/201424435}, \href
  {http://adsabs.harvard.edu/abs/2015A%26A...575A..48S} {575, A48}

\bibitem[\protect\citeauthoryear{{Simet}, {Battaglia}, {Mandelbaum}  \&
  {Seljak}}{{Simet} et~al.}{2017}]{simet17}
{Simet} M.,  {Battaglia} N.,  {Mandelbaum} R.,   {Seljak} U.,  2017, \mn@doi
  [\mnras] {10.1093/mnras/stw3322}, \href
  {http://adsabs.harvard.edu/abs/2017MNRAS.466.3663S} {466, 3663}

\bibitem[\protect\citeauthoryear{{Singh}, {Mandelbaum}  \& {More}}{{Singh}
  et~al.}{2015}]{singh15}
{Singh} S.,  {Mandelbaum} R.,   {More} S.,  2015, \mn@doi [\mnras]
  {10.1093/mnras/stv778}, \href
  {http://adsabs.harvard.edu/abs/2015MNRAS.450.2195S} {450, 2195}

\bibitem[\protect\citeauthoryear{{Singh}, {Mandelbaum}, {Seljak}, {Slosar}  \&
  {Vazquez Gonzalez}}{{Singh} et~al.}{2017}]{singh17}
{Singh} S.,  {Mandelbaum} R.,  {Seljak} U.,  {Slosar} A.,   {Vazquez Gonzalez}
  J.,  2017, \mn@doi [\mnras] {10.1093/mnras/stx1828}, \href
  {http://adsabs.harvard.edu/abs/2017MNRAS.471.3827S} {471, 3827}

\bibitem[\protect\citeauthoryear{{Smith} \& {Watts}}{{Smith} \&
  {Watts}}{2005}]{smith05}
{Smith} R.~E.,  {Watts} P.~I.~R.,  2005, \mn@doi [\mnras]
  {10.1111/j.1365-2966.2005.09053.x}, \href
  {http://adsabs.harvard.edu/abs/2005MNRAS.360..203S} {360, 203}

\bibitem[\protect\citeauthoryear{{Soussana} et~al.,}{{Soussana}
  et~al.}{2019}]{soussana19}
{Soussana} A.,  et~al., 2019, arXiv e-prints, \href
  {https://ui.adsabs.harvard.edu/abs/2019arXiv190811665S} {p. arXiv:1908.11665}

\bibitem[\protect\citeauthoryear{{Springel}, {White}, {Tormen}  \&
  {Kauffmann}}{{Springel} et~al.}{2001}]{springel01}
{Springel} V.,  {White} S.~D.~M.,  {Tormen} G.,   {Kauffmann} G.,  2001,
  \mn@doi [\mnras] {10.1046/j.1365-8711.2001.04912.x}, \href
  {http://adsabs.harvard.edu/abs/2001MNRAS.328..726S} {328, 726}

\bibitem[\protect\citeauthoryear{{Springel}, {White}  \&
  {Hernquist}}{{Springel} et~al.}{2004}]{springel04}
{Springel} V.,  {White} S.~D.~M.,   {Hernquist} L.,  2004, in {Ryder} S.,
  {Pisano} D.,  {Walker} M.,   {Freeman} K.,  eds,  IAU Symposium Vol. 220,
  Dark Matter in Galaxies. p.~421

\bibitem[\protect\citeauthoryear{{Takada} \& {Hu}}{{Takada} \&
  {Hu}}{2013}]{takada13}
{Takada} M.,  {Hu} W.,  2013, \mn@doi [\prd] {10.1103/PhysRevD.87.123504},
  \href {http://adsabs.harvard.edu/abs/2013PhRvD..87l3504T} {87, 123504}

\bibitem[\protect\citeauthoryear{{Takahashi}, {Sato}, {Nishimichi}, {Taruya}
  \& {Oguri}}{{Takahashi} et~al.}{2012}]{takahashi12}
{Takahashi} R.,  {Sato} M.,  {Nishimichi} T.,  {Taruya} A.,   {Oguri} M.,
  2012, \mn@doi [\apj] {10.1088/0004-637X/761/2/152}, \href
  {http://adsabs.harvard.edu/abs/2012ApJ...761..152T} {761, 152}

\bibitem[\protect\citeauthoryear{{Tempel} \& {Libeskind}}{{Tempel} \&
  {Libeskind}}{2013}]{tempel13}
{Tempel} E.,  {Libeskind} N.~I.,  2013, \mn@doi [\apjl]
  {10.1088/2041-8205/775/2/L42}, \href
  {http://adsabs.harvard.edu/abs/2013ApJ...775L..42T} {775, L42}

\bibitem[\protect\citeauthoryear{{Tenneti}, {Singh}, {Mandelbaum}, {Di Matteo},
  {Feng}  \& {Khandai}}{{Tenneti} et~al.}{2015a}]{tenneti15b}
{Tenneti} A.,  {Singh} S.,  {Mandelbaum} R.,  {Di Matteo} T.,  {Feng} Y.,
  {Khandai} N.,  2015a, \mn@doi [\mnras] {10.1093/mnras/stv272}, \href
  {http://adsabs.harvard.edu/abs/2015MNRAS.448.3522T} {448, 3522}

\bibitem[\protect\citeauthoryear{{Tenneti}, {Mandelbaum}, {Di Matteo},
  {Kiessling}  \& {Khandai}}{{Tenneti} et~al.}{2015b}]{tenneti15a}
{Tenneti} A.,  {Mandelbaum} R.,  {Di Matteo} T.,  {Kiessling} A.,   {Khandai}
  N.,  2015b, \mn@doi [\mnras] {10.1093/mnras/stv1625}, \href
  {http://adsabs.harvard.edu/abs/2015MNRAS.453..469T} {453, 469}

\bibitem[\protect\citeauthoryear{{Tenneti}, {Mandelbaum}  \& {Di
  Matteo}}{{Tenneti} et~al.}{2016}]{tenneti16}
{Tenneti} A.,  {Mandelbaum} R.,   {Di Matteo} T.,  2016, \mn@doi [\mnras]
  {10.1093/mnras/stw1823}, \href
  {http://adsabs.harvard.edu/abs/2016MNRAS.462.2668T} {462, 2668}

\bibitem[\protect\citeauthoryear{{Troxel} \& {Ishak}}{{Troxel} \&
  {Ishak}}{2015}]{troxel15}
{Troxel} M.~A.,  {Ishak} M.,  2015, \mn@doi [\physrep]
  {10.1016/j.physrep.2014.11.001}, \href
  {http://adsabs.harvard.edu/abs/2015PhR...558....1T} {558, 1}

\bibitem[\protect\citeauthoryear{{Troxel} et~al.,}{{Troxel}
  et~al.}{2018a}]{y1cosmicshear}
{Troxel} M.~A.,  et~al., 2018a, \mn@doi [\prd] {10.1103/PhysRevD.98.043528},
  \href {http://adsabs.harvard.edu/abs/2018PhRvD..98d3528T} {98, 043528}

\bibitem[\protect\citeauthoryear{{Troxel} et~al.,}{{Troxel}
  et~al.}{2018b}]{troxel18}
{Troxel} M.~A.,  et~al., 2018b, \mn@doi [\mnras] {10.1093/mnras/sty1889}, \href
  {https://ui.adsabs.harvard.edu/\#abs/2018MNRAS.479.4998T} {479, 4998}

\bibitem[\protect\citeauthoryear{{Tugendhat} \& {Sch{\"a}fer}}{{Tugendhat} \&
  {Sch{\"a}fer}}{2018}]{tugendhat17}
{Tugendhat} T.~M.,  {Sch{\"a}fer} B.~M.,  2018, \mn@doi [\mnras]
  {10.1093/mnras/sty323}, \href
  {https://ui.adsabs.harvard.edu/\#abs/2018MNRAS.476.3460T} {476, 3460}

\bibitem[\protect\citeauthoryear{{Velliscig} et~al.,}{{Velliscig}
  et~al.}{2015}]{velliscig15}
{Velliscig} M.,  et~al., 2015, \mn@doi [\mnras] {10.1093/mnras/stv1690}, \href
  {http://adsabs.harvard.edu/abs/2015MNRAS.453..721V} {453, 721}

\bibitem[\protect\citeauthoryear{{Vogelsberger}, {Genel}, {Sijacki}, {Torrey},
  {Springel}  \& {Hernquist}}{{Vogelsberger} et~al.}{2013}]{vogelsberger13}
{Vogelsberger} M.,  {Genel} S.,  {Sijacki} D.,  {Torrey} P.,  {Springel} V.,
  {Hernquist} L.,  2013, \mn@doi [\mnras] {10.1093/mnras/stt1789}, \href
  {http://adsabs.harvard.edu/abs/2013MNRAS.436.3031V} {436, 3031}

\bibitem[\protect\citeauthoryear{{Vogelsberger} et~al.,}{{Vogelsberger}
  et~al.}{2014}]{vogelsberger14}
{Vogelsberger} M.,  et~al., 2014, \mn@doi [\mnras] {10.1093/mnras/stu1536},
  \href {http://adsabs.harvard.edu/abs/2014MNRAS.444.1518V} {444, 1518}

\bibitem[\protect\citeauthoryear{{Welker}, {Power}, {Pichon}, {Dubois},
  {Devriendt}  \& {Codis}}{{Welker} et~al.}{2017}]{welker17}
{Welker} C.,  {Power} C.,  {Pichon} C.,  {Dubois} Y.,  {Devriendt} J.,
  {Codis} S.,  2017, arXiv e-prints, \href
  {https://ui.adsabs.harvard.edu/abs/2017arXiv171207818W} {p. arXiv:1712.07818}

\bibitem[\protect\citeauthoryear{{West} \& {Blakeslee}}{{West} \&
  {Blakeslee}}{2000}]{west00}
{West} M.~J.,  {Blakeslee} J.~P.,  2000, \mn@doi [\apj] {10.1086/318177}, \href
  {https://ui.adsabs.harvard.edu/\#abs/2000ApJ...543L..27W} {543, L27}

\bibitem[\protect\citeauthoryear{{Zentner}, {Kravtsov}, {Gnedin}  \&
  {Klypin}}{{Zentner} et~al.}{2005}]{zentner05}
{Zentner} A.~R.,  {Kravtsov} A.~V.,  {Gnedin} O.~Y.,   {Klypin} A.~A.,  2005,
  \mn@doi [\apj] {10.1086/431355}, \href
  {https://ui.adsabs.harvard.edu/\#abs/2005ApJ...629..219Z} {629, 219}

\bibitem[\protect\citeauthoryear{{Zuntz} et~al.,}{{Zuntz}
  et~al.}{2015}]{zuntz15}
{Zuntz} J.,  et~al., 2015, \mn@doi [Astronomy and Computing]
  {10.1016/j.ascom.2015.05.005}, \href
  {https://ui.adsabs.harvard.edu/\#abs/2015A&C....12...45Z} {12, 45}

\makeatother
\end{thebibliography}

\appendix
\section{Additional Alignment Correlation Functions}\label{app:wgp_wpp}

\begin{figure}
\includegraphics[width=\columnwidth]{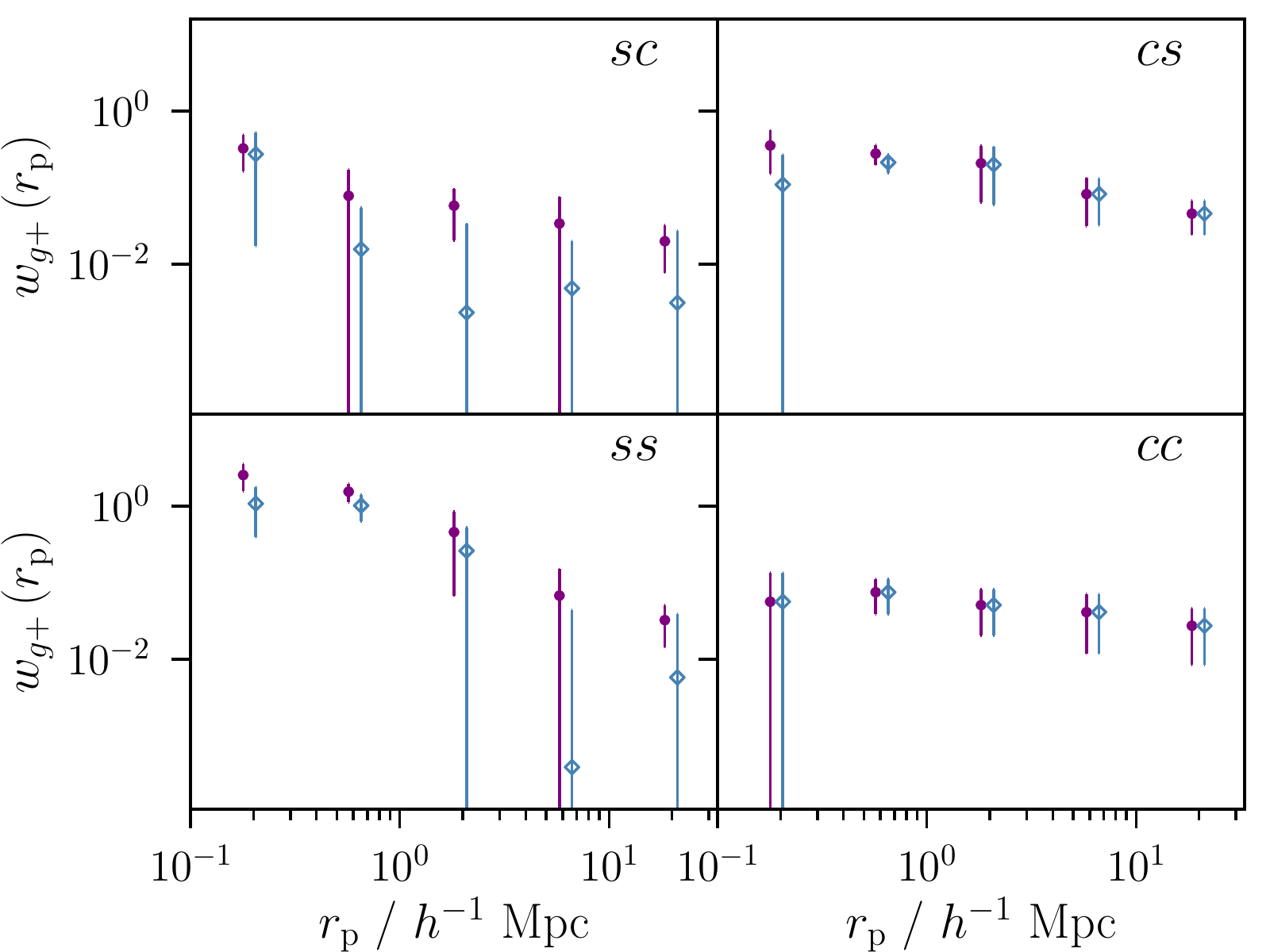}
\includegraphics[width=\columnwidth]{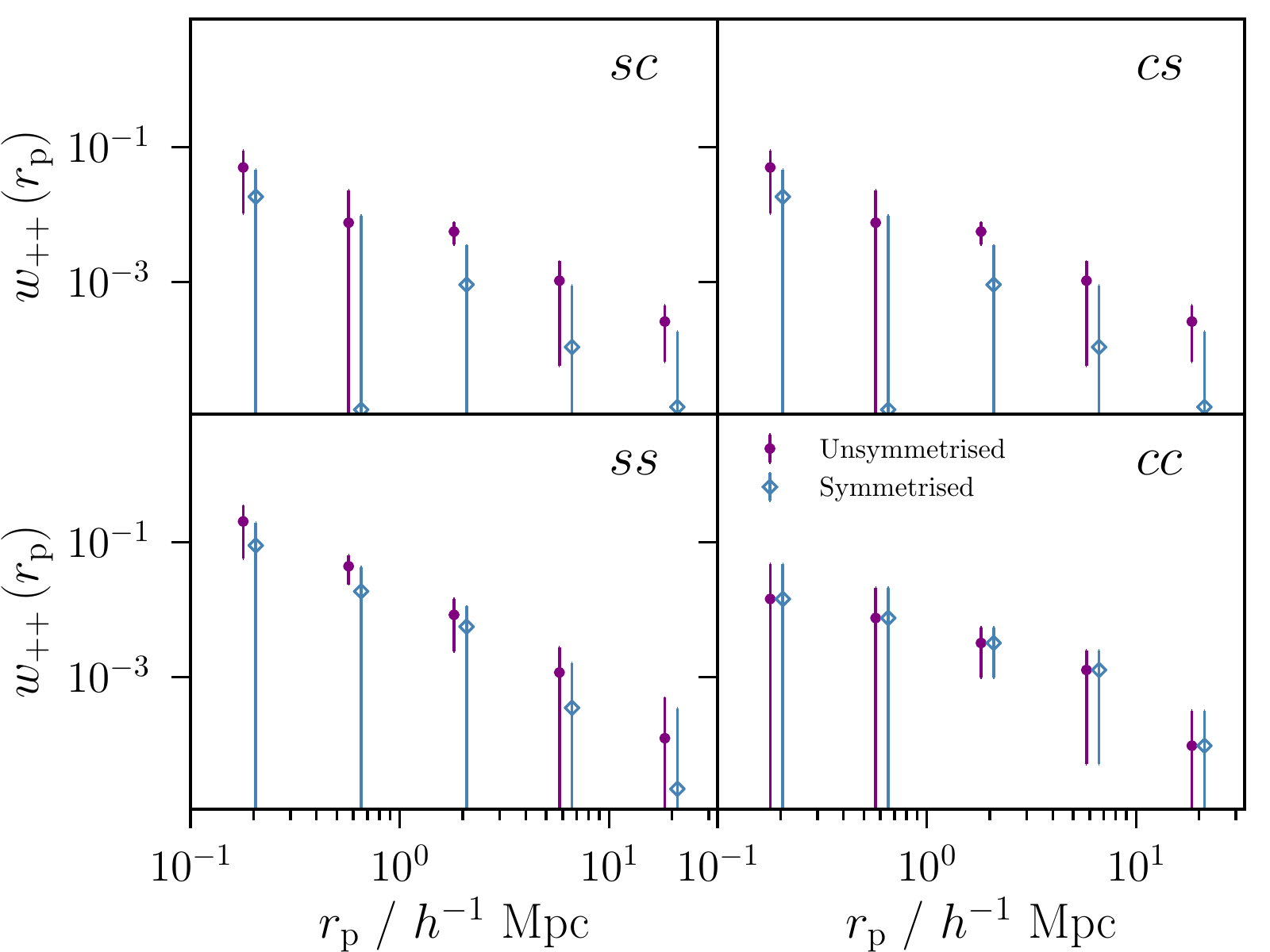}
\caption{Projected alignment correlation functions, as measured within the \mb~simulation volume
before and after halo symmetrisation. The two panels show (left) galaxy-shape and (right) shape-shape
measurements as a function of comoving separation perpendicular to the line of sight. 
Each of the four sub-panels shows a different combination of central and satellite samples.
In each case, the purple dots show the fiducial unsymmetrised measurements and
the open blue diamonds show the result of symmetrisation by rotating satellites within each halo
about the dark matter potential minimum. 
 }\label{results:main:wgp_wpp_cssplit}
\end{figure}

In this appendix we present some additional correlation functions;
these were not included in the main body of the paper due to the low
signal-to-noise of the measurements and qualitative agreement with
their (simpler) ED and EE counterparts.
They are shown in Figure \ref{results:main:wgp_wpp_cssplit} for completeness.
\newpage

As in the EE case, $w_{++}$ is symmetric, and as such the two upper panels 
of the bottom section of this figure are identical by construction.
Likewise, the purple and blue points in the $cc$ (bottom right) panels are
numerically identical (since symmetrisation leaves the positions and shapes of
central galaxies unchanged). 

\section{An Algorithm for Isotropising Three Dimensional Positions}\label{app:rotation_matrix}

The task of isotropising objects in three dimensional space, though apparently 
trivial at first glance, is very easy to mishandle. Beyond the mathematics of the per-object rotation itself, 
one must be careful that the rotation itself does not imprint some preferred direction in the
resulting distribution of object positions.
Defining a random rotation axis direction $\phi, \theta$ and a random rotation about that axis
$\alpha$, will leave one with a non-isotropic distribution of points
(since the process favours small rotation angles).
This is true even if $\phi$ and $\theta$ are correctly defined 
$\theta = 2 \pi u_2$ and $\phi = \arccos (2 u_1 - 1)$, where $u_1$ and $u_2$
are random draws from uniform distributions $u_i \in [0, 1]$.
In the following we present a recipe for transforming a distribution of points with some
preferred direction into an isotropic distribution about a centroid position 
$\mathbf{r}_\mathrm{cent}$.

As described in the main text of this paper, we first define a new position 
$\mathbf{r}' = (R, \phi', \theta')$,
where $\phi'$ and $\theta'$ are drawn from the distribution $U(0, 2\pi)$.
From these two positions, prior to and post rotation, one can calculate an orthogonal
three vector 
\begin{equation}
\mathbf{V} = \mathbf{r} \times \mathbf{r}' / N,
\end{equation}
\noindent
where $N = |\mathbf{r} \times \mathbf{r}'|$ is a normalisation constant.
This defines a axis, about which a rotation can be defined to carry a point from
$\mathbf{r}$ to position $\mathbf{r}'$.
That angle in the two dimensional plane defined by 
$\mathbf{r}$ and $\mathbf{r}'$ is given by

\begin{equation}
\alpha = \arccos \left ( \frac{\mathbf{r} \cdot \mathbf{r}'}{|\mathbf{r}| |\mathbf{r}'|} \right ).
\end{equation}

\newpage
\noindent
with the vector $\mathbf{V}=(v_x,v_y,v_z)$ and angle $\alpha$
we have the ingredients necessary to construct a rotation matrix.
In three dimensions this is given by:

\begin{widetext}
\begin{equation}
\mathbf{R}_\theta =
\left (
\begin{matrix}
\cos \alpha  + u_x u_x (1-\cos \alpha ) \;\;\;\;\; 
u_x u_y (1-\cos \alpha) - u_z \sin \alpha \;\;\;\;\; 
u_x u_z(1-\cos \alpha) + u_y \sin \alpha \\
u_y u_x (1-\cos \alpha) + u_z \sin \alpha \;\;\;\;\; 
\cos \alpha  + u_y u_y (1-\cos \alpha ) \;\;\;\;\; 
u_y u_z (1-\cos \alpha) - u_x \sin \alpha \\
u_z u_x (1-\cos \alpha) - u_y \sin \alpha  \;\;\;\;\; 
u_z u_y (1-cos \alpha) + u_x \sin \alpha  \;\;\;\;\; 
\cos \alpha + u_z u_z (1- \cos \alpha )  \\
\end{matrix}
\right )
\end{equation}
\end{widetext}

\noindent
This rotation matrix can then be applied identically to rotate the
shape vectors of a galaxy, such that the orientation relative to
the centre of rotation is maintained: $\mathbf{a}' = \mathbf{R}_\theta \cdot \mathbf{a}$.
An implementation of this algorithm and simple numerical examples of its application
can be found at 
\url{https://github.com/McWilliamsCenter/ia\_modelling\_public}.  

\begin{figure}
\includegraphics[width=\columnwidth]{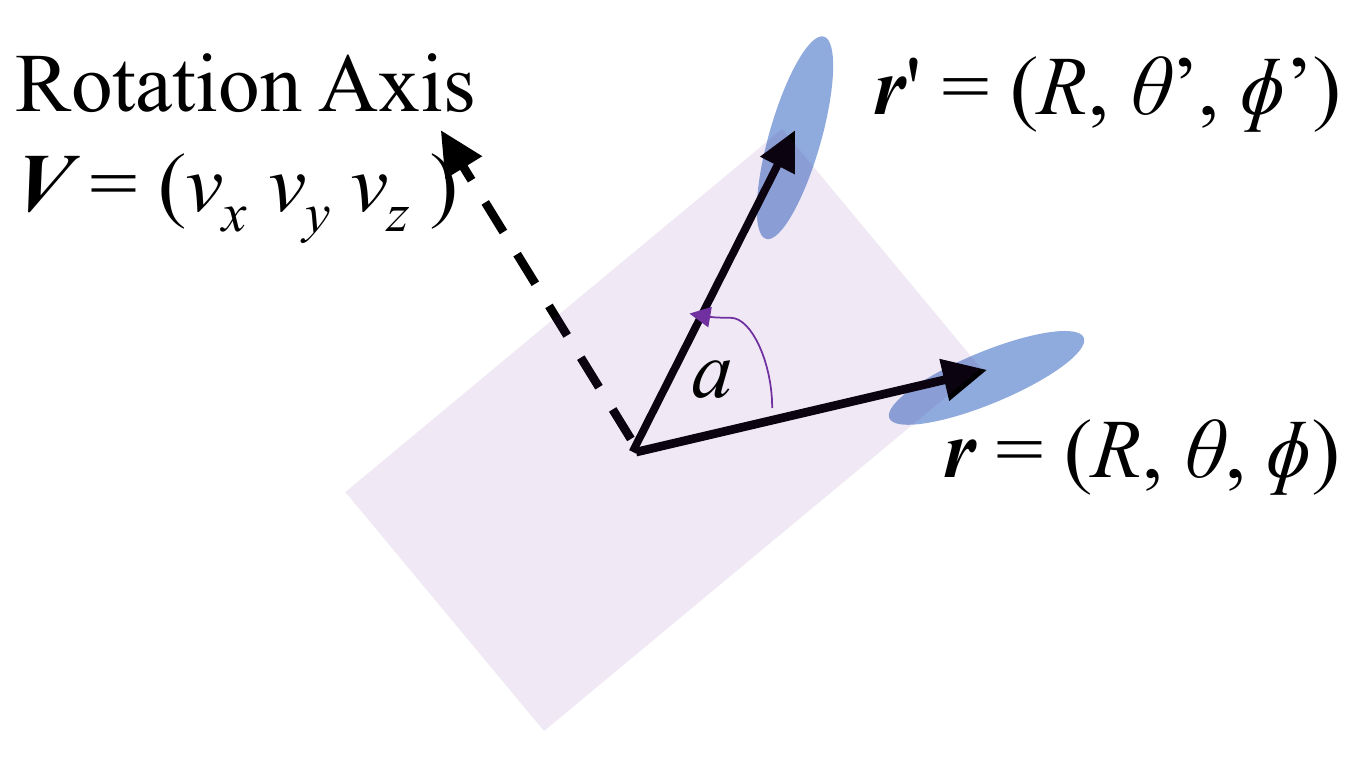}
\caption{Schematic diagram illustrating the symmetrisation algorithm described in Appendix \ref{app:rotation_matrix},
as applied to a single galaxy.
The galaxy with initial position $\mathbf{r}$ (the right-most blue oval) is
assigned a new position $\mathbf{r}'$ at the same distance from the halo centre.
The two position vectors $\mathbf{r}$ and $\mathbf{r}'$ define a rotation angle $\alpha$
in their two dimensional plane (shown in purple) and an orthogonal rotation axis $\mathbf{V}$.
These in turn can be used to calculate the rotation matrix that describes the 
mapping between the old and the new positions.}
\end{figure}

\section{Impact of Finite Simulation Volume}\label{app:truncation}

\begin{figure}
\includegraphics[width=\columnwidth]{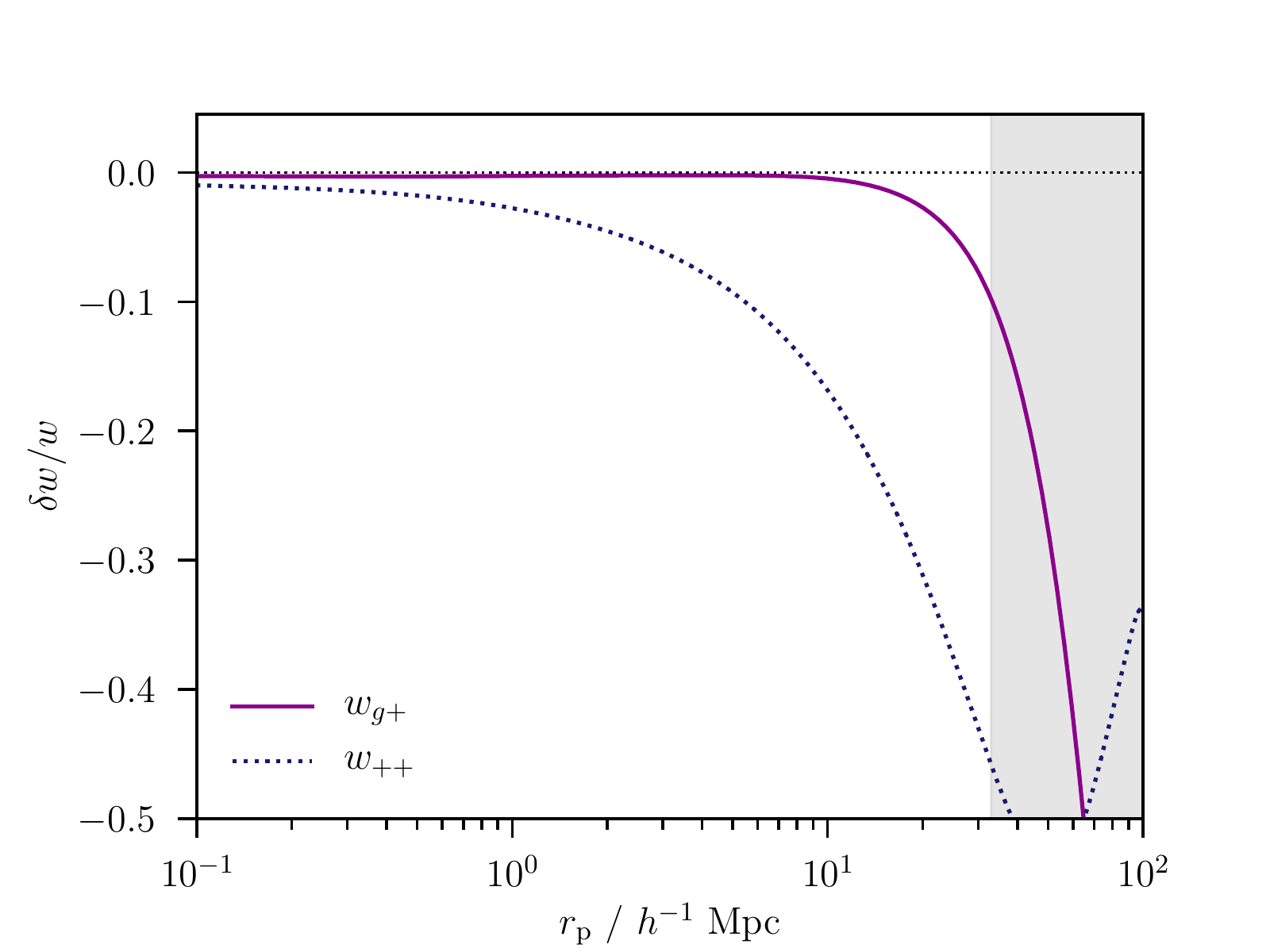}
\includegraphics[width=\columnwidth]{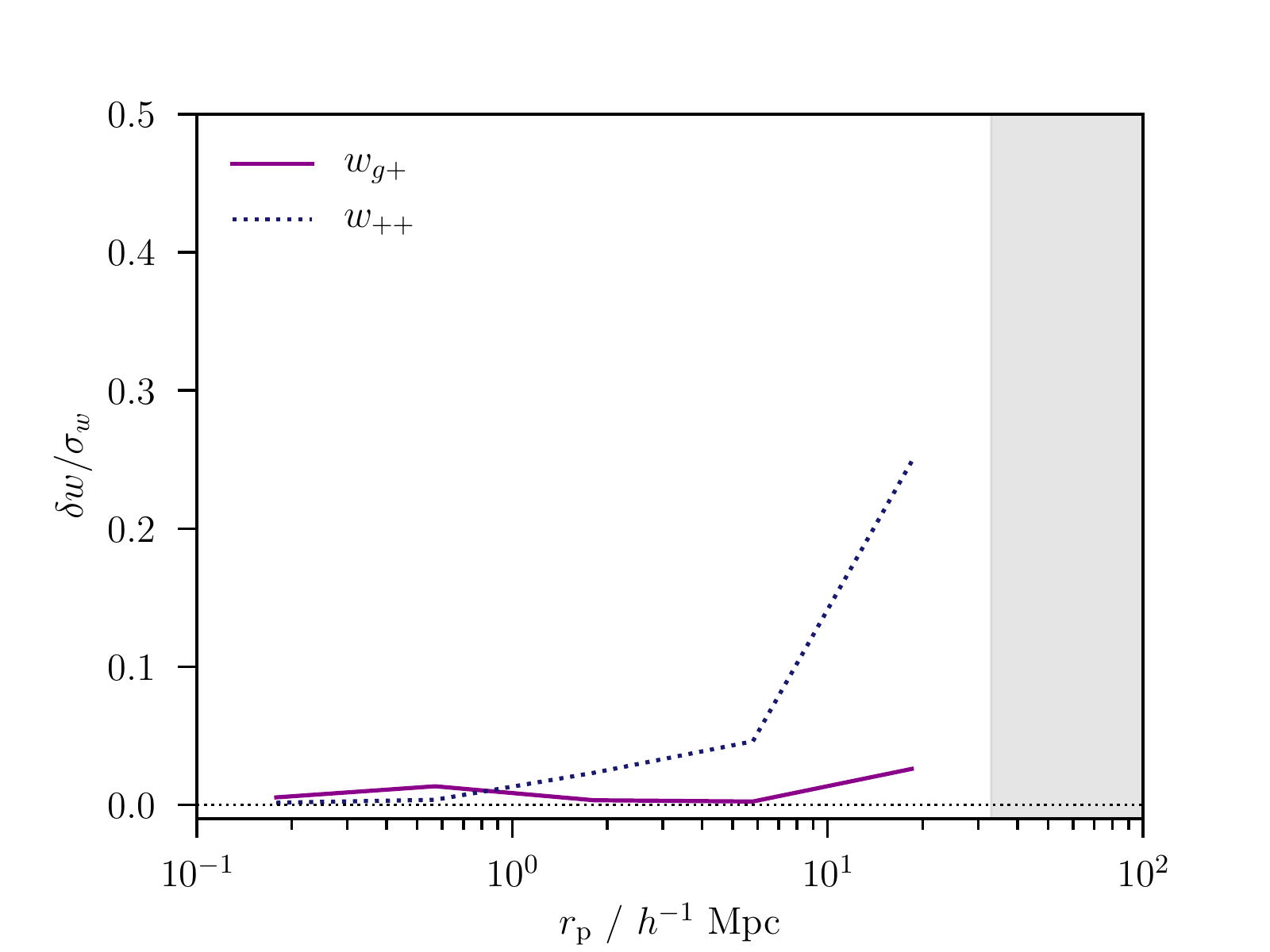}
\caption{\emph{Upper}: Fractional change in the projected correlation functions due to truncation
at $k_{\rm min} = 2 \pi / 100 h^{-1}$ Mpc. The grey shaded region shows the scales discarded in this
analysis. The sign
convention here is such that negative values translate into a suppression of power
in the truncated measurement.
\emph{Lower}: The change as a fraction of the jackknife $1\sigma$ error on each correlation
function, as measured from \mb. }\label{fig:pk_truncation}
\end{figure}

It is widely recognised in the literature that for a 
cosmological simulation of finite volume, measurements
on certain scales are unreliable. 
In general, the imposition of a sharp cutoff at the edges
of the box is equivalent to systematically cutting out
small-$k$ modes; on the largest scales, then, we expect
to see a deficit in power in the measured two-point
functions. 
In this work, in order to mitigate the impact of this effect,
we adopt a maximum scale of $r_\mathrm{p}=33 h^{-1}$ Mpc,
or a third of the \mb~box length. 
We test the validity of this choice as follows:
theory IA and galaxy clustering power spectra are first
generated
using \blockfont{CAMB} at a fiducial cosmology. We then
truncate those power spectra at a given $k_\mathrm{min}$, before 
propagating them through the appropriate Bessel integrals
to given theory predictions for $w_{g+}$, $w_{++}$ and $w_{gg}$.
The difference between these theory correlation functions 
with a $k_\mathrm{min}$ corresponding to the simulation
box size, and those extending to $k=10^{-5} h^{-1}$ Mpc
then gives us an estimate for the impact of this truncation
on different projected scales.  
The results for the lowest redshift snapshot 
are shown in Figure \ref{fig:pk_truncation}.

\noindent
The three coloured lines here show the three different types of 
correlation function. Naturally, given that they have different Bessel kernels,
we expect the scales on which the truncation impacts to differ slightly.
As one might expect, the impact of the missing $k$ modes is to suppress  
the correlation functions, and the impact is greatest on the largest scales.
Strikingly, due to the Bessel kernels used in converting the $++$
correlation into physical space, the impact is relatively large even on
scales $\sim$ a few Mpc.
The impact on $w_{g+}$ appears to be significantly smaller,
at the level of one percent or less on scales $r_\mathrm{p}<33 h^{-1}$ Mpc.

That the impact is clearly non-vanishing is not greatly surprising,
given the relatively small box size of \mb. It is, however, reassuring
that the suppression of the both IA correlation functions is comfortably 
less than the $1\sigma$ error bar on all scales considered.

\end{document}